\documentclass{aa}
\usepackage{natbib}
\usepackage{lscape} %landscape
\usepackage{graphicx}
\usepackage{txfonts}
\usepackage{hyperref}

\begin{document}

\title{Machine learning in APOGEE\thanks{Tables \ref{tab:assign}, \ref{tab:mean_spectra} and \ref{tab:std_spectra} are only available in electronic form
at the CDS via anonymous ftp to cdsarc.u-strasbg.fr (130.79.128.5) or via \href{http://cdsweb.u-strasbg.fr/cgi-bin/qcat?J/A+A/}{http://cdsweb.u-strasbg.fr/cgi-bin/qcat?J/A+A/}}}

   \subtitle{Unsupervised spectral classification with $K$-means}

        \author{
                                Rafael Garcia-Dias\inst{1,2}
                                \thanks{E-mail: rafaelagd@gmail.com},
                        \and
                                Carlos Allende Prieto \inst{1,2}
                        \and
                                Jorge S{\'a}nchez Almeida\inst{1,2}
                        \and
                                Ignacio Ordov{\'a}s-Pascual\inst{3}}

        % List of institutions
        \institute{Instituto de Astrof\'{i}sica de Canarias,  E-38200 La Laguna,
                           Tenerife, Spain\\
                \and
                           Departamento de astrof\'{i}sica, Universidad de La Laguna,
                           Tenerife, Spain\\
                \and
                   Instituto de F\'{i}sica de Cantabria (CSIC-UC), E-39005,
                   Santander, Spain\\}

   \date{Received xxx xx, xxxx; accepted xxx xx, xxxx}

  \abstract
   {The volume of data generated by astronomical surveys is growing rapidly. Traditional analysis techniques in spectroscopy either demand intensive human interaction or are computationally expensive. In this scenario, machine learning, and unsupervised clustering algorithms in particular, offer interesting alternatives. The Apache Point Observatory Galactic Evolution Experiment (APOGEE) offers a vast data set of near-infrared stellar spectra, which is perfect for testing such alternatives.}
   {Our research applies an unsupervised classification scheme based on $K$-means to the massive APOGEE data set. We explore whether the data are amenable to classification into discrete classes.}
   {We apply the $K$-means algorithm to 153,847 high resolution spectra ($R\approx22,500$). We discuss the main virtues and weaknesses of the algorithm, as well as our choice of parameters.}
   {We show that a classification based on normalised spectra captures the variations in stellar atmospheric parameters, chemical abundances, and rotational velocity, among other factors. The algorithm is able to separate the bulge and halo populations, and distinguish dwarfs, sub-giants, RC, and RGB stars. However, a discrete classification in flux space does not result in a neat organisation in the parameters' space.  Furthermore, the lack of obvious groups in flux space causes the results to be fairly sensitive to the initialisation, and disrupts the efficiency of commonly-used methods to select the optimal number of clusters. Our classification is publicly available, including extensive online material associated with the APOGEE Data Release 12 (DR12).}
   {Our description of the APOGEE database can help greatly with the identification of specific types of targets for various applications. We find a lack of obvious groups in flux space, and identify limitations of the $K$-means algorithm in dealing with this kind of data.}

        \keywords{methods: data analysis --
                          methods: numerical --
                          catalogues --
                          surveys --
                          techniques: spectroscopic --
                          Galaxy: stellar content
                      }

        \newcommand{\Teff}{$T_\mathrm{eff}$}
        \newcommand{\LOGG}{$\log g $}
        \newcommand{\MH}{[M/H]}
        \newcommand{\CM}{[C/M]}
        \newcommand{\NM}{[N/M]}
        \newcommand{\aM}{[$\alpha$/M]}
        \newcommand{\Al}{[Al/H]}
        \newcommand{\C}{[C/H]}
        \newcommand{\Na}{[Na/H]}
        \newcommand{\N}{[N/H]}
        \newcommand{\V}{[V/H]}
        \newcommand{\Ca}{[Ca/H]}
        \newcommand{\Mg}{[Mg/H]}
        \newcommand{\Ox}{[O/H]}
        \newcommand{\Si}{[Si/H]}
        \newcommand{\Su}{[S/H]}
        \newcommand{\Ti}{[Ti/H]}
        \newcommand{\Fe}{[Fe/H]}
        \newcommand{\K}{[K/H]}
        \newcommand{\Mn}{[Mn/H]}
        \newcommand{\mTeff}{T_\mathrm{eff}}
        \newcommand{\mLOGG}{\log g}
        \newcommand{\mMH}{\mathrm{[M/H]}}
        \newcommand{\mCM}{\mathrm{[C/M]}}
        \newcommand{\mNM}{\mathrm{[N/M]}}
        \newcommand{\maM}{[\alpha/\mathrm{M}]}
        \newcommand{\mAl}{\mathrm{[Al/H]}}
        \newcommand{\mMg}{\mathrm{[Mg/H]}}
        \newcommand{\mFe}{\mathrm{[Fe/H]}}
        \newcommand{\mSi}{\mathrm{[Si/H]}}

  \maketitle

%  \def\addOneNestingLevelStartLink{%
%    \gdef\Hy@StartlinkName##1##2{%
%      \sbox0{\Hy@StartlinkNameOrig{##1}{##2}}\usebox0
%      \global\let\Hy@StartlinkName\Hy@StartlinkNameOrig%
%    }%
%  }
%  \def\addOneNestingLevelEndLink{%
%    \gdef\pdfendlink{%
%      \sbox0{\pdfendlinkOrig}\usebox0%
%      \global\let\pdfendlink\pdfendlinkOrig%
%    }%
%  }
%  \let\Hy@StartlinkNameOrig\Hy@StartlinkName
%  \let\pdfendlinkOrig\pdfendlink
%  \else
%  \let\addOneNestingLevelStartLink\relax
%  \let\addOneNestingLevelEndLink\relax
%  \fi

%-------------------------------------------------------------------

\section{Introduction}

\qquad The volume of date generated by many existing and forthcoming astronomical instruments is simply too large for traditional analysis techniques. Two extreme cases are the Large Synoptic Survey Telescope (LSST; \citealt{LSST14}) and the Gaia mission \citep{Gaia15}.

Optimal use of modern astronomical instrumentation requires open and efficient access to the resulting observations. Such access is provided by large and well-organised databases, (e.g. the Hubble Space Telescope (HST), Gaia, or the Sloan Digital Sky Survey archives). As happens with data reduction, the exploitation of these vast data sets cannot be made using traditional tools (see e.g. the discussion by \citealt{Bailer-Jones02}). Classification is the  first step in any automated analysis. It can be used to identify and discard noisy data or to group like objects to follow a common interpretation pipeline. It is certainly needed when exploring new types of data and it is also an invaluable tool to identify rare objects, usually the most telling from a scientific point of view.

Numerous works have been done that explore the performance of the automatic MK classification of spectra (see e.g. \citealt{Bailer-Jones1998, Singh, Bailer-Jones2001, Rodriguez2004, Giridhar, Manteiga2009, Navarro2012}). The main approach followed in these works was to apply supervised learning training using labelled data. The unsupervised approach was also applied in works like \cite{vanderplas}, \cite{Daniel2011} and \cite{itamar}. In this work we focus on an unsupervised approach that does not aim to reproduce the MK classification.

Among all unsupervised classification methods, $K$-means (e.g. \citealt{macqueen67}, \citealt{everitt92}, \citealt{50years}) is a flexible clustering algorithm that has being extensively used in the literature. We have already employed $K$-means in several applications, including the identification of similar targets to average and reduce noise \citep{sanchez09}, the classification of one million galaxy spectra representative of the local universe \citep{sanchez10},
a systematic search for rare extremely metal-poor galaxies \citep{Morales-Luis11, sanchez16}, and the classification of the large stellar spectra data set available from the Sloan Digital Sky Survey, in particular data from the Sloan Extension for Galactic Understanding and Exploration (SEGUE; \citealt{sanchez13}). In this work we show the virtues and limitations of $K$-means in this context, making a first step in the search for alternatives. This work is also the first to perform classification on APOGEE.

In this paper, we turn our attention to high-resolution stellar spectroscopy, and in particular to the Apache Point Galactic Evolution Experiment (APOGEE), part of the Sloan Digital Sky Survey (\citealt{eisenstein01}; \citealt{blanton17}). We examine whether or not the massive APOGEE data set is amenable to a sensible unsupervised classification scheme based on $K$-means. Section 2 describes the APOGEE spectroscopic data in detail, including the APOGEE Stellar Parameters and Chemical Abundances Pipeline (ASPCAP; \citealt{garcia-perez16}). Section 3 is devoted to the details of the classification algorithm, and Section 4 describes its application to the APOGEE data, preceded by numerical experiments based on simulated data. Section 5 discusses the main results, and Section 6 summarises the conclusions.

\section{Data set}

\qquad APOGEE makes use of a novel fibre-fed high-resolution $H-$band spectrograph to obtain simultaneously up to 300 stellar spectra \citep{wilson10, wilson12}. The APOGEE spectrograph is usually coupled to the Sloan Foundation 2.5-m telescope at the Apache Point Observatory, but has also been linked to the New Mexico State University 1-m telescope at the same location. The project has already obtained spectra for more than 300,000 stars in the Milky Way, focusing on red giants and therefore covering a broad range of galactocentric distances. Working in the near-IR, between 1.5 and 1.7 $\mu$m, APOGEE can access regions of the Galaxy heavily obscured by dust, such as the mid-plane of the Galaxy, or the bulge and the Galactic bar near the centre \citep{majewski16}.

APOGEE spectra are processed by a custom-made data pipeline that extracts the spectra, calibrates them, and  corrects telluric absorption and sky emission lines before measuring radial velocities \citep{Nidever15}. The pipeline ASPCAP performs an automated analysis based on model atmospheres, delivering atmospheric parameters and chemical abundances for the majority of the observed stars.\footnote{Approximately 93 per cent of the spectra in APOGEE DR12 have uncalibrated atmospheric parameters, \MH, \aM, \NM\, and \CM\, determined. The calibrated values are defined to $\approx$ 63 per cent of the spectra.} The atmospheric model grid boundaries in effective temperature are 3500 and 8000 K, in \LOGG the boundaries are 0 and 5, and in \MH\, they are -2.5 and 0.5 dex. More details about the grid can be found in Table 2 of \citealt{Holtzman15}.

The APOGEE pipelines are in constant evolution and the data set continues to grow. In this work, we have adopted the data made publicly available in DR12\footnote{The catalogue is available at \href{http://data.sdss3.org/sas/dr12/apogee/spectro/redux/r5/allStar-v603.fits}{allstar file}.}, the final data release from SDSS-III (\citealt{Alam15}; \citealt{Holtzman15}). This data set includes over 150,000 stars observed between 2011 and 2014. The resolving power of the APOGEE data is  $R\equiv \lambda/\delta \lambda\simeq 22,500$, and the typical signal-to-noise ratio exceeds 100 per half a resolution element. In addition, we used  quality and target flags\footnote{The flags were extracted from the objects \textsc{targflags}, \textsc{starflags}, \textsc{andflags} and \textsc{aspcapflags}.
A complete description can be found  in \href{http://www.sdss.org/dr13/algorithms/bitmasks/}{bitmasks documentation}.}, and the uncalibrated parameters derived by ASPCAP\footnote{This parameters are accessible through the objects \textsc{fparam} and \textsc{felem}, see \href{https://data.sdss.org/datamodel/files/APOGEE_REDUX/APRED_VERS/APSTAR_VERS/ASPCAP_VERS/RESULTS_VERS/allStar.html}{data model documentation}.} in order to evaluate the result of the classification (Section \ref{sec:description}). Besides sky coordinates and atmospheric parameters (temperature, surface gravity, and micro turbulence), the data set includes metallicities, $\alpha$-element abundance, and individual chemical abundances for 15 elements.\footnote{Al, Ca, C, Fe, K, Mg, Mn, Na, Ni, N, O, Si, S, Ti, and V.} As described in \citealt{Holtzman15}, the DR12 results were calibrated using star clusters' data in order to eliminate abundance trends with temperature and systematic differences with the literature. Since calibrated parameters are not available for all stars in DR12, we chose to use the uncalibrated parameters and chemical abundances. This choice should not affect the interpretation of our results; we are not interested in absolute values for each object, but in relative differences among spectra with intrinsically different shapes. In addition, using the uncalibrated data we can arguably better understand ASPCAP.

\section{Classification algorithm}
\label{sec:algorithm}
\qquad Cluster analysis aims to organise a collection of objects into classes based on a similarity criterion, such that objects in the same class are more alike than objects in different classes. There is a numerous set of cluster algorithms available in the literature (e.g. \citealt{everitt92}), but in general, all involve the following main steps: (1) Feature selection, the identification of the features that better represent the objects in the data set; (2)  choosing a feature proximity indicator, the figure of merit that optimally defines the similarity between objects in the data set; (3) establishing the grouping criterion - meaning the clustering algorithm itself, and (4) cluster validation, an evaluation of the output quality.

In the feature selection phase, we excluded all pixels potentially affected by sky emission and telluric absorption. Standard $K$-means algorithms are designed such that all input objects must have the same dimensions, and therefore we have to consider the same pixels in all spectra. For the vast majority of APOGEE observations 35 fibres are devoted to observe warm stars, measuring telluric absorption, 35 fibres to observe the sky, pointing them to blank regions in the sky, and 230 fibres to acquire science spectra. To determine which are the pixels more affected by sky emission and telluric absorption, we have taken the average of the normalised sky, and the telluric spectra for all fields in APOGEE DR12, and used them to identify and exclude in our analysis all pixels for which the mean sky count is above 1 per cent of the maximum mean normalised sky count. We have also excluded all pixels at which the mean normalised telluric spectrum falls more than five per cent below the continuum.

\begin{figure}
\centering
\includegraphics[width=.48\textwidth]{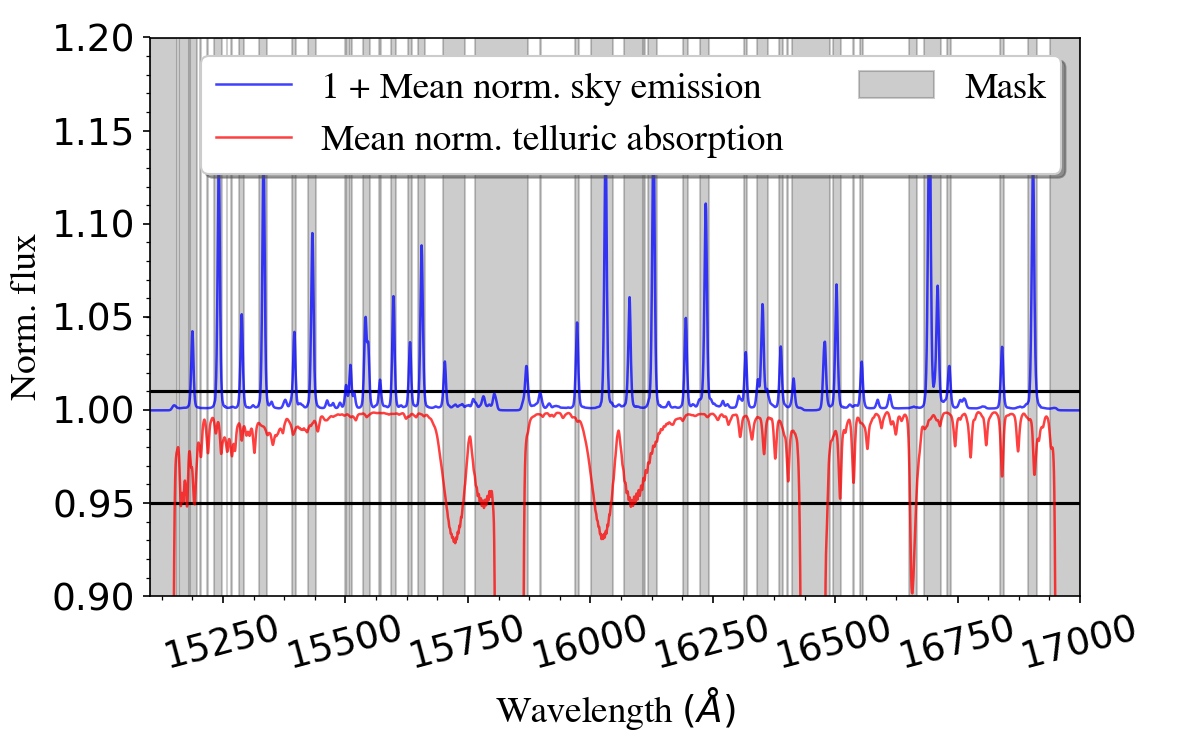}
\caption{\label{fig:Sky_Tel} Mean sky normalised emissions (blue line) and telluric absorption (red line) spectra for the 153,847 spectra in the sample. Mean sky normalised emissions fluxes are displaced by one unit to help visualisation. Black lines define the cut applied to each spectrum. Grey shades highlight the areas excluded from the $K$-means classification.}
\end{figure}

Figure \ref{fig:Sky_Tel} shows the mean sky and telluric spectra, the cuts applied, and the regions excluded from the spectra used in the $K$-means classification. In this figure we have displaced vertically the mean sky spectrum for clarity. Since stars have different heliocentric velocities, the spectra were corrected for Doppler shifts, and therefore they are affected by sky emission and telluric absorption at different wavelengths for different stars. This can be seen in Figure \ref{fig:Sky_Tel} from the width of the mean normalised telluric lines and sky emissions lines. From the $8575$ original wavelength pixels, we kept $4838$ pixels, or $56$ per cent of the APOGEE spectral coverage. All the spectra were also normalised using a fourth degree polynomial regression for each of the three chips in the APOGEE spectrograph. We have also removed values in the normalised flux higher than 1.02 (i.e. two per cent above the pseudo continuum level), setting the flux value to 1.02, avoiding any remaining problem with sky emission lines.

The chosen feature proximity metric was the Euclidean distance. That is the most straightforward possibility, since the objects to be classified are normalised spectra, which can be regarded as data points in an N-dimensional space. It also has the advantage of being easily interpreted and having a low computational cost.

The grouping criterion is the way one assigns each object to a certain cluster and is how groups are designed. For example, groups can be selected in a single partition, that is to say, all clusters are simple partitions, hierarchically equivalent samples, otherwise they would be hierarchical clusters that have a structure with clusters and sub-clusters. Furthermore, clustering is said to be hard if it assigns each object to a single cluster, in opposition to soft clustering where the objects are assigned as having a non-zero probability of belonging to more than one cluster.

In this work we explore the use of $K$-means \citep{macqueen67}, a partitional hard clustering algorithm. It is one of the most popular clustering algorithms, mainly because it is easy to implement and its computational cost scales linearly with the number of objects to be classified. The fundamental steps in $K$-means are (1) to choose the number of clusters $K$; (2) define $K$ initial cluster centres; (3) assign each object in the sample to the closest cluster; (4) recompute cluster centres as the centroid of the objects assigned to each cluster; (5) repeat steps 3 and 4 until a convergence criterion is met. Usually the convergence criterion is either a decrease of the within-cluster variance under a threshold or a minimal re-assignation between two consecutive iterations. Here we adopt the criterion of having less than one per cent of re-assignation  between two consecutive iterations.

Initialisation also can be done in different ways. The simplest is to randomly choose objects in the entire sample, but if the data set has an over-abundance of a particular kind of object, the clusters would over-sample those objects. In order to avoid this, we initialise in an iterative fashion; we carry out a couple of $K$-means iterations with $K = 10$, randomly choose an object in the most abundant cluster as  initial centre, discard all objects in this cluster and repeat the process until the desired number of initial cluster centres is reached. During the process, if more than 95 per cent of the objects are discarded, we select the remaining cluster centres randomly in the whole sample. In this work we have translated the algorithm presented by \citealt{sanchez10} from IDL\footnote{\href{http://www.harrisgeospatial.com/ProductsandSolutions/GeospatialProducts/IDL.aspx}{http://www.harrisgeospatial.com/ProductsandSolutions/GeospatialProducts/}} to Python \footnote{\href{www.python.org}{www.python.org}}. Besides serial and parallel performance optimisation, no major modifications were made. Using Python we achieved a simpler and faster code, which also has the advantage of being  available in an open source platform.

We have compared our results with the results using scipy\footnote{\href{www.scipy.org}{www.scipy.org}} and scikit learn\footnote{\href{http://scikit-learn.org}{http://scikit-learn.org}} algorithms. The results are qualitatively equivalent. The advantage of using our own code is that we are coherent with previous works in the literature  \citep{sanchez09, sanchez10, Morales-Luis11, sanchez13, sanchez16}.

A major drawback in any clustering classification is that the algorithm will always return partitions regardless of the existence of clusters or not. In addition, the algorithm does not guarantee convergence to a global solution. Moreover, many implementations require choosing the number of clusters. In order to overcome these problems, or even just to find out how serious they are, we apply cluster validation techniques. We are interested in verifying whether the data have intrinsic clusters, whether there is an optimal number of clusters, and whether the clusters derived in flux space exist in parameters' space.

\subsection{Choosing the number of clusters}
\label{subsec:find_K}
\begin{figure}
                \centering
                \includegraphics[width=.48\textwidth]{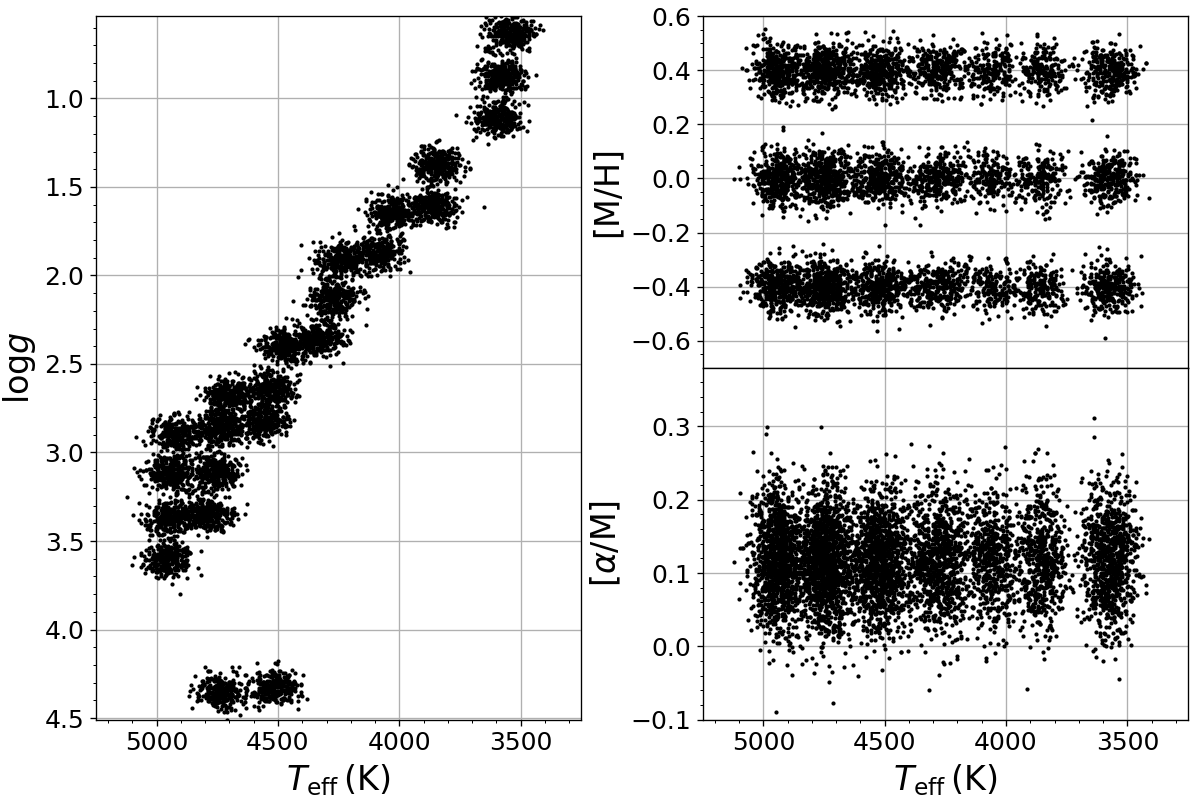}
                \caption{\label{fig:synt_parms} Atmospheric parameters for the synthetic data set. Left panel shows effective temperature and surface gravity for the synthetic spectra. The right top panel presents the projection of the clusters in the $\mTeff - \mMH$ plane, while the right bottom panel shows the plane $\mTeff - \maM$.}
\end{figure}

Choosing the optimal number of clusters is a critical step in $K$-means classification. There is no universal criterion to do it, although many heuristic criteria have being  developed over the last fifty years \citep{tibshirani2001estimating}. In an attempt to select the most suitable criteria for our problem we built a testbed data set with 6900 synthetic spectra spread over 69 well-defined clusters in surface gravity (\LOGG, in cgs units), temperature (\Teff), $\alpha$ abundance (\aM), and metallicity (\MH), as shown in Figure \ref{fig:synt_parms}. The brackets, for two given elements X and Y, [X/Y] is defined as: $$
[\mathrm{X}/\mathrm{Y}]= \log_{10}{\left({\frac{N_{\mathrm{X}}}{N_{\mathrm{Y}}}}\right)_{\mathrm{star}}}-\log_{10}{\left({\frac{N_{\mathrm{X}}}{N_{\mathrm{Y}}}}\right)_{\odot}},
$$ where $N_{X}$ and $N_{Y}$ are the number of X and Y nuclei per unit volume, respectively. Metallicity is a measure of all the chemical elements heavier than He, assuming they vary in the same proportions with respect to the solar values. Analogously, \aM\, is a measure of all $\alpha$-elements (O, Ne, Mg, Si, S, Ar, Ca, and Ti) assuming they vary in union. The centres of the clusters were chosen based on the most dense regions in the HR diagram of the empirical data set with parameters from DR12. The parameters for each spectrum were randomly chosen around each cluster centre, following a normal distribution with $\sigma_{\mTeff} = 50 $K and $\sigma_{\mLOGG} = \sigma_{\mMH} = \sigma_{\maM} = 0.05$. The synthetic spectra were built using the code \textsc{FERRE}\footnote{\href{https://github.com/callendeprieto/ferre}{https://github.com/callendeprieto/ferre}.}, interpolating in a grid of theoretical models (\citealt{ferreII, ferreIII, zamora2015}). We use a model grid with seven parameters per spectrum, microturbulence velocity $(\xi_\mathrm{v})$, carbon abundance (\CM), nitrogen abundance (\NM), mean $\alpha-$elements abundance (\aM), metallicity (\MH), surface gravity (\LOGG), and effective temperature (\Teff). But the parameters $\xi_\mathrm{v},\, \mCM,\, \mNM$ were fixed to the mean values\footnote{$\langle \xi_\mathrm{v} \rangle = 0.169\,km\,s^{-1},\, \langle \mCM \rangle = 0.122\,,\, \langle [N/M] \rangle = 0.227.$} of the stars in the DR12 sample for all spectra. In order to explore the best-case scenario we have not added any noise to the spectra.

We applied $K$-means to the simulated data set ten times, with $K$ varying from 5 to 100. We then applied four different statistical criteria trying to recover the optimal number of clusters, knowing that the actual number is 69. We tried the KL index \citep{DIFF_ref}, the gap statistic \citep{tibshirani2001estimating}, the CH index \citep{CH_ref}, and the silhouette index \citep{silhouette}.
These indexes were selected for being the  most widely and successfully used in the literature. None of the chosen criteria was able to identify the right number of clusters, with the CH index being the only one capable of giving consistent results over different initializations, finding $K = 9 \pm 1.8$, far from the true value of 69. The other methods found a  $\sigma_{K} > 12$ over the ten different runs, while randomly selecting ten numbers in this range would result in $\sigma_{K} \approx 25$. A possible explanation for this failure is  that, despite the clusters being  well-defined in parameters' space, the classification is made in flux space, where the separation between classes seems to be more subtle.

\begin{figure}
        \centering
        \includegraphics[width=0.48\textwidth]{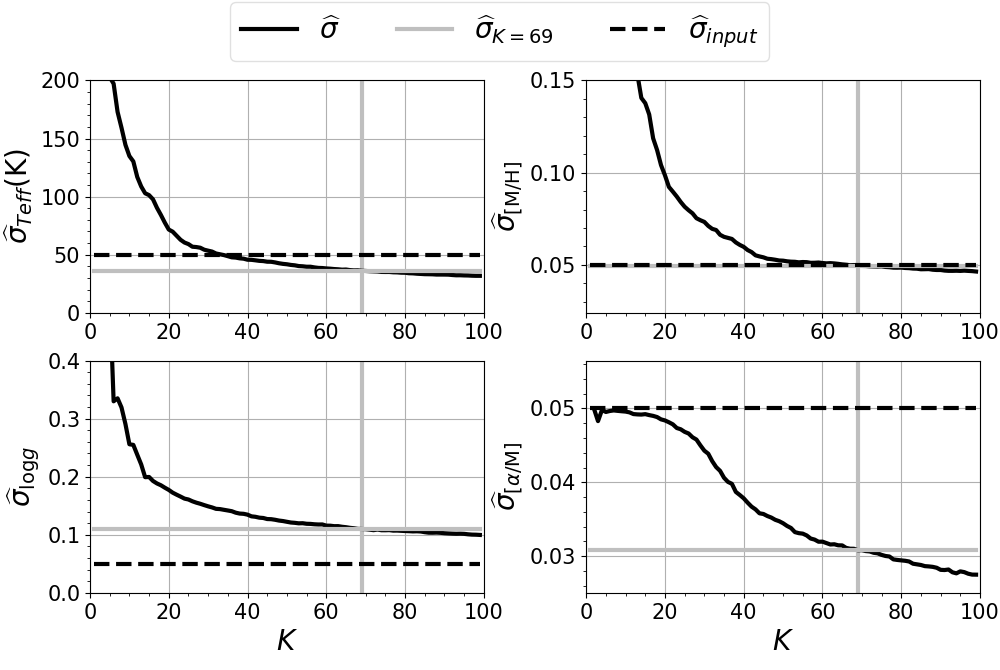}
\caption{\label{fig:mean_std_synt} Variation of median standard deviation as function of the number of clusters $K$ in the synthetic data set. The solid black lines represent the median standard deviation of the classes in a run. The solid horizontal grey lines show the median standard deviation at $K = 69$. Dashed black lines show the input standard deviation. The top left panel refers to $\alpha$ abundances \aM, the top right to metallicity \MH, bottom left to  logarithmic surface gravity \LOGG\,, and the bottom right to effective temperature \Teff.}
\end{figure}

In the absence of better criteria, we have chosen the numbers of clusters based on the within-class standard deviation of the atmospheric parameters and chemical abundances. Figure \ref{fig:mean_std_synt} shows the variation of the median $\sigma$ values for each of the four main input parameters. We use the notation $\widehat{X}$ meaning the median of $X$. It is important to use medians instead of means in order to avoid the predominance of the fewer classes which gather faulty and unusual spectra. Especially when we start to work with the observed data set, with classes having few spectra ($< 30$) and a large dispersion in atmospheric parameters and chemical abundances. We see a decrease in $\widehat{\sigma}_{X}$ as $K$ grows for all quantities. This means that dividing the spectra in flux space into more classes results also in  finer partitions in atmospheric parameters and abundances spaces. Therefore, we can choose $K$ based on a threshold value for $\sigma$. The extreme case would be to increase $K$ until having one star per class, reaching the minimum variation. However, since the computational cost scales with $K$ and we also lose generality when increasing $K$, we should choose $K$ making a compromise between accuracy, agility, and generality.

We know the $\widehat{\sigma}_{X}$ values and $K$ for the synthetic data set. Therefore we can verify how much we can trust the variance for the choice of $K$. Figure \ref{fig:mean_std_synt} shows that when $K = 69$ we have exactly the input metallicity dispersion; $\widehat{\sigma}_{\log(g)}$ is highly above the input level, while $\widehat{\sigma}_{\alpha}$ and $\widehat{\sigma}_{Teff}$ are both below the input level.
The figure also shows the slope ($|\partial \widehat{\sigma}_{X}/\partial K|$) of the curves decreases rapidly for $K \gtrsim 50$. Therefore, increasing $K$ does not produce a significant change in $\widehat{\sigma}_{\mTeff}$ and $\widehat{\sigma}_{\mLOGG}$ for $K  \gtrsim 50$. The plots also reveal a different sensitivity for each parameter.

\begin{figure}
\centering
\includegraphics[width=0.48\textwidth]{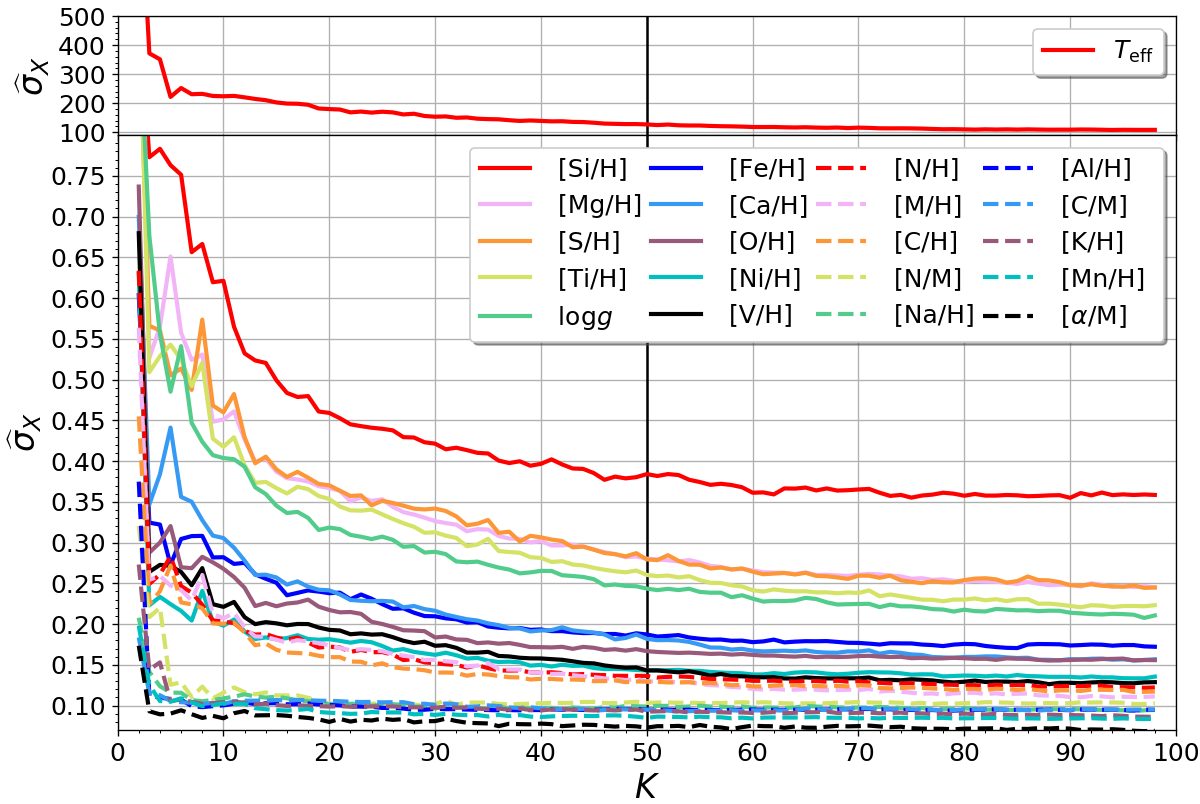}
\caption{\label{fig:mean_std_real} Variation of median standard deviation as function of the number of clusters $K$ in the real data set. Top panel refers to effective temperature \Teff, while the bottom panel to the variation of median standard deviation for the other 20 parameters available in DR12 as indicated in the legend box.}
\end{figure}

The actual APOGEE data set behaves in a similar way. Figure \ref{fig:mean_std_real} shows how $K$ affects the median of the standard deviation of \Teff, \LOGG, \MH\, and the abundances of carbon, nitrogen, and $\alpha$-elements with respect to metallicity, and the same for the abundances of the chemical elements Al, Ca, C, Fe, K, Mg, Mn, Na, Ni, N, O, Si, S, Ti, and V. From these plots we have chosen $K = 50$ as the number of clusters to be used throughout the paper, since beyond that value increasing $K$ does not reduce significantly the within-cluster parameters' dispersion.

\begin{figure*}
\begin{minipage}{\textwidth}
        \centering
        \includegraphics[width= 0.92\textwidth]{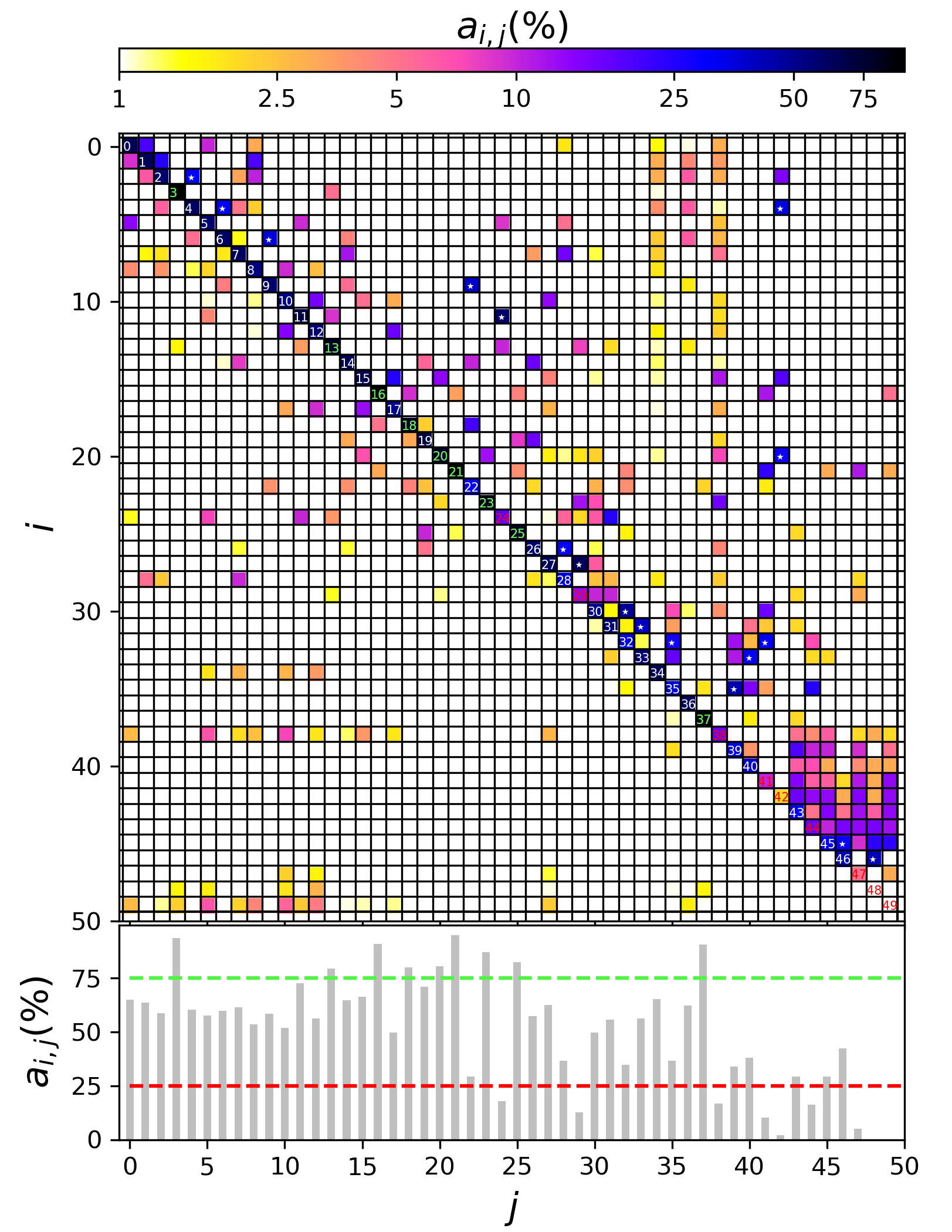}
        \caption{\label{fig:mean_confusion} Top panel shows the mean coincidence matrix comparing the chosen classification with the other 99 performed classifications. Elements on the mean diagonal represent the coincidence ratio of a  class and can be interpreted as the stability of the class. The elements in the diagonal are labelled with their corresponding class number and highlighted in green if the class has a coincidence ratio above 75 per cent or in red if the class has coincidence ratio below 25 per cent. Elements off the diagonal can be interpreted as the confusion rate between two classes. We highlight confusion rates above 25 per cent with white stars. The bottom panel presents a histogram of the coincidence ratios corresponding to the diagonal of the coincidence matrix. A green dashed line marks the 75 per cent level, while a red line marks the 25 per cent level. }
\end{minipage}
\end{figure*}

\begin{figure}
  \centering
  \includegraphics[width=0.48\textwidth]{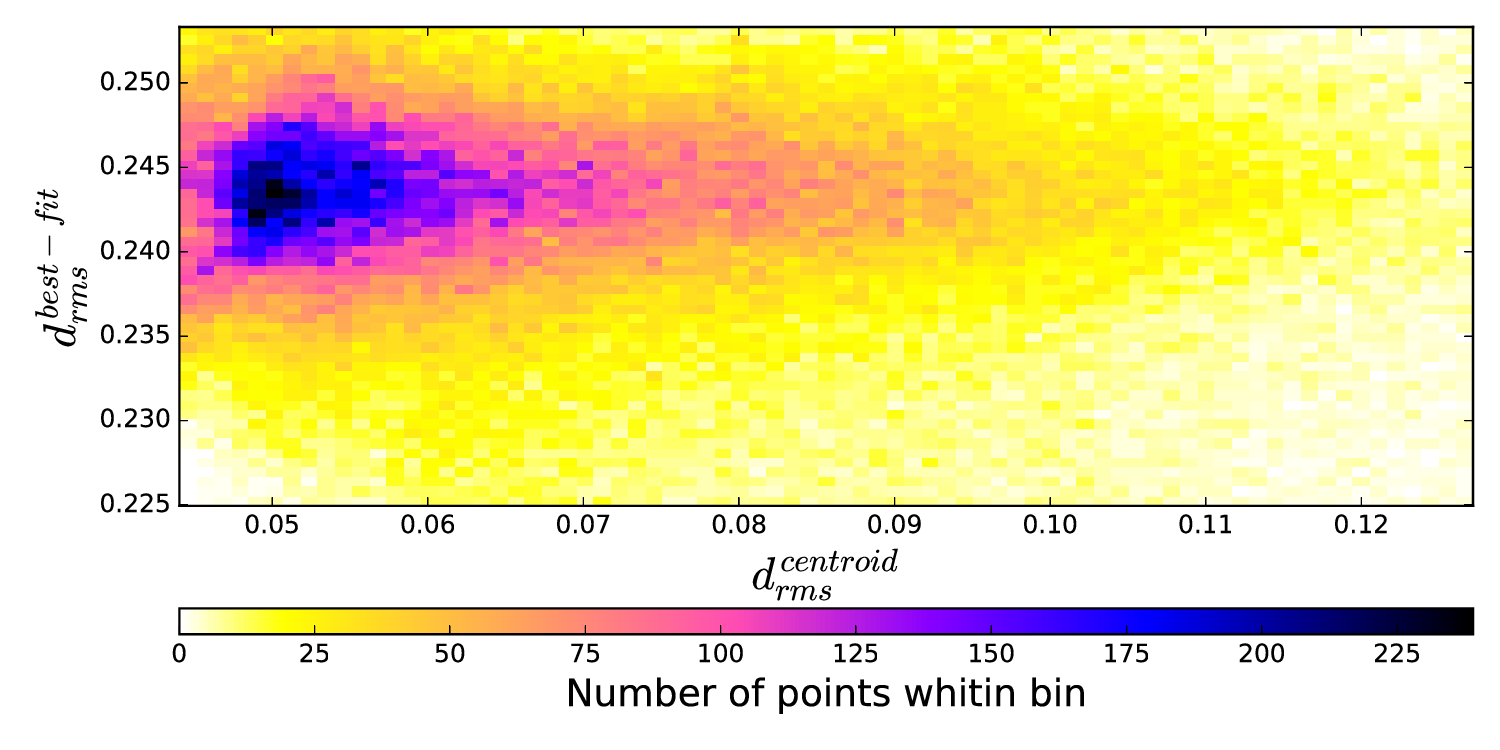}
  \caption{\label{fig:chi2} Two-dimensional histogram comparing distances from each spectra to its best fit with the distance from each spectra to its class centroid. Each pixel in the image is colour-coded according to the number of spectra in that region, as indicated by the colour bar.}
\end{figure}

\begin{figure*}
\begin{minipage}{\textwidth}
                \centering
                \includegraphics[width=0.98\textwidth]{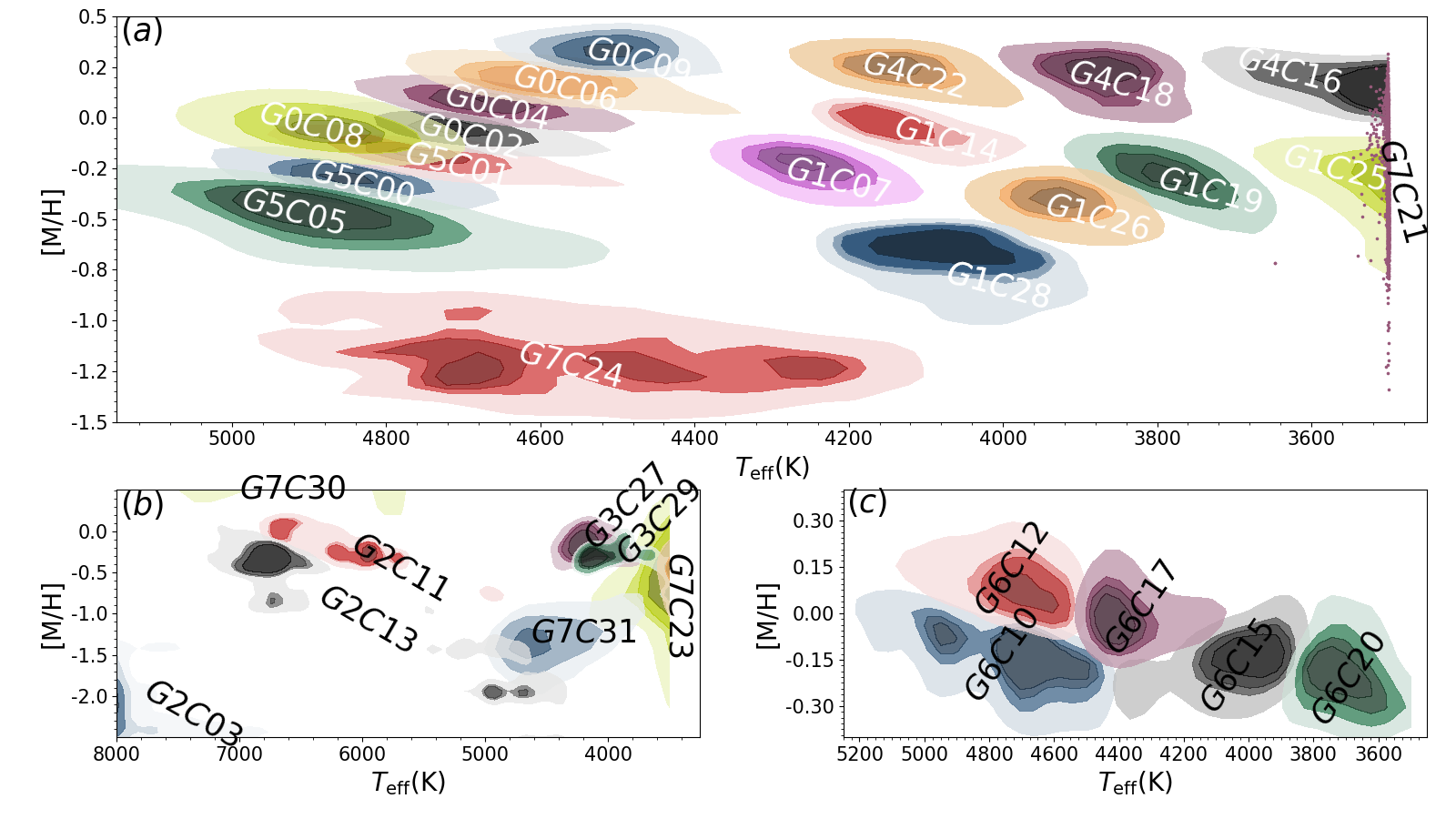}
                \caption{\label{fig:Groups_MH_Teff} Contour diagrams in the $\mTeff - \mMH$ plane. Different colours are used to distinguish different classes. Each class is represented by four colour shades; from dark to light, the shades enclose 15, 30, 45, and 68.3 per cent of the data points in the class. The groups are separated into three panels minimising the superposition of classes. Panel $(a)$ shows groups 0, 1, 4, 5, and two classes of group 7, panel $(b)$ groups 2, 3, and three classes of group 7 and panel $(c)$ shows group 6. In these panels each class is flagged with a floating label in the form G\textsc{x}C\textsc{xx}, C referring to class and G to its group. Class 21 is represented as a scatter plot, since it is too concentrated to present visible contours on this scale.}
\end{minipage}
\end{figure*}

\begin{figure}
        \centering
        \includegraphics[width=0.48\textwidth]{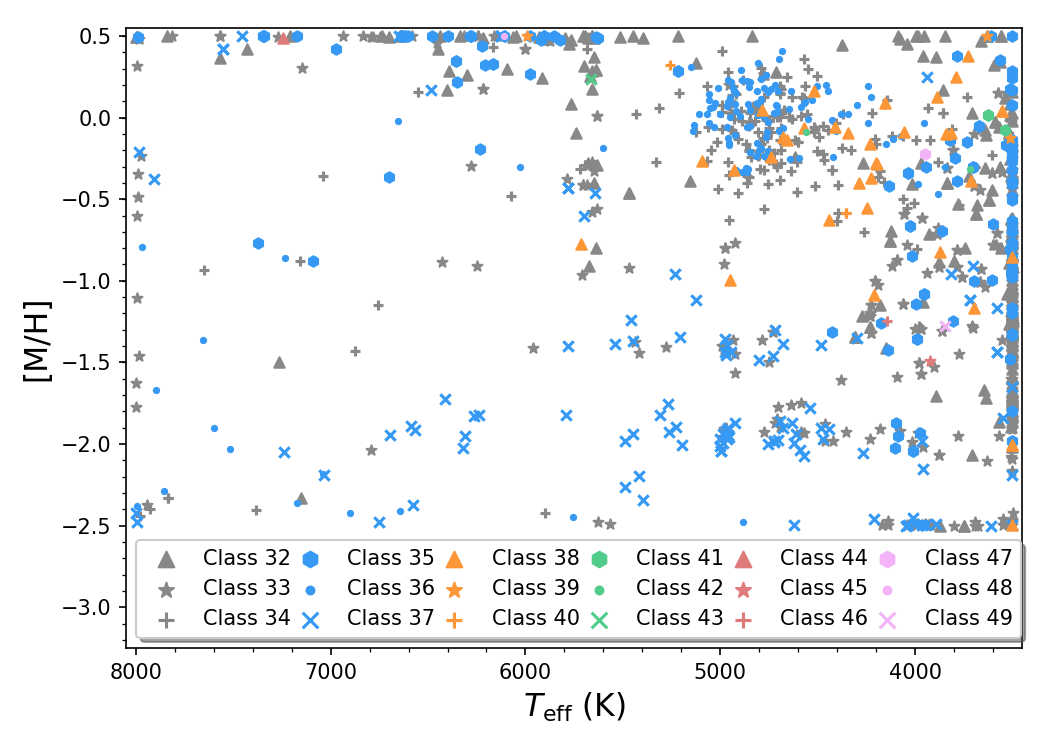}
        \caption{\label{fig:MH_Teff_G8} Scatter plot of \Teff\, against \MH\, for classes in group 8. The classes are identified as shown in the legend. The stars in this group are scattered throughout the plane.}
\end{figure}

\begin{figure*}
        \centering
        \includegraphics[width=0.98\textwidth]{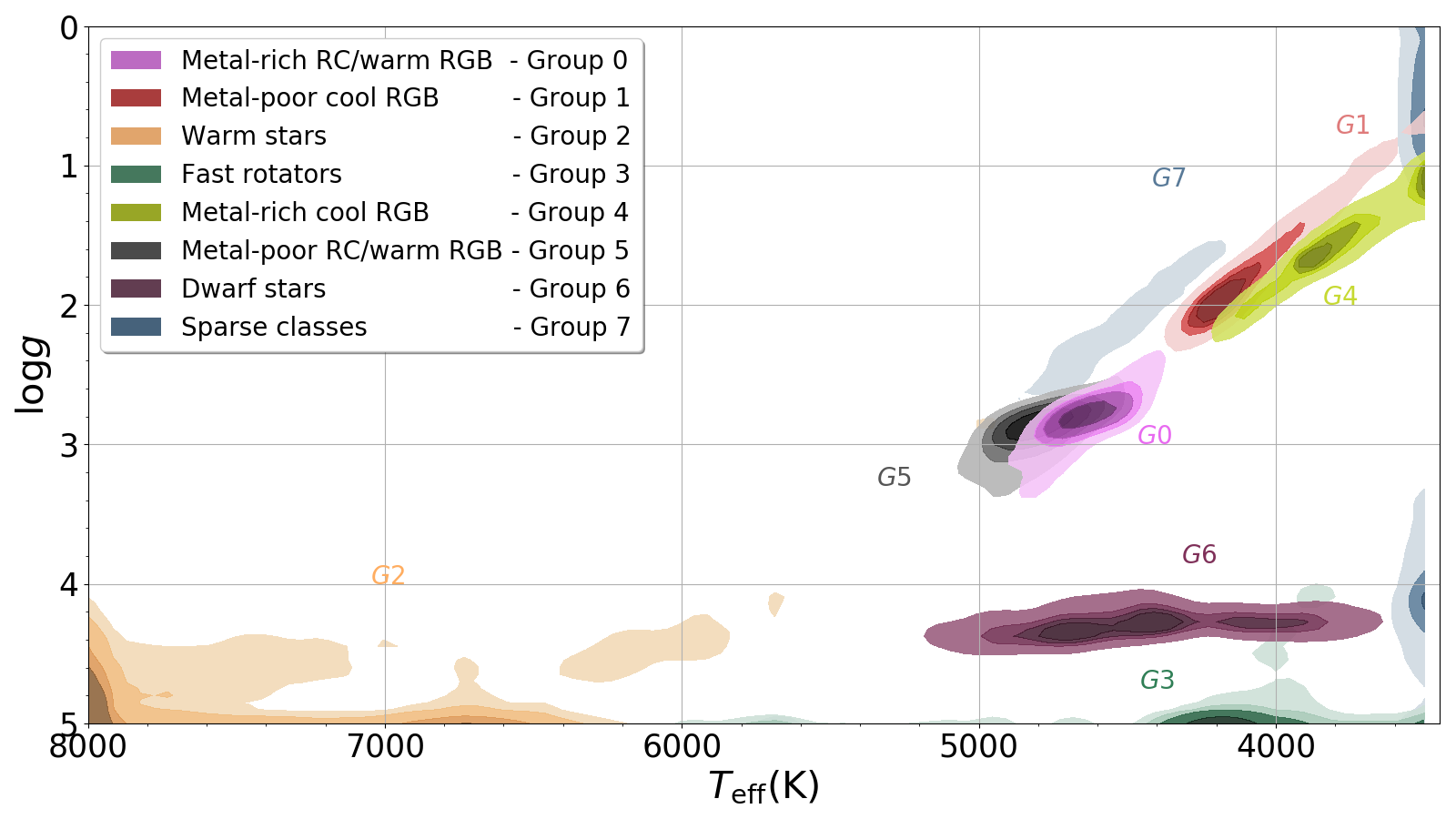}
        \caption{\label{fig:Groups_LOGG_Teff} Contour diagram for the groups in the $\mTeff - \mLOGG$ plane. Each group is represented by a different colour. Colour shades enclose, from dark to light, 15, 30, 45, and 68.3 per cent of the objects in each group.}
\end{figure*}

\begin{figure*}
\begin{minipage}{\textwidth}
        \centering
        \includegraphics[angle=0, width=0.97\textwidth] {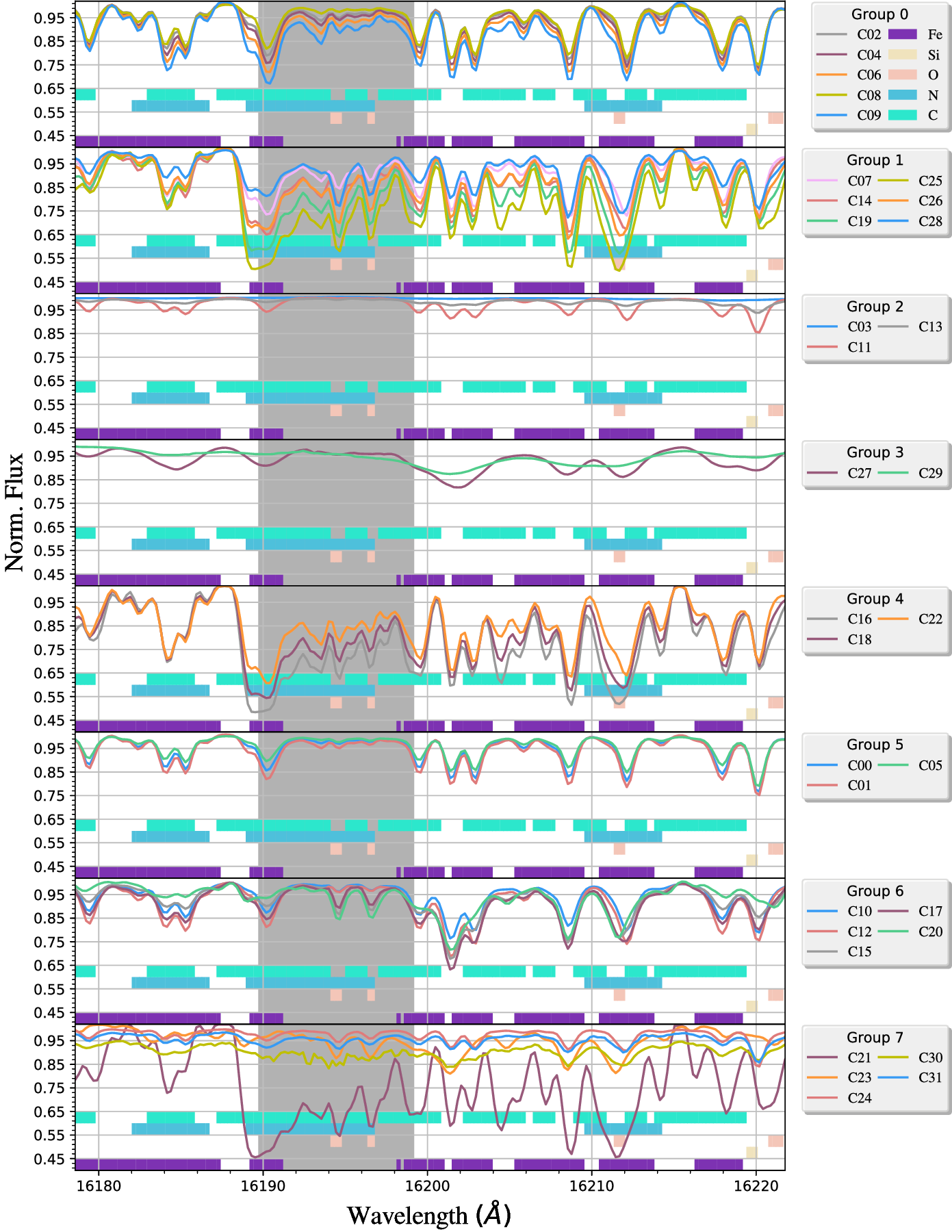}
        \caption{\label{fig:spectra_pile} Mean spectra of the classes in the wavelength range from 16178 to 16222 \AA, where the differences among classes are particularly enhanced. Top to bottom: Panels show the mean spectra for classes belonging to groups from 0 to 7. Each mean spectrum is drawn with the same colours used in Figure \ref{fig:Groups_MH_Teff}. In all panels we plot the spectral windows used in ASPCAP to determine the chemical abundances of stars. Each set of element windows is colour-coded as indicated in the legend of the first panel.}
\end{minipage}
\end{figure*}

\subsection{Repeatability of the classification}
\label{subsec:repeat}

\qquad The randomized initialization of $K$-means implies that different runs generate slightly different results. In order to evaluate the repeatability of the process we define a coincidence index $\varepsilon$, which measures the ratio of coincidence between two different classifications based on the number of spectra in equivalent classes, as described in \citealt{sanchez10}. We note that the label assigned to a class can vary over the classifications, even when the class remains with essentially the same objects. Therefore, when comparing two different classifications, we first need to cross identify the classes. For example, let $X$ be a set of $N$ objects, $X = \{\vec{x}_0, \vec{x}_1,..., \vec{x}_N\}$ classified in $K$ clusters, with two different initializations. Each initialization generates a set of clusters, say $\Omega = \{\omega_0, \omega_1,..., \omega_K\}$ in one classification and $\Gamma = \{\gamma_0, \gamma_1,..., \gamma_K\}$ in a second classification. In each classification we label the classes ensuring that the number of objects in the $i$th class ($n_{i}$) follows the rule $n_{i} \geq n_{i+1}$. To build a comparison between clusters we define a coincidence matrix $ \mathbf{A}_{K,K}$, with the elements $a_{i,j}$ being the number of objects in cluster $\omega_i$ that are also in cluster $\gamma_j$.
\begin{equation}
\centering
a_{i,j} = \sum_{\iota \epsilon \omega_i} \delta_{\iota}^{j}\textrm{, where } \delta_{\iota}^{j} = \begin{cases} 1, & \textrm{if $\vec{x}_\iota$ is in cluster $\gamma_j$} \\ 0, & \textrm{if it is not.} \end{cases}
\end{equation}
Thus, we match the $j$th cluster in $\Gamma$ to the cluster in $\Omega$ having the maximum number of coincidences with it, $j_{match} = argmax\{a_{0, j}, a_{1, j},..., a_{i, j}\}$, always ensuring no cluster in $\Omega$ is assigned to more than one cluster in $\Gamma$. Then we use the matches to transform the matrix $\mathbf{A}$ into $\mathbf{A}'$ permuting its columns to have their largest numbers in the diagonal. The elements of the diagonal of $\mathbf{A}'$ ($a'_{i,j}$, with $i = j$) are counts of the number of agreement between the two classifications, while the other elements ($a'_{i,j}$, with $i \neq j$) are counts of the number of confusions between the two classifications. The trace of $\mathbf{A}'$ divided by the total number of classified objects gives an estimate of the mean overall coincidence rate between the two classifications, $ \bar{\varepsilon}_{total} = \textrm{Tr} \left\{\mathbf{A}'\right\}/N $. By defining the mean normalised coincidence matrix between a chosen classification ($\bar{\mathbf{A}}'_{chosen}$) and a set of $\eta$ classifications with the same $K$ as $\bar{\mathbf{A}}'_{chosen}$, the diagonal elements will give the mean coincidence ratio of each class over the $\eta$ classifications, which is a measure of how stable  the classes in the chosen classification are. Likewise, the elements out of the diagonal measure the mean confusion ratio between different classes.

\subsubsection{Synthetic data set}

We performed a series of classifications for the synthetic data set varying the number of clusters from $K = 5$ to $100$. For each value of $K$ we initialized the classification with ten different random seeds, the same ten seeds for all values of $K$. In order to avoid some possible bias caused by choosing a particular reference, the coincidence ratio was measured for every pair of classifications having the same $K$. For the expected number of clusters in the synthetic data set ($K=69$) the mean coincidence ratio is $\bar{\varepsilon}(K=69) = 74.7 \pm 6.2$ per cent. The mean coincidence ratio computed for all runs with $K = 5$ to $100$ for the synthetic data set is $75.1 \pm 8.4$ per cent.

\subsubsection{DR12 data set}

\qquad Under equivalent conditions, that is, comparing all combinations of the ten classifications per value of $K$, with $K$ from 5 to 100, the DR12 data set had a mean coincidence ratio of $\bar{\varepsilon} = 77.9 \pm 7.8$ per cent.
When we consider only the $K=50$, for which we performed 100 classifications with different random initialization, and using the chosen classification (see Sec. \ref{subsec:chosen}) as reference, the mean coincidence ratio is found to be $\bar{\varepsilon}(K=50) = 79.6 \pm 2.6 $ per cent.

To understand what a mean coincidence ratio of 79.6 per cent means, we measured the mean difference between the matching classes over the 100 classifications, and compared this with the mean within the cluster dispersion of the chosen classification (see Appendix \ref{sec:app0} for more details). We found that the variations of the class centroid over the 100 classifications amount to $6.4 \pm 3.3$ per cent of the average mean internal dispersion of its corresponding class in the chosen classification. That is to say, even for runs with different classifications, for about 25 per cent of the spectra (coincidence of 75 per cent) the main classes end up having their centres displaced by about 6 per cent of the internal dispersion of its class in the $4838$-dimensional flux space. As we show in Section \ref{subsec:chosen}, the confusion occurs mainly between classes sharing borders in the space $\mTeff - \mLOGG - \mMH$. Except for some outlier classes, the shapes of the classes are very similar over different classifications.

\subsection{Chosen classification}
\label{subsec:chosen}

\qquad After running $K$-means a hundred times with $K = 50$, we chose the classification with the lowest sum of squared error (SSE). As we are working with the Euclidean metric, the SSE is computed as
\begin{equation}
\centering
\mathrm{SSE} = \sum_{i=1}^{K} \sum_{\iota \epsilon \omega_i}||\vec{x_{\iota}} -\vec{\mu_i}||^2, \textrm{ where } \vec{\mu}_i = \frac{1}{n_i} \sum_{\iota \epsilon \omega_i} \vec{x}_\iota,
\end{equation}
where $x_\iota$ is the $\iota$th spectrum in cluster $\omega_i$ and $\mu_i$ the centroid of the class $i$. The chosen run has an SSE 9 per cent smaller than the average SSE over all classifications. As mentioned in \ref{subsec:repeat}, the coincidence ratio is measured by the number of spectra sharing the same class over two distinct classifications. Comparing the chosen classification with the other 99 runs, the average coincidence ratio is $ 79.6 \pm 2.6 $ per cent, which can be considered a high repeatability rate. Also the mean variation of the centres of the most popular classes, containing 99 per cent of the objects, is $\approx 2.4 $ per cent of the mean within-cluster variation of the classes in the chosen classification. Again, this is a comparison between the standard deviation of the centroids over the 100 classification with the internal standard deviation of the main classes in the chosen classification. In this case the number falls from 6.4 to 2.4\% because we are only taking into account the classes containing 99\% of the spectra in the sample, classes from 0 to 31.

In Figure \ref{fig:mean_confusion} we plot $\mathbf{A}'_{chosen}$, comparing the chosen classification with the other 99 classifications. The elements of $\mathbf{A}'_{chosen}$ are represented by a colour scale in a 2D histogram; the bottom panel in this figure shows a histogram with the main diagonal values of $\mathbf{A}'_{chosen}$. This plot will be useful in Section \ref{sec:description}, where we will describe each group of classes and comment on the stability of each class. From now on, we will refer to the elements in the main diagonal of $\mathbf{A}'$ as coincidence rates and to its other elements as the confusion rates.

In Figure \ref{fig:chi2} we show a comparison between the root mean squared distances for each spectrum in the sample to its best fit spectrum with respect to the centroid of its assigned class. The plot shows the centroids are n average approximately five times closer to the spectra than its best fit. These higher distances between the spectra and the models are due to systematic differences between synthetic spectra based on model atmospheres and real spectra.\footnote{This can also be seen in panel F of the summarised plots in the appendix. For instance, \href{https://garciadias.github.io/APOGEE/group0/class2/index.html}{Class 02}  present these systematic differences near 16205 \AA\, and 16215 \AA.}

Table \ref{tab:sens} shows a comparison between the standard deviation within clusters ($\widehat{\sigma}$) and the overall standard deviation ($\sigma_{random}$), corresponding to clusters randomly built. For example, \Teff\, and \LOGG\, have a $\widehat{\sigma}$ about 3.6 and 4.2 times smaller than their corresponding $\sigma_{random}$, respectively. This means that the algorithm is especially sensitive to \Teff\, and \LOGG.
In Table \ref{tab:sens} we also highlight the parameters that present $\widehat{\sigma}$ at least two times smaller than its $\sigma_{random}$. They are \Teff, \LOGG, \MH, \Ca, \C, \Mg, \N, \Si, \Su\, and \Ti. Since these are the most sensitive parameters to $K$-means, we will focus mainly on them in order to interpret the classes in the next section.

\begin{table}
\centering
\caption{\label{tab:sens} Comparison of the internal median standard deviation (third column) with the overall standard deviation (second column) for each parameter. The fourth column displays the ratio of these quantities. We highlight the parameters that have internal median standard deviation at least two times smaller than the overall standard deviation.}
\begin{tabular}{cccc}
\centering
 Parameter& $\sigma_{random}$&$\widehat{\sigma}^{K=50}$&$\sigma_{random}/\widehat{\sigma}^{K=50}$ \\ \hline
\Teff\, (K)              &        553        & 152  & \textbf{3.6}\\
$\log g$                 &        1.17       & 0.28 & \textbf{4.2}\\
$\mathrm{[M/H]}$         &        0.35       & 0.17 & \textbf{2.1}\\
$\mathrm{[C/M]}$         &        0.12       & 0.11 &         1.1 \\
$\mathrm{[N/M]}$         &        0.18       & 0.12 &         1.5 \\
$\mathrm{[\alpha/M]}$    &        0.10       & 0.08 &         1.3 \\
$\mathrm{[Al/H]}$        &        0.13       & 0.10 &         1.3 \\
$\mathrm{[Ca/H]}$        &        0.48       & 0.22 & \textbf{2.2}\\
$\mathrm{[C/H]}$         &        0.31       & 0.15 & \textbf{2.1}\\
$\mathrm{[Fe/H]}$        &        0.38       & 0.23 &         1.7 \\
$\mathrm{[K/H]}$         &        0.12       & 0.10 &         1.2 \\
$\mathrm{[Mg/H]}$        &        0.75       & 0.35 & \textbf{2.1}\\
$\mathrm{[Mn/H]}$        &        0.15       & 0.09 &         1.6 \\
$\mathrm{[Na/H]}$        &        0.15       & 0.10 &         1.5 \\
$\mathrm{[Ni/H]}$        &        0.29       & 0.18 &         1.6 \\
$\mathrm{[N/H]}$         &        0.32       & 0.16 & \textbf{2.0}\\
$\mathrm{[O/H]}$         &        0.39       & 0.21 &         1.9 \\
$\mathrm{[Si/H]}$        &        1.00       & 0.43 & \textbf{2.3}\\
$\mathrm{[S/H]}$         &        0.77       & 0.35 & \textbf{2.2}\\
$\mathrm{[Ti/H]}$        &        0.71       & 0.33 & \textbf{2.1}\\
$\mathrm{[V/H]}$         &        0.36       & 0.19 &         1.9
\end{tabular}
\end{table}

\section{Results}
\label{sec:description}

\qquad After visual inspection we divided all the classes into nine groups sharing similar properties. Here we describe in detail each group, giving a summary of their classes' mean properties. In Figure \ref{fig:Groups_MH_Teff} we present contour plots in $\mTeff - \mMH$ space. We highlight regions enclosing progressively 15, 30, 45, and 68.3 per cent of the stars in each class, with the colour shades varying from strong to light respectively. Class 21 is too concentrated to have its contours seen at this scale, so it is represented by purple dots in the figure. In some cases the separation of the contours is too tight and only three contours are visible. The figure is divided into three panels, aiming to minimise the superposition of classes. We use different colours to help identifying borders between classes. Some classes have the same colour, but there is no overlap between classes with the same colour. Classes are identified with labels.
For the labels we use the abbreviations $G$ for group and $C$ for its associated classes. Classes in group 8 have few objects, which are sparsely distributed in the $\mTeff - \mMH$\, plane, making this plot very noisy and hard to read; for these objects we present a scatter plot in Figure \ref{fig:MH_Teff_G8}. Figure \ref{fig:Groups_LOGG_Teff} shows the main distribution of the groups in the $\mTeff - \mLOGG$ plane. Besides the differences found in the $\mTeff - \mMH$ space, we also found some other particularities in the classes and groups, some of them based in the spatial distribution (RA - DEC), global chemical abundances, or spectral fluxes.

In Figure \ref{fig:spectra_pile} we present the mean spectra, in a limited spectral window, for all classes in groups 0 to 7. Each panel in this figure shows the mean spectrum of the classes in each group colour-coded as in Figure \ref{fig:Groups_MH_Teff}. In order to offer the highest contrast between the classes' mean spectra, we chose the spectral coverage which maximises the cumulative variance over the first 32 classes in a 150-pixels-long window. The grey shades in the background of these plots highlight the masked pixels (those discarded from the classification, as discussed in Section \ref{sec:algorithm}). Besides the description presented in this section, we include complementary material with detailed plots for many of the DR12 available features in the supplementary online material described in the Appendix. Table \ref{tab:desc} gives a short description for each class and provides links to the online material. With these figures the reader can find more details about the atmospheric parameters, spatial distributions, and chemical abundances for each class presented.

Tables \ref{tab:quantiles0} and \ref{tab:quantiles1} present the median values for the atmospheric parameters and all the individual chemical elements in each class. The error bars presented in the tables, as well as those shown in the next sections, were calculated by taking the interval around the median, which encloses 68.3 per cent of the points in each class.

\subsection[G0: Metal-rich RC/warm RGB]{Metal-rich RC/warm RGB - Group 0 (Classes 2, 4, 6, 8 and 9)}
\label{subsec:G0}

\begin{figure}
\centering
\includegraphics[width=0.48\textwidth]{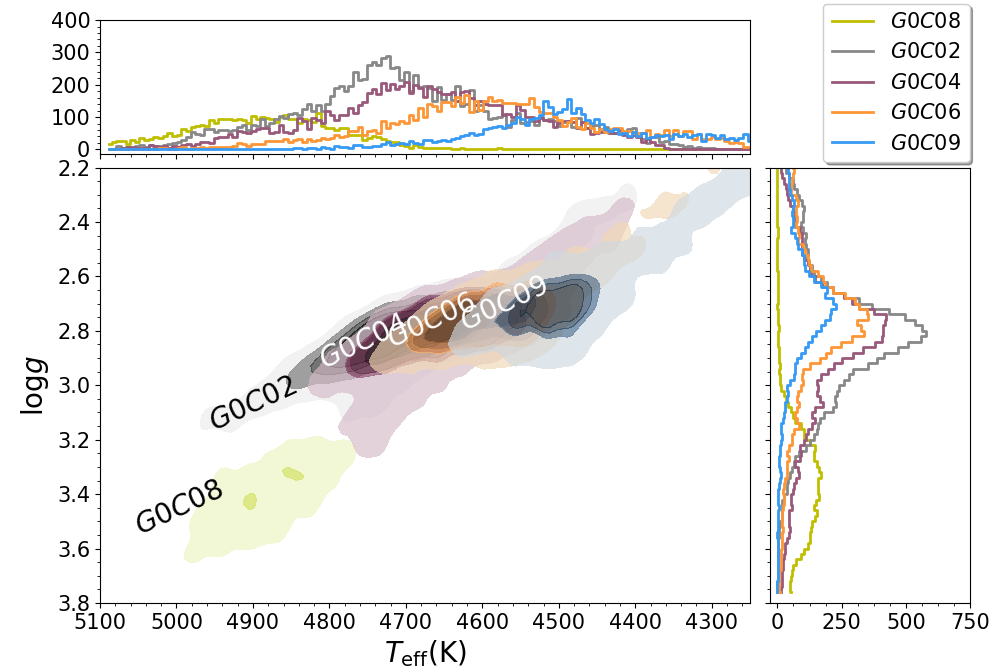}
\caption{\label{fig:G0_LOGG} $\mTeff - \mLOGG$ distribution for classes in group 0. The same rules and colours from Figure \ref{fig:Groups_MH_Teff} were applied to contours here. Top and right panels show histograms of the distributions of \Teff\, and \LOGG, respectively. The histogram line colours match the colours of the contours.}
\end{figure}

\begin{figure}
\centering
\includegraphics[width= 0.48\textwidth]{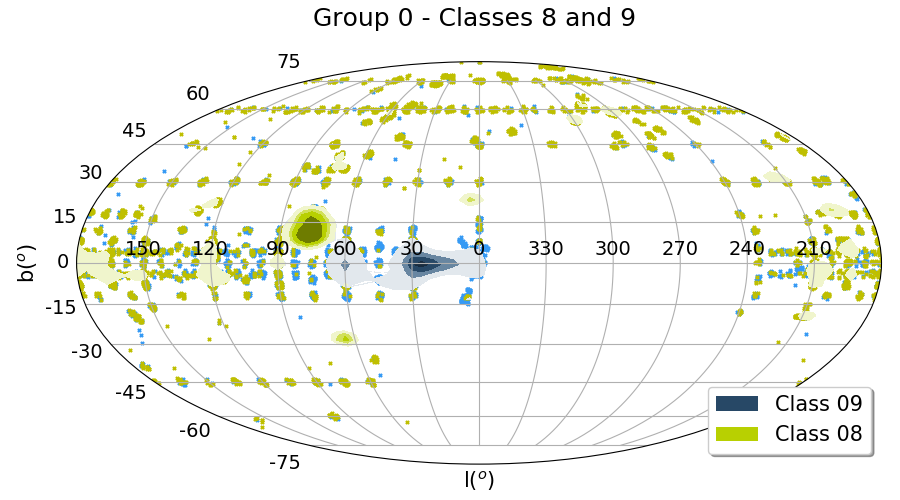}
\caption{\label{fig:projection_G0} Mollweide's projection of the Galactic coordinates distribution of classes 8 and 9. Yellow and blue contours enclose 68.3 per cent of the stars in classes 8 and 9, respectively. Yellow squares represent stars in class 8 and blue squares represent stars in class 9 out of the regions containing 68.3 per cent of the points. The contour shades follow the same rule as in Figure \ref{fig:Groups_MH_Teff}.}
\end{figure}

\qquad From the distribution of \LOGG\, and \Teff\, values in Figure \ref{fig:G0_LOGG} one can spot this group among the red clump (RC) stars and at the warmest end of the red giant branch (RGB) \citep{binney}. Comparing these classes with \citealt{bovy14}'s catalogue of red clump stars, we found that 31, 26, 26, 1, and 21 per cent of the stars in classes 2, 4, 6, 8, and 9, respectively, belong to the red clump. The classes increase in metallicity in the sense $-0.07\pm^{0.10}_{0.11} = \widehat{\mMH}_{c2} < \widehat{\mMH}_{c8} < \widehat{\mMH}_{c4} < \widehat{\mMH}_{c6} < \widehat{\mMH}_{c9} = 0.30\pm^{0.09}_{0.12}$. As metallicity increases, the position of the RC moves towards cooler regions in the plane \Teff - \LOGG, as shown in Figure \ref{fig:G0_LOGG}. Chemical abundances for individual elements also vary inside this group; for example, \Si\, varies as follows: $-0.22\pm^{0.20}_{0.30} = \widehat{\mSi}_{c2}  < \widehat{\mSi}_{c8}  < \widehat{\mSi}_{c4}  < \widehat{\mSi}_{c6} < \widehat{\mSi}_{c9} = 0.26 \pm^{0.18}_{0.19}$. This group is similar to group 5 in terms of atmospheric parameters, but classes here are more metal rich.
For this group there is some confusion among classes, as shown in Figure \ref{fig:mean_confusion}. About 30 per cent of the spectra belonging to class 4 in the chosen classification are assigned to class 2 in other classifications.

Classes 2 and 8 are similar in  chemical abundances, but differ in \LOGG. Besides metallicity differences, classes 8 and 9 also differ in their spatial distribution over the Galactic plane, as shown in Figure \ref{fig:projection_G0}. While stars in class 8, with lower \MH, lie preferentially at higher galactic longitudes, stars in class 9, which are cooler and more metal rich, are mainly towards the galactic centre. In general the fittings for class 9 are poor, the spectral lines are deeper than the chosen models. Classes 2, 4, and 6 follow approximately the same spatial distribution of the APOGEE sample.

In the top panel of Figure \ref{fig:spectra_pile} we have a comparison of the mean spectra for all the classes in this group. For group 0, we see that their mean spectra are very similar in shape, but with different line strengths ($s$). The intensity of lines grows in the sense $s_{c8} < s_{c2} < s_{c4} < s_{c6} < s_{c9}$, following their median temperatures.
Together, these classes include $\approx 27$ per cent of the spectra in DR12.

\subsection[G1: Metal poor cool RGB]{Metal poor cool RGB - Group 1 (Classes 7, 14, 19, 25, 26 and 28)}
\label{subsec:G1}

\begin{figure}
\centering
\includegraphics[width=0.48\textwidth]{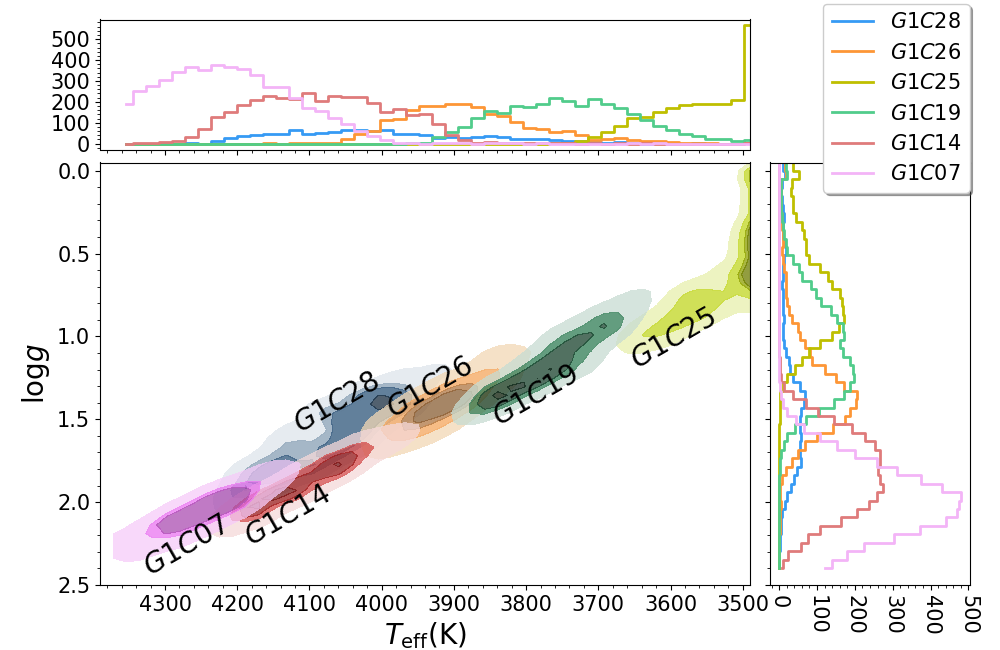}
\caption{\label{fig:G1_LOGG} The distribution in $\mTeff - \mLOGG$  for classes in group 1. The same rules and colours from Figure \ref{fig:Groups_MH_Teff} were applied to the contours here. Top and right panels show histograms of the distributions of \Teff\, and \LOGG, respectively. The colours of the histogram match the colours of the contours.}
\end{figure}

\qquad As shown in Figure \ref{fig:G1_LOGG}, the classes in group 1 are composed of cooler stars in the RGB ($3500 \lesssim  \widehat{\mTeff} \lesssim 4200 $ K and $0.79 \lesssim \widehat{\mLOGG} \lesssim 2.03 $ ) \citep{binney}. All classes are mainly formed of low latitude stars, composed of a mixture of thin and thick disk population, except for class 28 which is mainly projected towards the Galactic centre and with high $\alpha$ abundances, $\widehat{\maM} = 0.24\pm^{0.04}_{0.11}$.  All of them are classes composed of stars in the RGB, but with increasing metallicities,
\footnote{$ -0.81\pm^{0.19}_{0.33} = \widehat{\mMH}_{c28} < \widehat{\mMH}_{c26} < \widehat{\mMH}_{c19} < \widehat{\mMH}_{c25} < \widehat{\mMH}_{c7} < \widehat{\mMH}_{c14} = -0.09 \pm 0.13.$}
surface gravities, \footnote{ $ 0.79\pm^{0.25}_{0.37} = \widehat{\mLOGG}_{c25} < \widehat{\mLOGG}_{c19} < \widehat{\mLOGG}_{c26} < \widehat{\mLOGG}_{c28} < \widehat{\mLOGG}_{c14} < \widehat{\mLOGG}_{c7} = 2.03 \pm 0.22.$}
and temperatures.\footnote{$ 3561\pm^{84}_{60} = \widehat{\mTeff}_{c25} < \widehat{\mTeff}_{c19} < \widehat{\mTeff}_{c26} < \widehat{\mTeff}_{c28} < \widehat{\mTeff}_{c14} < \widehat{\mTeff}_{c7} = 4236\pm^{97}_{100}$ K.}

Concerning the stability of the classes, class 25 is very stable, having a mean coincidence ratio of 82 per cent. As shown in Figures \ref{fig:Groups_MH_Teff} and \ref{fig:G1_LOGG}, this class consists of giant stars at the tip of the RGB. Confusion higher than 10 per cent occurs between classes inside the group. The highest confusion rates are 12 per cent and 16 per cent between class 7 and classes 14 and 28, respectively, 16 per cent between classes 14 and 26, 16 per cent between classes 19 and 26, and 30 per cent between classes 26 and 28. Again, classes overlapping in the 3D space $\mTeff - \mLOGG - \mMH$ present the highest degrees of confusion. Between classes in this group and  other classes out of the group, the confusion rate is above 5 per cent only between class 14 and class 22 (10 per cent).

Tables \ref{tab:quantiles0} and \ref{tab:quantiles1} show the classes in this group are selecting stars within narrow distributions of the parameters, including the abundances. They typically have $\bar{\sigma}_{\mTeff} \approx 100$ K, $\bar{\sigma}_{\mLOGG} \approx 0.30,$  and, for example, in class 14, the within class dispersion of the parameter can reach $\bar{\sigma}_{X} \leq 0.1$  for \aM, \NM, \CM, \Na, \Mn\, and \K.

Class 28 is particularly spread in $\widehat{\mCM} =  -0.09 \pm^{0.15}_{0.30}$, $\widehat{\mFe} =  -1.14 \pm^{0.42}_{0.73}$  , and $\widehat{\mAl} =  -0.10 \pm^{0.16}_{0.31}$. In Figure \ref{fig:spectra_pile}, second panel from top to bottom, we see the mean spectra of the stars in this group. As in group 0, we see very similar spectral shapes, but with different line strengths.

\subsection[G2: Warm Stars]{Warm stars - Group 2 (Classes 3, 11, and 13)}
\label{subsec:G2}

\qquad This group assembles the warmest stars in DR12. The sample includes 15,233 spectra flagged as telluric standards, warm objects ideal for characterising the telluric lines that plague the IR, of which 67 per cent are in class 3, 16 per cent in class 11, and 12 per cent in class 13. According to target-selection flags, 96 per cent of the 10,628 objects in class 3 are telluric standards, while classes 11 and 13 have up to 50 per cent of stars of this kind. The differences between the classes in this group are mainly found in \Teff\, and \MH, as seen in panel b of Figure \ref{fig:Groups_MH_Teff}; class 3 is the warmest, containing A and B type stars, according to a match with the SIMBAD catalogue \citep{simbad}, while classes 11 and 13 are RGB stars, cooler and richer in metals compared with class 3 (see Table \ref{tab:quantiles0}). The third panel in Figure \ref{fig:spectra_pile} shows the differences between the mean spectra of the classes in group 2. The mean spectrum of class 3 is almost featureless, while the mean spectrum in class 13 has the strongest lines in the group. Moreover, there is a difference in their spatial distribution; while class 3 mainly occupies low latitudes, classes 11 and 13 are found primarily out of the Galactic plane and towards the Galactic centre.

\begin{figure}
\centering
\includegraphics[width=0.48\textwidth]{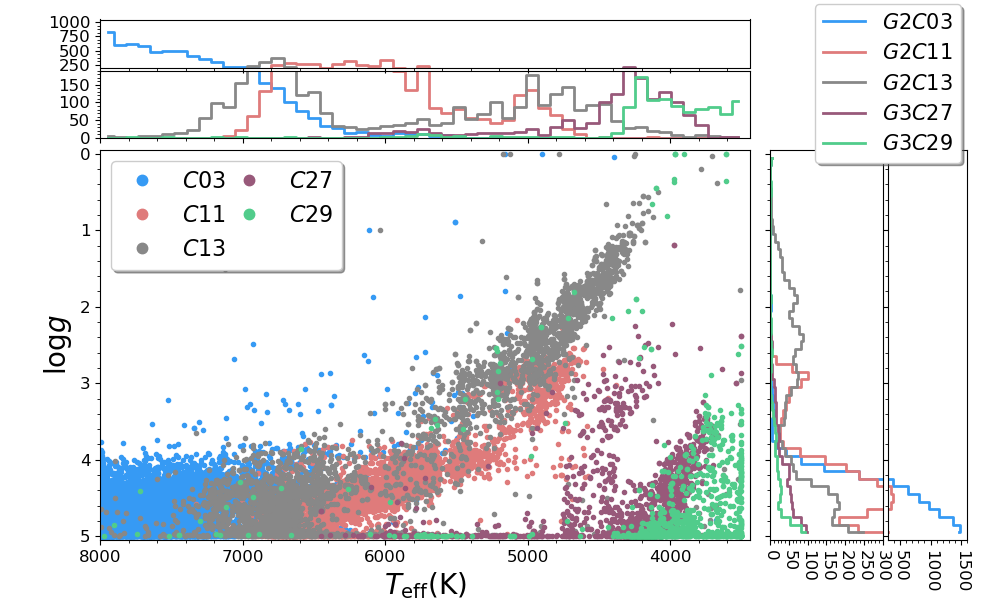}
\caption{\label{fig:Group_2_3_LOGG_TEFF} Scatter plot for $\mTeff - \mLOGG$ distributions of the classes in groups 2 and 3. Top and right panels show histograms of the distributions of \Teff\, and \LOGG, respectively. To aid visualisation, both panels are split into two plots with different scales. The histogram line colours match the colours of the scatter plot.}
\end{figure}

As we would expect, since they are easily distinguishable even by eye, classes in this group are among the most stable classes in the classification, with mean coincidence rates of 94 per cent, 73 per cent, and 80 per cent for classes 3, 11, and 13, respectively. As class 11 is cooler than classes 3 and 13, it has the highest mean confusion rate with other classes (for example it has about 10 per cent mean confusion with classes 5 and 24). Classes 5 and 24 are among the most metal-poor in the classification, emphasising the role that the degeneracy between \Teff\, and \MH\, plays in the determination of the stellar parameters.

All the chemical elements have very wide distributions except for  \K\, in class 11. Nevertheless, the atmospheric parameters of the stars in this group are out of the DR12 model grid, and thus it should be seen as a failure of the model fittings, as suggested by the ASPCAP flag \textit{star warn} found in $\approx$ 35 per cent of the objects in this class.

\subsection[G3: Fast rotators]{Fast rotators - Group 3 (Classes 27 and 29)}
\label{subsec:G3}

\qquad This group is formed by fast rotating stars. For both classes the ASPCAP models poorly fit their spectra. As a consequence of this, some artefacts are observed in their abundances, for example, the abundances of \CM, \aM, \Al, \K, \Na\, and \Si\, are not continuous; they appear in clumps, having gaps of at least 0.2  in abundance between them.

In terms of atmospheric parameters, this group is very close to group 6 (dwarfs), but their spectra are remarkably different. The spectra of group 3 have fewer, shallower, and broader lines than those found in group 6, as can be seen in the fourth and seventh panel in Figure \ref{fig:spectra_pile}. This shows that the algorithm is sensitive to rotation, since it is able to split the stars affected by \LOGG\, line broadening from those affected by rotational line broadening. On the other hand, ASPCAP determines that the great majority of the stars in this group have \LOGG\, greater than 4.9 (see Figure \ref{fig:Group_2_3_LOGG_TEFF}), but since the rate of stars flagged with a fast rotation warning are 81 per cent and 93 per cent for classes 27 and 29, respectively, we cannot trust these determinations. The rate of stars flagged with a rotation warning in the entire DR12 data set is 7 per cent.

Class 29 is the most unstable of the classes, excluding the outliers (see Section \ref{subsec:G8}). It has a confusion rate of 62.8  per cent with class 27, which means that for some classifications class 29 dissolves mainly in classes 13, 23, 27 and 29. Class 27 is more stable, with 63 per cent of coincidence, having some degree of confusion with class 10 (13 per cent), which has the shallower lines in group 6.

About one quarter of the stars in class 27 and about half of the stars in class 29 are either young embedded cluster members or known calibration cluster members. Statistically we expect fast rotating stars to be younger than those that rotate more slowly \citep{van13}. In addition, the great majority of stars form in star clusters, dispersing latter on, and thus the fastest rotating stars are expected to be in young embedded clusters.

\subsection[G4: Metal-rich cool RGB]{Metal-rich cool RGB - Group 4 (Classes 16, 18 and 22)}
\label{subsec:G4}

\qquad Group 4 classes include metal rich stars covering the RGB with effective temperatures from 3620 to 4140 K, and with metallicities from 0.17 to 0.22  in the order $\widehat{\mMH}_{c16} < \widehat{\mMH}_{c18} < \widehat{\mMH}_{c22}$. Some stars in this group are near the edge of the model grid, at $\mFe = 0.50$ (36 per cent in class 16, 26 per cent in class 18, and 24 per cent in class 22). That also happens in \Teff\, for class 16, which has 43 per cent of the stars cooler than 3600 K.

\begin{figure}
\centering
\includegraphics[width=0.48\textwidth]{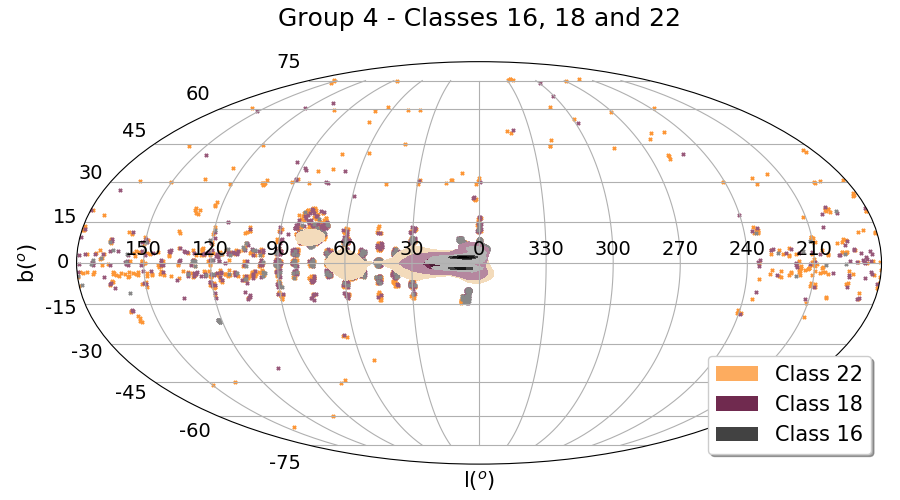}
\caption{\label{fig:projection_G4} Galactic coordinates in Mollweide's projection for objects in classes 22 (orange triangles and contours), 18 (purple triangles and contours), and 16 (grey circles and contours), all belonging to group 4. The contour shades follow the same rule as in Figure \ref{fig:Groups_MH_Teff}.}
\end{figure}

The stars in these classes are very concentrated in the Galactic disk, with \aM\, close to the solar value. As shown in Figure \ref{fig:projection_G4}, the spatial distribution of class 16 is more concentrated towards the Galactic centre than classes 18 and 22.

\begin{figure}
\centering
\includegraphics[width=0.48\textwidth]{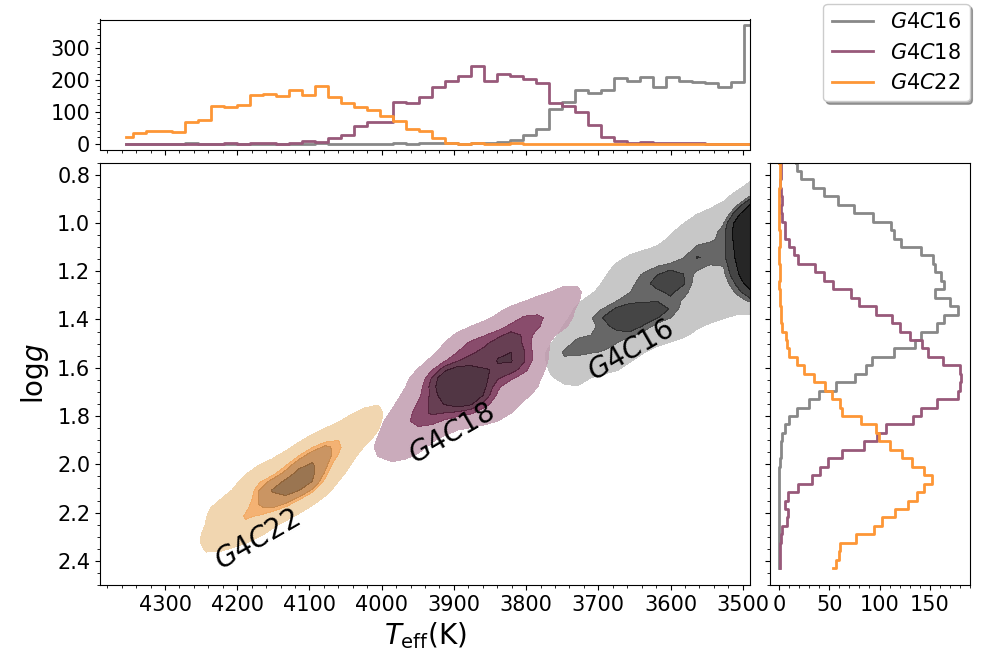}
\caption{\label{fig:G4_LOGG} $\mTeff - \mLOGG$ distribution for classes in group 4. The same rules and colours from Figure \ref{fig:Groups_MH_Teff} were applied to contours here. Top and right panels show histograms of the distributions of \Teff\, and \LOGG, respectively. The colours of the histograms match the colours of the contours.}
\end{figure}

Classes 16 and 18 are very stable, with a coincidence rate of 91 per cent and 80 per cent, respectively. Class 22 is much less stable having a coincidence rate of 29 per cent. The highest degree of confusion for class 22 occurs with class 9 (38 per cent), but classes 14 and 18 also contaminate class 22. Those three classes, 9, 14, and 18, share borders with class 22 in the space $\mTeff - \mMH$, as shown in Figure \ref{fig:Groups_MH_Teff}, and also with superposition in \LOGG, as can be seen by comparing Figures \ref{fig:G1_LOGG} and \ref{fig:G4_LOGG}. Once again, we see that the overlap in the space $\mTeff - \mMH - \mLOGG$ is the main cause of confusion between classes.
The abundance distributions for these classes are narrow, as reflected in Tables \ref{tab:quantiles0} and \ref{tab:quantiles1}.

\subsection[G5: Metal-poor RC/warm RGB]{Metal-poor RC/warm RGB - Group 5 (Classes 0, 1 and 5)}
\label{subsec:G5}

\begin{figure}
\centering
\includegraphics[width=0.48\textwidth]{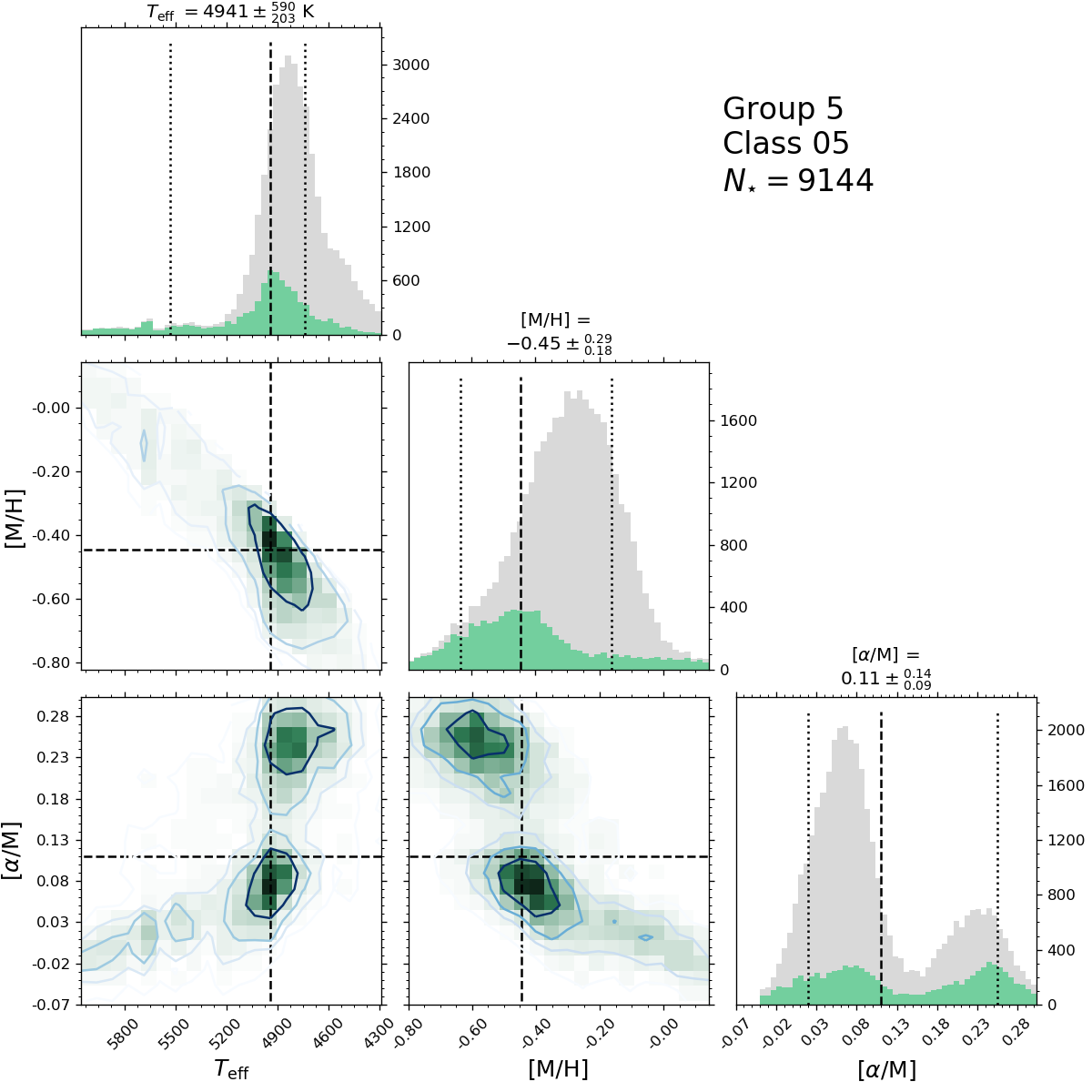}
\caption{\label{fig:class_5_corner} Properties of class 5 (group 5), which contains 9,144 stars ($N_\star$). The panels in the uppermost diagonal contain histograms for \Teff, \MH\, and \aM, from left to right, respectively. In these plots vertical black dashed lines show the median value and the limits enclosing 68.3 per cent of the data points around the median value. The green histograms correspond to the objects in class 5 and the grey histogram shows the distribution of the whole group 5. As indicated by labels in the axes, the other three panels show 2D histograms for $\mTeff - \mMH$, $\mTeff - \maM$ and $\maM - \mMH$. From outside to inside the contours enclose 68.3, 45, 30, and 15 per cent of the objects in the class.}
\end{figure}

\begin{figure}
\centering
\includegraphics[width=0.48\textwidth]{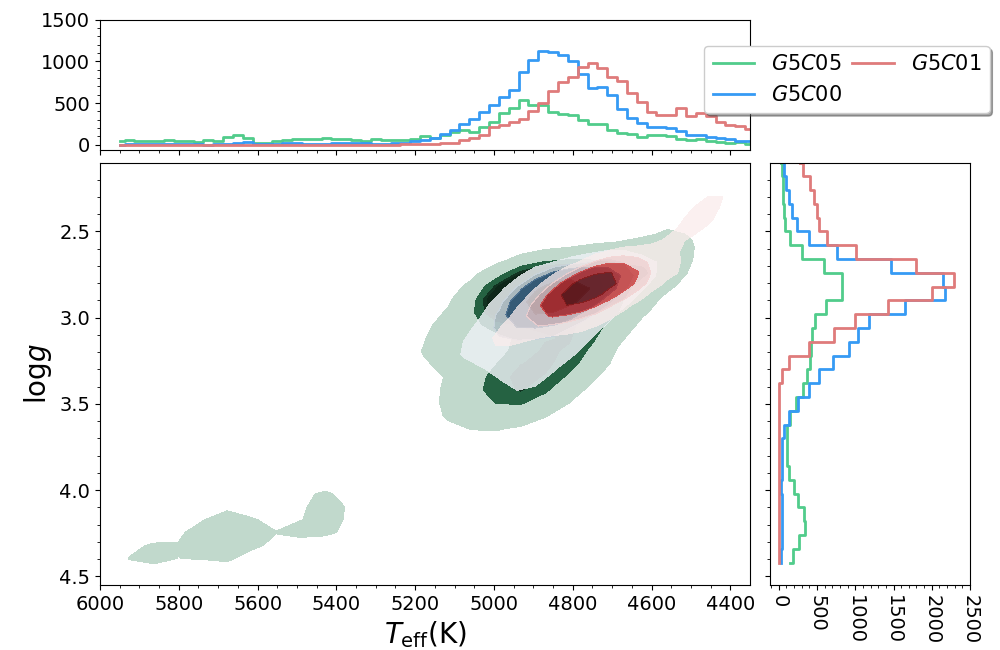}
\caption{\label{fig:Group_5_LOGG_TEFF} $\mTeff - \mLOGG$ distribution for classes in group 5. The same rules and colours from Figure \ref{fig:Groups_MH_Teff} were applied to contours here. Top and right panels show histograms of the distributions of \Teff\, and \LOGG, respectively. The colours of the histograms match the colours of the contours.}
\end{figure}

\qquad Just like group 0, this group is made of classes that include stars from the RC and the warmest end of the RGB. For classes 0, 1, and 5 the ratios of red clump stars are 30, 31, and 16 per cent according to a comparison with \citealt{bovy14}. In comparison with group 0, this group is more metal-poor, with $ - 0.45 \lesssim \widehat{\mMH} \lesssim -0.22 $. The group lacks stars in the direction of the Galactic centre, being homogeneously distributed in all other directions. Relative to group 0, group 5 is more dense in regions with Galactic latitudes higher than 30 degrees. All three classes are a mixture of thin and thick disk populations, but class 5 is more populated by high \aM\, stars than other classes in the group, as shown in Figure \ref{fig:class_5_corner}.

As shown in Figure \ref{fig:Group_5_LOGG_TEFF}, class 5 almost completely overlaps with classes 0 and 1 in  $\mTeff - \mLOGG$ space. The median temperatures of class 0 stars are about 150 K warmer than class 1 stars. Class 5 is particularly broad in \Teff\, and \LOGG, covering temperatures from 4125 to 7170 K, with a median value of $\widehat{\mTeff} = 4942 \pm^{584}_{202}$ K and $\mLOGG = 3.16  \pm^{+1.04}_{-0.38}$. Figure \ref{fig:class_5_corner} shows the distribution of the stellar parameters in the planes $\mTeff - \mMH$, $\mTeff - \maM$ and $\maM - \mMH$. The dispersion there is likely to be an artefact due to the degeneracy between \Teff\, and \MH\, in the ASPCAP parameter determination pipeline. Also the class is broadly spread in $\widehat{\mSi} = -1.38 \pm^{0.96}_{1.38}$, which may also be an artefact of ASPCAP. In this range of atmospheric parameters the pipeline is probably confusing warmer temperatures with lower metallicities, as discussed in \cite{Holtzman15}.

\subsection[G6: Dwarfs stars]{Dwarfs stars - Group 6 (Classes 10, 12, 15, 17 and 20)}
\label{subsec:G6}

\begin{figure}
\centering
\includegraphics[width=0.48\textwidth]{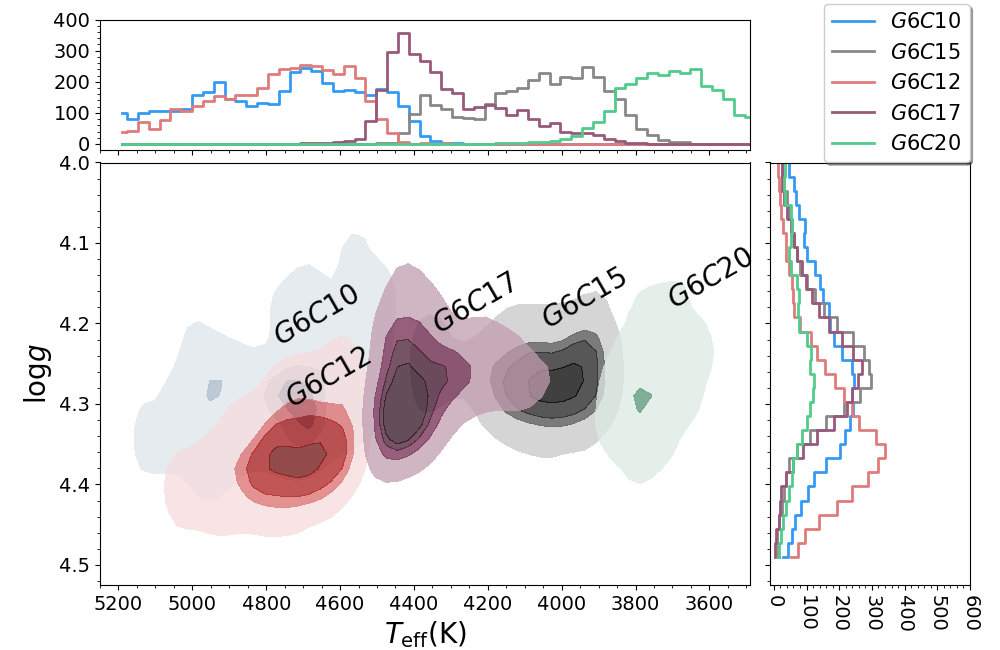}
\caption{\label{fig:Group_6_LOGG_TEFF} $\mTeff - \mLOGG$ distribution for classes in group 6. The same rules and colours from Figure \ref{fig:Groups_MH_Teff} were applied to contours here. Top and right panels show histograms of the distributions of \Teff\, and \LOGG, respectively. The histogram line colours match the colours of the contours.}
\end{figure}

\qquad With \LOGG\, ranging from 4.23 to 4.35, group 6 has only dwarf stars. The classes differ because of their different temperatures and abundance patterns. Figure \ref{fig:Group_6_LOGG_TEFF} shows the distribution of \LOGG\, and \Teff\, for this group.

Class 12 is over-abundant in Mg ($\widehat{\mMg} = +0.38 \pm^{0.32}_{0.28}$ ), and classes 15 and 20 have low \aM, especially in \Ca\, and \Ox. Some bimodality is found for \Al\, and \K\, for classes 15 and 20. However, 99 per cent of the objects in the group have their chemical abundances flagged with a warning and are not reliable, so this strange behaviour is likely to be an artefact of ASPCAP.

In Figure \ref{fig:spectra_pile} we see the FeI line around 16210 \AA\, is blended with the CN and CO lines near it for classes 15 and 20. In other regions of the spectra, blends like this are present. This is caused by the enhancement of molecular lines at low \Teff\, values.

Class 20 presents two separate blobs of \aM\, abundances, one around solar values and the other around $\widehat{\maM} = -0.3$, but almost 70 per cent of the stars in this class are flagged with the star warning, so abundance determination for these stars is not reliable.
The abundance distributions of these classes are very narrow, as shown in Tables \ref{tab:quantiles0} and \ref{tab:quantiles1}.

The classes here are relatively stable. Class 17 is the most unstable (50 per cent of mean coincidence rate), but has a significant degree of confusion only with classes 10, 12, and 15. Class 20 is the most stable in the group with a mean coincidence rate of 81 per cent. Other significant confusion rates are found only between classes inside the group, showing that the classes are stable as a group.

\subsection[G7: Sparse classes]{Sparse classes - Group 7 (Classes 21, 23, 24, 30 and 31)}
\label{subsec:G7}

\begin{figure}
  \centering
  \includegraphics[width=0.48\textwidth]{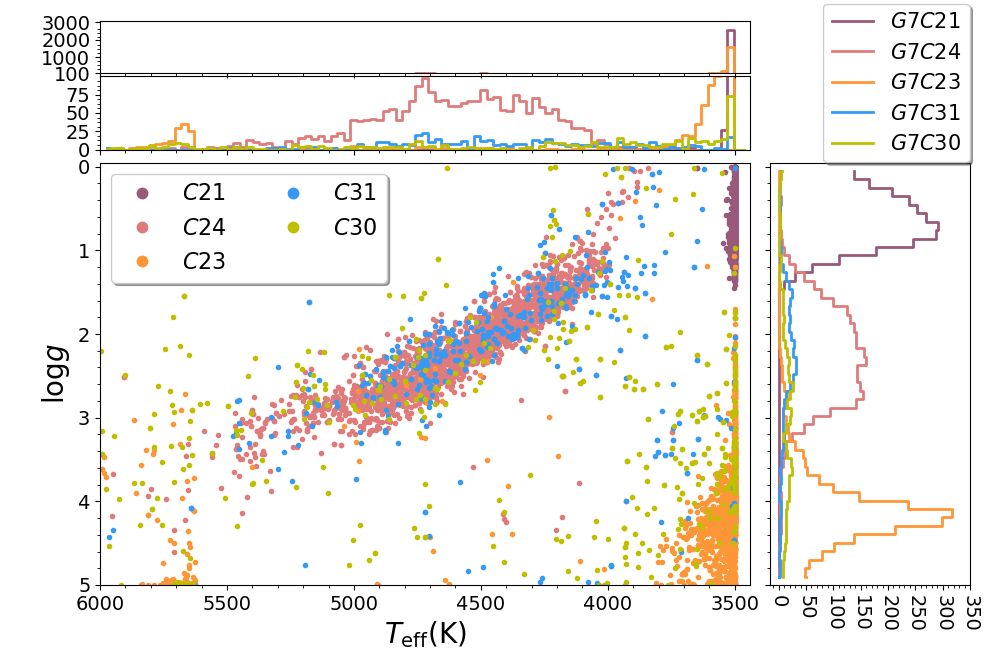}
  \caption{\label{fig:Group_7_LOGG_TEFF} Scatter plot for \Teff\, versus \LOGG\, of the classes in group 7. Top and right panels show histograms of the distributions of \Teff\, and \LOGG, respectively. Top panel is divided in two plots with different scales. The colours of the lines in the histograms match the colours of the scatter plot, as indicated in the legends.}
\end{figure}

\qquad This group is formed by the most peculiar classes, with a number of objects corresponding to at least 0.5 per cent of the whole DR12 sample. The group is very diverse, so in this case we describe each class individually. All classes that represent less than 0.5 per cent of the sample are treated as outliers and are discussed in Section \ref{subsec:G8}. Figure \ref{fig:Group_7_LOGG_TEFF} shows the \Teff - \LOGG\, distribution for the group.

\subsubsection{M-giants/Bulge - Class 21}
\begin{figure}
\centering
\includegraphics[width=0.48\textwidth]{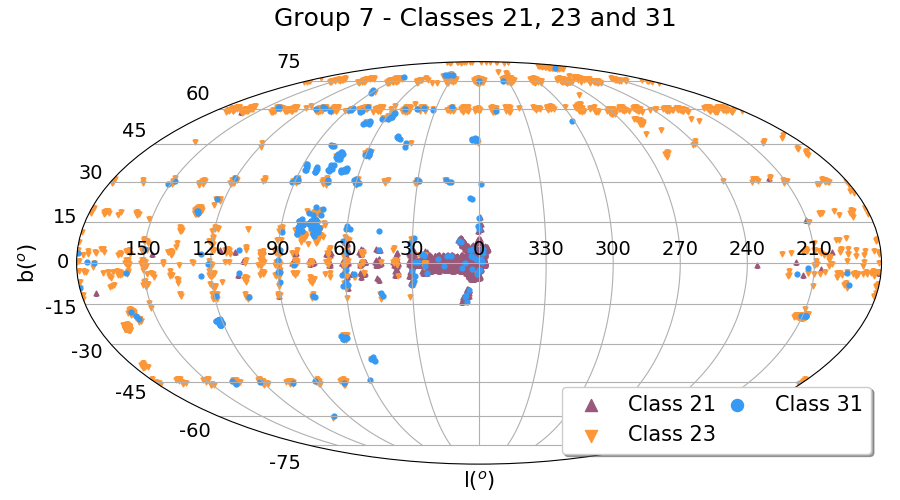}
\caption{\label{fig:projection_G7} Galactic coordinates distribution of classes 21 (purple triangles), 23 (orange triangles) and 31 (blue circles).}
\end{figure}

Ninety-seven per cent of the stars in class 21 are at the edge of the model grid in \Teff. That is to say, their temperatures are likely to be lower than the minimum \Teff\, of the models in the spectral library. The class presents other anomalies; except for \CM, \NM, \aM, \Al, \K, \Mn\, and \Na, all other abundances are also at the edge of the model grid. Lacking sufficiently cool spectra, ASPCAP probably tries to change the abundances until reaching its limits. For these stars, the problem has been corrected in DR13 \citep{DR13}.
This class is the most stable class with a coincidence rate of 95 per cent. Figure \ref{fig:spectra_pile}, bottom panel, shows that the mean spectra of this class looks totally different from the other classes, with very strong molecular bands, so $K$-means easily identifies  these spectra as a class.
Spatially, the stars are concentrated at low latitude, specially towards the galactic centre, as shown in Figure \ref{fig:projection_G7}. This class also gathers 23 per cent of the bulge targets in DR12, according to its target flags.

\subsubsection{Metal-poor M dwarfs - Class 23}
This class is dominated by metal-poor ($\widehat{\mMH} \approx -0.54$ ) M dwarfs. The distribution of \aM\, is divided into four clumps, showing there is some problem with the determination of these abundances, since very similar spectra correspond to differences of 0.25 in \aM. The mean spectrum is similar to that of class 20, but with cooler stars; here more than 60 per cent of the stars are at the minimum $\mTeff = 3500$ K. This similarity in their spectra causes a mean confusion rate with class 20 of 12 per cent. However, class 23 is quite stable, with a mean coincidence rate of 87 per cent.
Similar to what happened to class 21, this class has many anomalies in its parameters, gaps in chemical abundances, and a high concentration at the borders of the model grid. This can also be related to limitations in ASPCAP.
As shown in Figure \ref{fig:projection_G7}, there seems to be no anisotropy in this class. It approximately follows the spatial distribution of APOGEE.

\subsubsection{K-giants from the Halo - Class 24}
This is a very metal-poor class with stars lying over the whole RGB, $\widehat{\mTeff} = 4583 \pm^{322}_{330}$ K and $\widehat{\mLOGG} = 2.22 \pm^{0.60}_{0.54}$, as shown in Figure \ref{fig:Group_7_LOGG_TEFF}. With a median metallicity of $\widehat{\mMH} = -1.20 \pm^{0.22}_{0.25}$  it is one of the most metal-poor classes in the classification, certainly the most well-defined class among the metal-poor ones. This class is also $\alpha$ enhanced, with $\widehat{\maM} = 0.24 \pm 0.07$. We find that 593 out of 2388 ($\approx 25$ per cent) of these objects are globular cluster members used in APOGEE's calibration. Its spacial distribution is more dense in Galactic latitudes above 30$^o$.
Class 24 has a very low stability, having a coincidence rate of 18 per cent. Its stars are classified as class 11 members 59 per cent of the time.

\subsubsection{M31 GCs - Class 30}
In APOGEE DR12, 236 integrated spectra of Globular Clusters (GCs) in M31 were observed; each of these spectra appears as duplicate in the dataset. In order to remove the contamination from the unresolved M31 stellar population in these spectra, 141 background spectra near to the clusters were obtained \citep{Zasowski13}. Altogether they add up to 613 spectra in the region of M31. This class has the largest number of objects in this region, 171, with 33 background spectra and 69 duplicated GCs spectra. In general the spectra present high absorption in the continuum, as shown for the mean behaviour by the yellow line in the bottom panel of Figure \ref{fig:spectra_pile}. Its spectra are poorly fitted by the ASPCAP, and their wide chemical abundances and atmospheric parameters distributions (see Tables \ref{tab:quantiles0} and \ref{tab:quantiles1}) should not be trusted since they are all flagged with ASPCAP warnings. \citealt{Sakari16} have determined the abundance for 25 of the GCs in DR12 (eight are in this class) and we refer to their work as a better source of chemical abundances for these objects. This group also has 62 stars in embedded clusters, two member candidates of the GC Palomar 1, six bulge giants, and many metal-poor RGB stars. The class has 562 spectra, from which 93 per cent are flagged with star warnings, so the ASPCAP values cannot be trusted.

\subsubsection{M31 GCs/high persistence - Class 31}
Class 31 also has some spectra in the region of M31 (84 out of 613), from which 20 are background spectra and 64 are duplicated spectra of 32 clusters. In this class the spectra seem to be less affected by continuum absorption. As shown by the light blue circles in Figure \ref{fig:projection_G7}, this class has a peculiar spacial distribution, being  more dense in $ 60^o \leq l \leq 90^o $ and $0^o \leq b \leq 45^o $. Further investigation is needed to determine why the stars in that direction have these characteristics. In this class there are 88 calibration cluster members and 38 spectra that overlap with the Kepler mission sample. Comparing the position of the stars of this class in Figure \ref{fig:projection_G7} with Figure 2 in \citealt{Zasowski13} one sees the position of these objects match the locus of observation targets of the halo population, the Kepler mission, and some of the calibration cluster. Thirty-five per cent (170) of the spectra in this class are flagged with a warning.

Thirty one per cent of the stars in this class are flagged as \textit{high persistence} observations. Persistence refers to the latent image of a previous exposure appearing in subsequent images, due to a slow release of an  appreciable fraction of accumulated charge in the previous exposure over the subsequent ones. It affects the bluest chip particularly \citep{Nidever15}. The intensity of the persistence effect depends on the brightness of the spectra and their history of previous observations. In DR12, a flag is used to inform the relevance of the persistence effect on each spectra \citep{Holtzman15}. Some of the affected spectra by persistence present an obvious excess/deficit of flux in the blue chip. This behaviour is flagged as a \textit{positive/negative jump in blue chip}.

\subsection[G8: Outliers]{Group 8 and outliers}
\label{subsec:G8}

\begin{figure}
        \centering
        \includegraphics[width=0.48\textwidth]{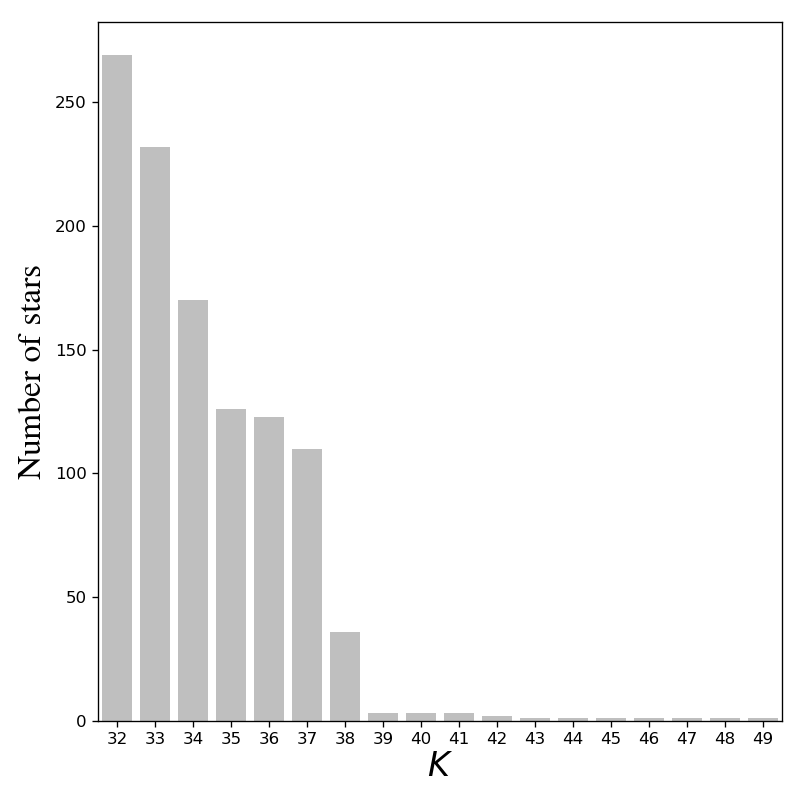}
        \caption{\label{fig:count_G8} Number of objects in outlier classes.}
\end{figure}

\begin{figure*}
\begin{minipage}{\textwidth}
        \centering
        \includegraphics[width=\textwidth]{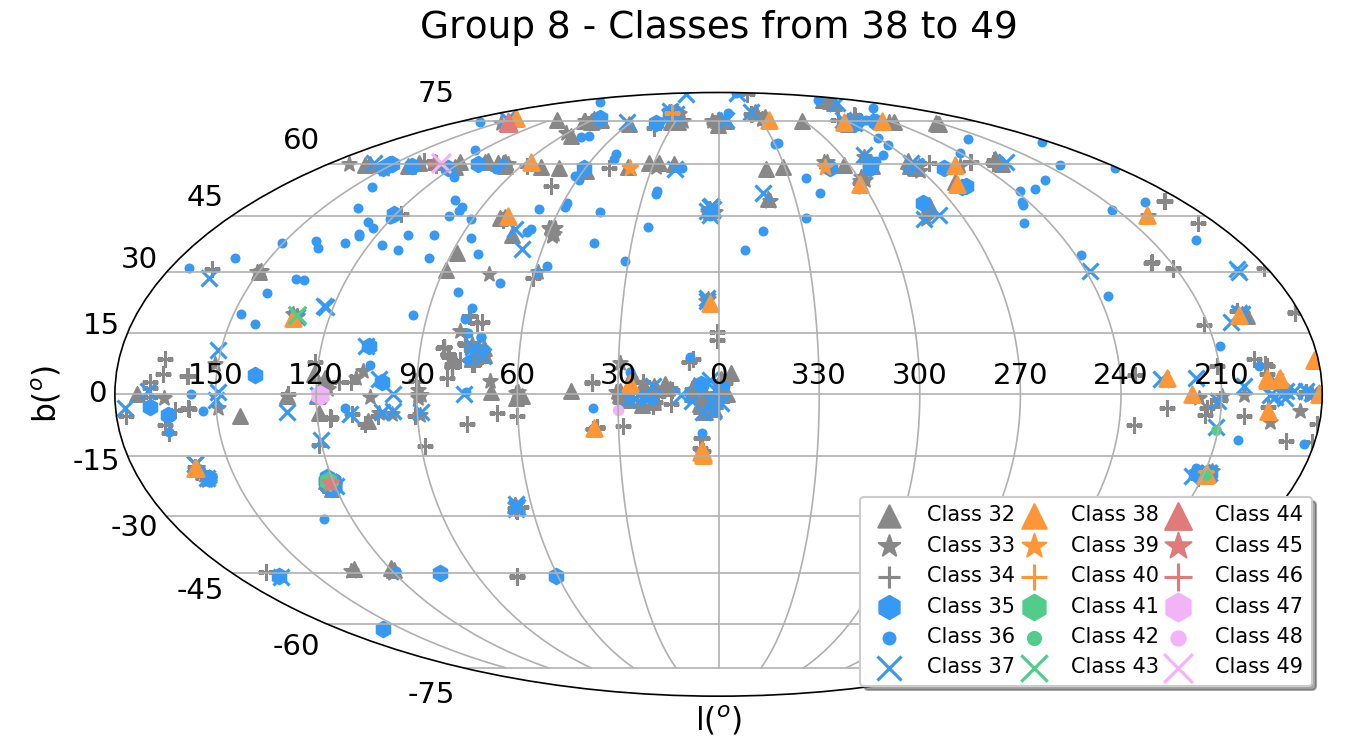}
        \caption{\label{fig:projection_G8} Galactic coordinates distribution of targets in group 8.}
\end{minipage}
\end{figure*}

\begin{figure*}
\begin{minipage}{\textwidth}
        \centering
        \includegraphics[width=0.99\textwidth]{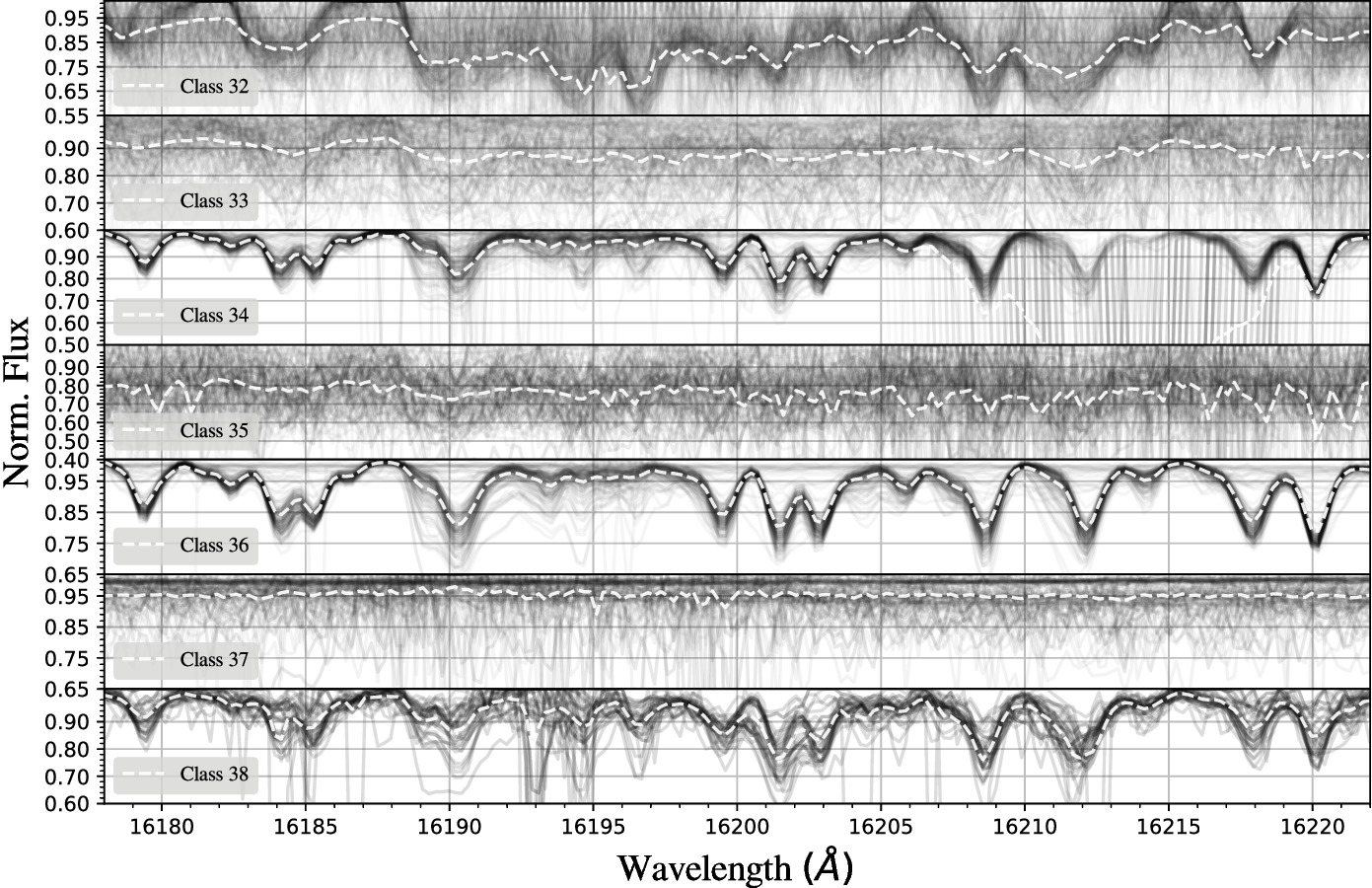}
        \caption{\label{fig:outliers_32-38} Spectra of the objects in classes from 32 to 38. Each spectra is plotted as a semi-transparent line, in a way that the darkest regions represent the most dense regions in this flux window. The wavelength coverage here is the same of figure \ref{fig:spectra_pile}.}
\end{minipage}
\end{figure*}

\qquad Ninety-nine per cent of the stars in APOGEE are in the classes presented in sections  \ref{subsec:G0} to \ref{subsec:G7}. We briefly discuss the remaining  1 per cent. In addition, we also investigate the outliers of the main classes, that is, those spectra in classes from 0 to 31 for which the distance to the class mean spectrum is larger than 3-$\sigma$. Figure \ref{fig:count_G8} shows the number of spectra in the classes of group 8. Figure \ref{fig:projection_G8} shows the spatial distribution of these classes. In this figure the classes are represented by different symbols and colours. Figure \ref{fig:outliers_32-38} shows the spectra in classes from 32 to 38 in the same wavelength window as Figure \ref{fig:spectra_pile}; we plot the spectra as semi-transparent black lines to highlight the locations where the spectra are more similar to each other. In Figure \ref{fig:outliers_32-38} the mean spectrum of each class is drawn as a white dashed line.

\subsubsection{Bulge giants - Class 32}
\qquad This class has 269 spectra, from which 71 are of supergiant stars in the bulge, 33 bulge giants, and 44 spectra in the region of M31 (20 of the background and 24 of 12 duplicated GCs). Forty-one per cent of the spectra in this class are flagged as having a negative jump in blue chip, 19 per cent of them as having high persistence, and 99 per cent of them are flagged as \textit{star bad}, assigned if there is warning about any of the following issues: \Teff, \LOGG, model fitting $\chi^2$, rotation, S/N (signal-to-noise ratio), and if the difference between photometric and spectroscopic temperature is greater than 500K.

\subsubsection{M31 GCs/high persistence - Class 33} Class 33 has 116 spectra in the region of M31, 18 background spectra, and 98 spectra from 49 GCs. There are 39 spectra flagged as emission line stars in DR12, eight of which are in this class. Figure \ref{fig:outliers_32-38} second panel from the top shows all 232 spectra overlapped. Emission lines are not visible in this figure because all spectra were truncated at 1.02 of the normalised flux. In spite of this constraint, the algorithm is able to identify emission lines since they affect the form of the continuum around them. Ninety-five per cent of the 232 spectra in this class are flagged as \textit{star bad}, 56 per cent are flagged with the rotation warning, and 33 per cent of these spectra are flagged as high persistence spectra. In Figure \ref{fig:outliers_32-38}, second panel from the top, we see there is no clear resemblance between the spectra in the class. The pixels with lower dispersion seem to be emission-dominated by lines, suggesting the spectra are either actually emission line stars or have some problem with the sky subtraction.

\subsubsection{Bad pixels - Class 34}
\qquad Seventy six per cent of the 170 stars in this class are flagged as high persistence observations. In Figure \ref{fig:outliers_32-38}, third panel from the top, we see they are mainly giant stars whose spectra have sequences of bad pixels, as those seen between 16,205 and 16,220 \AA.

\subsubsection{M31 GCs/high persistence - Class 35}
\qquad Class 35 has 88 spectra in the region of M31, 38 background spectra, and 50 spectra from 25 duplicated GCs. These 88 spectra are 70 per cent of the 126 spectra in the class. There are 99.2 per cent of the objects in this class flagged as \textit{star bad}, and 94 per cent of them have signal to noise lower than 30. As we see in Figure \ref{fig:outliers_32-38}, central panel, all the spectra are very noisy.

\subsubsection{1m Telescope - Class 36}
\qquad There is 817 spectra observed with the 1m telescope in DR12, and 93 of these are in this class. With 123 spectra, it corresponds to 76 per cent of the spectra in the class. Apart from a few cases, the spectra seem to contain sequences of a few bad pixels like the ones seen in class 34, Fig. \ref{fig:outliers_32-38}, but in different regions of the spectrum.

\subsubsection{Emission line stars/M31 GCs - Class 37}
\qquad This class has 13 emission line stars. There are also 11 spectra in the M31 region, one spectrum from the background, and ten spectra of five GCs. There are six objects identified by SIMBAD as galaxies.

\subsubsection{Negative flux - Class 38} This class has 36 spectra, from which eight are embedded cluster members, four are Sagittarius dwarf galaxy members, and one is an integrated spectra of the Pal1 GC. Eighty three per cent of the spectra in the class have pixels with negative counts.

\subsubsection{Classes from 39 to 49}
\qquad Except for class 42, all classes here have extreme negative flux values in some pixels. These negative counts imply high Euclidean distances between these spectra and those restricted to positive fluxes. Therefore they are segregated within these classes. Here we give a brief description of these objects.

\begin{itemize}
\item \textbf{Class 39:} Three noisy spectra, one of them flagged as an embedded cluster member.
\item \textbf{Class 40:} Two duplicated spectra of a globular cluster in M31 and one spectrum of the background in the M31 region.
\item \textbf{Class 42:} Two stars with a very similar pattern of sequences of pixels with flux equal to zero.
\item \textbf{Class 43:} One spectrum of the Pal1 globular cluster. This spectrum has deep asymmetric lines.
\item \textbf{Class 44:} One noisy spectrum with negative spikes.
\item \textbf{Class 45:} One background spectrum in the region of M31.
\item \textbf{Class 46:} One stellar spectrum with broad absorption lines.
\item \textbf{Class 47:} One spectrum with great negative spikes.
\item \textbf{Class 48:} One spectrum with high persistence and a positive jump in the blue chip.
\item \textbf{Class 49:} One noisy spectrum with wide absorption lines.
\end{itemize}
\subsubsection{Outliers in classes.} For the first 32 classes, we define the outliers as the spectra with a distance outside of the 3$\sigma$ interval around the mean spectrum of their class. It corresponds, on average, to $1.7 \pm 0.6 $ per cent of the objects in the classes. Exploring their target flags, we notice some phenomena as having high persistence, a positive jump in blue chip, emission lines, sequences of bad pixels, and many stars with signal to noise below 70.

\section{Summary and conclusion}

\subsection{Main results}

\qquad We performed an automated unsupervised classification of 153,847 APOGEE spectra included in SDSS DR12, using $K$-means. We classified the spectra into 50 classes, which were afterwards sorted manually into nine major groups. By construction, each class collects spectra that are very similar. The resulting classes and groups are interpreted using the physical parameters inferred by the APOGEE Stellar Parameters and Chemical Abundances Pipeline (ASPCAP). We found that classes were divided mainly according to their \Teff, \LOGG\, and \MH, and less strongly by other characteristics, such as elemental abundances or the quality of the spectra. Groups from 0 to 7 include 32 classes containing 99.3 per cent of the spectra in DR12. The identified groups can be described as follows:

\begin{itemize}
\item \textbf{Group 0:} Includes five classes dominated by red clump (RC) stars and the warmest end of the red giant branch (RGB) with different chemical abundances;
\item \textbf{Group 1:} Composed of six classes with stars from the RGB, cooler than those in group 0, and mainly separated from each other by their chemical abundances;
\item \textbf{Group 2:} Made up of three classes mainly populated by warm dwarfs, warm subgiant stars, and some A- and B-type stars used for telluric correction;
\item \textbf{Group 3:} Composed of two classes with fast rotating stars. Due to the strong line broadening, they are among the most poorly-fitted spectra in the survey;
\item \textbf{Group 4:} Has two classes covering almost the same range of \Teff\, and \LOGG\, as group 1, RGB stars, but with higher metallicities;
\item \textbf{Group 5:} Contains three classes formed by stars from the RC and the warm end of RGB, with stellar populations from both the thin and thick disk;
\item \textbf{Group 6:} Formed of five classes composed of dwarf stars over a wide range of temperatures;
\item \textbf{Group 7:} Including five classes with peculiar stars;
\item \textbf{Group 8:} Collects 18 classes with all the outliers of the classification, less than 1 per cent of the spectra in SDSS DR12.
\end{itemize}

\subsection{Uses of the classification}

\quad As with any classification, this work can be used to provide an overview of the APOGEE DR12 data set, which simplifies the visualisation and highlights some features of the survey. For example, we can easily see that class 3, composed of very warm stars with almost featureless spectra, has an unexpectedly well-behaved distribution of values for \CM, \NM, \aM, \Mn\, and \Na. It also easily identifies strange behaviours such as the bimodality in \K\, for class 15, the gaps in metallicity found in class 11, and the similarity in parameters of stars with very different spectra, as is the case for classes 20 and 27.

We provide extensive online and appendix material in order to encourage the search for features that may be interesting for specific purposes. For example, the catalogue provides a set of standard spectral templates that could be applied in stellar populations synthesis for galaxies. The mean spectrum (centroids) of the classes are arguably more reliable  templates than the traditional synthetic models of standard MK type stars. However, the application should be restricted to those classes with a high number of members and low internal dispersion. Moreover, calibration of the  atmospheric parameters and abundances is required, since the ones presented here are based on uncalibrated parameters.

The centroids of the classes are also useful to find substantial differences between the spectra and their best fit model found by ASPCAP. Since the classes are a collection of very similar spectra, the comparison between the class' mean and the mean of their best fit model can underline systematic differences between spectra and models. This comparison will be implemented soon and made available in a future publication.

Some classes have a different spatial distribution without an obvious reason, for example, classes in group 2 differ in their spatial distribution, something unexpected since the main difference among them is the \Teff\, of their member stars. Class 31 has an especially peculiar distribution, occupying mainly the region with $60^\mathrm{o} \leq l \leq 90^\mathrm{o} $ and $0^o \leq b \leq 45^o $. The reason is unclear. Further investigations must be carried out to find out the cause of this spatial segregations. Other spatial distributions are less surprising, for example, classes in group 4 are concentrated in the disk. This is to be expected, since their metallicity and \aM\, distributions match those expected for red giants that are part of the thin disk population. Classes 24 and 28, formed by metal-poor stars with high $\alpha$-element abundances, corresponding to the halo population, are expected to be out of the galactic disk, as we found. Class 21 can be interpreted as the population of the bulge, with high $\alpha$-element abundances and high metallicity, and is also expected to have a preferential spatial distribution like the one observed. These are the most evident examples of spatial segregation, but others can be found among the classes.

Finally, the extensive online and appendix material we provide can be used to explore deeper aspects of DR12 APOGEE. We encourage the use of Tables \ref{tab:desc} and \ref{tab:assign} for the reader to explore the results of the classification.

\subsection{Additional issues}

\quad In this work we face the problem of determining the optimal number of clusters for the $K$-means classification. In our case, none of the standard criteria provided a reliable answer. That is probably a consequence of the continuous nature of the dataset. In general, there are no sharp changes in the spectral properties of the stars. Indexes like CH and KL are mathematically proven to work in data sets with well separated clusters, but perform poorly in overlapping clusters or continuous distributions. In this case, $K$-means provides a way of artificially dividing a continuous space into meaningful slices, maximising the similarity among objects in the same class. Thus, the number of classes can be tuned according to the degree of within-class compactness we are interested in, as shown in Section \ref{subsec:find_K}.

Another consequence of applying $K$-means to a continuous data set is a significant observed degree of confusion between classes sharing borders in the space $\mTeff - \mLOGG - \mMH$. However, these issues are not restricted to $K$-means. Any analysis tool, independently of whether it is supervised or not, will face the intrinsic degeneracy of these quantities in the stellar spectra. Soft clustering algorithms such as fuzzy $K$-means or density based algorithms such as Gaussian mixture models or DBSCAN could provide a more natural way to deal with this kind of problem, but would not solve the overlap of the classes in the space of parameters.

We have shown how the random seed used by the algorithm affects its solution. Although there is no unique solution, the variations are negligible compared to the internal dispersion of the classes. In addition, we show how the centroids of the classes are much closer to the spectra in the class than their corresponding best fit models. This suggests that $K$-means can be used to identify the systematic deficiencies of the modelling  adopted in the determination of physical parameters and abundances with ASPCAP, and improve the agreement with the data.

Although the within-class dispersions in the parameter space are larger than the typical uncertainties derived from this kind of data, $K$-means provides good insight into the general characteristics of the spectra in the data set. In this sense, $K$-means is not the optimal algorithm to be used for parameter determination, but can be useful in an early analysis of the data, helping to design solutions and map the general behaviour of the data set.

$K$-means essentially performs hyperspherical cuts in the N-dimensional space. Future works in unsupervised spectral classification should address the issues presented in this section and search for algorithms that can more generically divide the space taking into account its density distribution. Also a soft clustering approach can arguably produce a more reliable classification. However, more complex algorithms are also more computationally expensive, therefore any further application has to address the scalability problem.

\subsection{Conclusions}

\quad As exemplified in this work, $K$-means provides an easy way to divide complex problems into smaller pieces, which are simpler to solve. The version of ASPCAP used in DR12 was designed to work optimally on K and early-M giant stars. For dwarfs, warmer ($T_{\rm eff} >$ 6000 K), cooler ($T_{\rm eff} <$ 3800 K), or metal-poor stars ([M/H] $< -1$), the results are less accurate. Prior to a model-atmospheres spectral analysis, $K$-means can provide guidance on the most natural groups in the data set. This can be very useful to design a pipeline that treats differently the distinct groups of objects, which is  necessary for groups such as 2, 6, 7, and 8, for example.

\citealt{NFLT} puts forward what is known as the 'no free lunch' theorem for machine learning. That is to say, there is no best machine learning algorithm; it is always a matter of which one is better suited to the specific features of a given problem. Knowing the problem, we can only presume which kind of algorithm is most suitable for solving it, but finding the best solution always requires testing some algorithms and tuning their parameters. This work adds to previous applications of $K$-means \citep{sanchez09, sanchez10, Morales-Luis11, sanchez13, sanchez16} consolidating a guideline for the use of this algorithm in the analysis of spectroscopic data, and providing a new perspective for the APOGEE data.

In this work we made a serious effort to organise the spectra into classes and groups according to the similarity within their spectra. This classification is completely independent of any atmospheric and spectroscopic model. It provides a useful way to explore the data in APOGEE, since it allows a quick identification of the main different types of objects in the survey.

\begin{landscape}
\renewcommand{\arraystretch}{1.35}
\begin{table}
	\centering
	\caption{\label{tab:quantiles0} Median atmospheric parameters and chemical abundances for the 32 most populated classes. Last column contains the number of stars in each class.}
	\begin{tabular}{cccccccccccc}
		Class & $T_\mathrm{eff}$ (K) &  $\log g$  &  [M/H]  &  [C/M]  &  [N/M]  &  [$\alpha$/M]  &  [Al/H]  &  [Ca/H]  &  [C/H]  &  [Fe/H]  & $N_{\star}$ \\ \hline 	00 & $4853\pm^{ 144}_{ 159}$ & $+2.93\pm^{0.34}_{0.19}$ & $-0.31\pm^{0.11}_{0.12}$ & $+0.03\pm^{0.09}_{0.08}$ & $+0.04\pm^{0.09}_{0.11}$ & $+0.08\pm^{0.13}_{0.04}$ & $+0.02\pm^{0.10}_{0.09}$ & $+0.04\pm^{0.09}_{0.12}$ & $+0.14\pm^{0.11}_{0.07}$ & $-0.29\pm^{0.22}_{0.23}$ & 15066 \\ 
	01 & $4731\pm^{ 144}_{ 231}$ & $+2.81\pm^{0.21}_{0.35}$ & $-0.22\pm^{0.11}_{0.15}$ & $-0.01\pm^{0.10}_{0.08}$ & $+0.10\pm^{0.09}_{0.08}$ & $+0.07\pm^{0.12}_{0.04}$ & $-0.01\pm^{0.10}_{0.08}$ & $+0.10\pm^{0.09}_{0.09}$ & $+0.12\pm^{0.10}_{0.06}$ & $-0.23\pm^{0.18}_{0.21}$ & 14177 \\ 
	02 & $4712\pm^{ 130}_{ 175}$ & $+2.83\pm^{0.25}_{0.23}$ & $-0.07\pm^{0.10}_{0.11}$ & $-0.05\pm^{0.10}_{0.07}$ & $+0.15\pm^{0.08}_{0.08}$ & $+0.05\pm^{0.05}_{0.03}$ & $-0.05\pm^{0.10}_{0.08}$ & $+0.15\pm^{0.08}_{0.08}$ & $+0.07\pm^{0.07}_{0.04}$ & $-0.09\pm^{0.16}_{0.16}$ & 12482 \\ 
	03 & $7814\pm^{ 171}_{ 617}$ & $+4.76\pm^{0.22}_{0.35}$ & $-2.03\pm^{0.73}_{0.35}$ & $-0.18\pm^{0.44}_{0.46}$ & $+0.18\pm^{0.45}_{0.59}$ & $-0.26\pm^{0.34}_{0.41}$ & $-1.00\pm^{1.73}_{0.00}$ & $+0.42\pm^{0.58}_{1.17}$ & $+0.31\pm^{0.65}_{0.84}$ & $+0.12\pm^{0.38}_{1.88}$ & 10628 \\ 
	04 & $4679\pm^{ 137}_{ 145}$ & $+2.84\pm^{0.34}_{0.21}$ & $+0.07\pm^{0.10}_{0.11}$ & $-0.04\pm^{0.09}_{0.08}$ & $+0.19\pm^{0.09}_{0.09}$ & $+0.04\pm^{0.04}_{0.03}$ & $-0.04\pm^{0.09}_{0.09}$ & $+0.19\pm^{0.09}_{0.10}$ & $+0.05\pm^{0.06}_{0.04}$ & $+0.07\pm^{0.17}_{0.15}$ & 10253 \\ 
	05 & $4941\pm^{ 590}_{ 203}$ & $+3.16\pm^{1.04}_{0.38}$ & $-0.45\pm^{0.28}_{0.19}$ & $+0.08\pm^{0.11}_{0.12}$ & $+0.00\pm^{0.19}_{0.19}$ & $+0.11\pm^{0.14}_{0.09}$ & $+0.07\pm^{0.13}_{0.12}$ & $-0.01\pm^{0.20}_{0.18}$ & $+0.18\pm^{0.14}_{0.14}$ & $-0.35\pm^{0.37}_{0.44}$ &  9144 \\ 
	06 & $4589\pm^{ 136}_{ 158}$ & $+2.76\pm^{0.28}_{0.27}$ & $+0.17\pm^{0.10}_{0.12}$ & $-0.03\pm^{0.07}_{0.08}$ & $+0.24\pm^{0.09}_{0.08}$ & $+0.04\pm^{0.03}_{0.03}$ & $-0.03\pm^{0.08}_{0.08}$ & $+0.24\pm^{0.09}_{0.09}$ & $+0.04\pm^{0.05}_{0.04}$ & $+0.19\pm^{0.19}_{0.16}$ &  8096 \\ 
	07 & $4236\pm^{  97}_{ 101}$ & $+2.03\pm^{0.23}_{0.22}$ & $-0.29\pm^{0.14}_{0.15}$ & $-0.02\pm^{0.10}_{0.08}$ & $+0.17\pm^{0.07}_{0.09}$ & $+0.09\pm^{0.12}_{0.04}$ & $-0.02\pm^{0.11}_{0.09}$ & $+0.17\pm^{0.07}_{0.10}$ & $+0.12\pm^{0.12}_{0.05}$ & $-0.34\pm^{0.17}_{0.21}$ &  5325 \\ 
	08 & $4919\pm^{ 270}_{ 111}$ & $+3.45\pm^{0.42}_{0.22}$ & $-0.02\pm^{0.15}_{0.11}$ & $+0.02\pm^{0.08}_{0.09}$ & $+0.03\pm^{0.11}_{0.14}$ & $+0.05\pm^{0.07}_{0.03}$ & $+0.02\pm^{0.09}_{0.09}$ & $+0.01\pm^{0.12}_{0.14}$ & $+0.08\pm^{0.08}_{0.05}$ & $+0.03\pm^{0.20}_{0.15}$ &  5054 \\ 
	09 & $4495\pm^{ 108}_{ 145}$ & $+2.70\pm^{0.19}_{0.28}$ & $+0.30\pm^{0.09}_{0.12}$ & $-0.00\pm^{0.05}_{0.08}$ & $+0.30\pm^{0.10}_{0.08}$ & $+0.04\pm^{0.03}_{0.03}$ & $-0.00\pm^{0.05}_{0.08}$ & $+0.29\pm^{0.10}_{0.08}$ & $+0.03\pm^{0.04}_{0.04}$ & $+0.39\pm^{0.11}_{0.18}$ &  4820 \\ 
	10 & $4791\pm^{ 359}_{ 242}$ & $+4.27\pm^{0.14}_{0.18}$ & $-0.15\pm^{0.14}_{0.17}$ & $-0.08\pm^{0.11}_{0.15}$ & $+0.04\pm^{0.22}_{0.16}$ & $+0.04\pm^{0.08}_{0.04}$ & $-0.07\pm^{0.11}_{0.14}$ & $+0.01\pm^{0.20}_{0.16}$ & $+0.07\pm^{0.10}_{0.08}$ & $-0.08\pm^{0.16}_{0.19}$ &  4771 \\ 
	11 & $6125\pm^{ 478}_{ 698}$ & $+4.47\pm^{0.50}_{0.52}$ & $-0.24\pm^{0.27}_{0.37}$ & $+0.06\pm^{0.17}_{0.18}$ & $+0.26\pm^{0.54}_{0.55}$ & $+0.01\pm^{0.12}_{0.05}$ & $-0.04\pm^{0.31}_{0.96}$ & $+0.74\pm^{0.26}_{0.77}$ & $+0.05\pm^{0.12}_{0.07}$ & $-0.20\pm^{0.27}_{0.73}$ &  4271 \\ 
	12 & $4761\pm^{ 241}_{ 166}$ & $+4.35\pm^{0.08}_{0.12}$ & $+0.13\pm^{0.13}_{0.11}$ & $-0.05\pm^{0.07}_{0.14}$ & $+0.06\pm^{0.13}_{0.11}$ & $+0.00\pm^{0.03}_{0.02}$ & $-0.04\pm^{0.07}_{0.12}$ & $+0.04\pm^{0.12}_{0.12}$ & $+0.03\pm^{0.06}_{0.04}$ & $+0.19\pm^{0.22}_{0.16}$ &  4051 \\ 
	13 & $6396\pm^{ 482}_{1681}$ & $+4.35\pm^{0.61}_{2.10}$ & $-0.88\pm^{0.57}_{1.06}$ & $+0.07\pm^{0.33}_{0.33}$ & $+0.33\pm^{0.48}_{0.62}$ & $+0.01\pm^{0.26}_{0.25}$ & $-0.93\pm^{1.51}_{0.07}$ & $+1.00\pm^{0.00}_{0.63}$ & $+0.08\pm^{0.32}_{0.18}$ & $-1.26\pm^{1.10}_{1.24}$ &  3696 \\ 
	14 & $4095\pm^{  98}_{ 107}$ & $+1.85\pm^{0.24}_{0.24}$ & $-0.09\pm^{0.13}_{0.14}$ & $-0.03\pm^{0.09}_{0.08}$ & $+0.19\pm^{0.06}_{0.07}$ & $+0.06\pm^{0.10}_{0.04}$ & $-0.03\pm^{0.09}_{0.09}$ & $+0.19\pm^{0.06}_{0.07}$ & $+0.08\pm^{0.10}_{0.05}$ & $-0.11\pm^{0.15}_{0.17}$ &  3545 \\ 
	15 & $4065\pm^{ 241}_{ 159}$ & $+4.25\pm^{0.08}_{0.13}$ & $-0.18\pm^{0.13}_{0.15}$ & $+0.02\pm^{0.03}_{0.18}$ & $-0.30\pm^{0.37}_{0.31}$ & $-0.05\pm^{0.06}_{0.04}$ & $+0.01\pm^{0.03}_{0.14}$ & $-0.24\pm^{0.31}_{0.30}$ & $-0.05\pm^{0.08}_{0.05}$ & $-0.05\pm^{0.17}_{0.18}$ &  3450 \\ 
	16 & $3620\pm^{ 104}_{ 101}$ & $+1.30\pm^{0.24}_{0.26}$ & $+0.17\pm^{0.13}_{0.15}$ & $+0.04\pm^{0.02}_{0.03}$ & $+0.14\pm^{0.08}_{0.07}$ & $+0.00\pm^{0.03}_{0.02}$ & $+0.00\pm^{0.03}_{0.03}$ & $+0.16\pm^{0.08}_{0.07}$ & $-0.01\pm^{0.04}_{0.02}$ & $+0.41\pm^{0.09}_{0.25}$ &  3157 \\ 
	17 & $4361\pm^{ 104}_{ 221}$ & $+4.26\pm^{0.08}_{0.11}$ & $+0.01\pm^{0.13}_{0.09}$ & $-0.01\pm^{0.04}_{0.17}$ & $+0.01\pm^{0.13}_{0.34}$ & $-0.02\pm^{0.04}_{0.04}$ & $-0.02\pm^{0.05}_{0.14}$ & $-0.01\pm^{0.13}_{0.30}$ & $-0.02\pm^{0.06}_{0.05}$ & $+0.14\pm^{0.24}_{0.15}$ &  3109 \\ 
	18 & $3868\pm^{  95}_{  88}$ & $+1.65\pm^{0.22}_{0.25}$ & $+0.18\pm^{0.15}_{0.18}$ & $+0.02\pm^{0.03}_{0.05}$ & $+0.21\pm^{0.08}_{0.08}$ & $+0.02\pm^{0.03}_{0.02}$ & $+0.00\pm^{0.03}_{0.05}$ & $+0.21\pm^{0.08}_{0.08}$ & $+0.01\pm^{0.04}_{0.03}$ & $+0.32\pm^{0.18}_{0.27}$ &  2955 \\ 
	19 & $3753\pm^{  95}_{  95}$ & $+1.12\pm^{0.26}_{0.33}$ & $-0.34\pm^{0.17}_{0.24}$ & $-0.02\pm^{0.13}_{0.08}$ & $+0.12\pm^{0.09}_{0.11}$ & $+0.08\pm^{0.15}_{0.05}$ & $-0.04\pm^{0.13}_{0.09}$ & $+0.14\pm^{0.09}_{0.11}$ & $+0.10\pm^{0.15}_{0.06}$ & $-0.35\pm^{0.21}_{0.33}$ &  2874 \\ 
	20 & $3724\pm^{ 131}_{ 128}$ & $+4.23\pm^{0.16}_{0.37}$ & $-0.22\pm^{0.19}_{0.16}$ & $+0.02\pm^{0.03}_{0.24}$ & $-0.24\pm^{0.45}_{0.54}$ & $-0.07\pm^{0.04}_{0.21}$ & $+0.02\pm^{0.04}_{0.21}$ & $-0.14\pm^{0.49}_{0.49}$ & $-0.07\pm^{0.05}_{0.21}$ & $-0.06\pm^{0.19}_{0.20}$ &  2771 \\ 
	21 & $3500\pm^{   1}_{   0}$ & $+0.58\pm^{0.32}_{0.35}$ & $-0.21\pm^{0.20}_{0.27}$ & $+0.19\pm^{0.04}_{0.16}$ & $+0.25\pm^{0.07}_{0.08}$ & $+0.27\pm^{0.06}_{0.18}$ & $+0.09\pm^{0.06}_{0.11}$ & $+0.29\pm^{0.07}_{0.08}$ & $+0.31\pm^{0.04}_{0.22}$ & $+0.22\pm^{0.27}_{0.30}$ &  2556 \\ 
	22 & $4137\pm^{ 114}_{ 102}$ & $+2.09\pm^{0.27}_{0.24}$ & $+0.23\pm^{0.13}_{0.14}$ & $+0.01\pm^{0.04}_{0.07}$ & $+0.27\pm^{0.10}_{0.08}$ & $+0.03\pm^{0.04}_{0.03}$ & $-0.00\pm^{0.04}_{0.08}$ & $+0.26\pm^{0.10}_{0.08}$ & $+0.03\pm^{0.04}_{0.04}$ & $+0.33\pm^{0.17}_{0.22}$ &  2494 \\ 
	23 & $3503\pm^{ 108}_{   3}$ & $+4.15\pm^{0.56}_{0.40}$ & $-0.54\pm^{0.24}_{0.55}$ & $+0.07\pm^{0.26}_{0.23}$ & $+0.28\pm^{0.60}_{0.65}$ & $-0.08\pm^{0.23}_{0.22}$ & $+0.07\pm^{0.27}_{0.06}$ & $+0.26\pm^{0.57}_{0.64}$ & $-0.05\pm^{0.25}_{0.08}$ & $-0.14\pm^{0.23}_{0.25}$ &  2425 \\ 
	24 & $4582\pm^{ 330}_{ 322}$ & $+2.22\pm^{0.54}_{0.60}$ & $-1.20\pm^{0.25}_{0.22}$ & $-0.08\pm^{0.32}_{0.25}$ & $+0.21\pm^{0.45}_{0.24}$ & $+0.24\pm^{0.07}_{0.07}$ & $-0.14\pm^{0.36}_{0.25}$ & $+0.26\pm^{0.41}_{0.29}$ & $+0.30\pm^{0.09}_{0.09}$ & $-1.47\pm^{0.87}_{1.02}$ &  2388 \\ 
	25 & $3561\pm^{  83}_{  59}$ & $+0.79\pm^{0.25}_{0.38}$ & $-0.31\pm^{0.17}_{0.29}$ & $+0.01\pm^{0.16}_{0.06}$ & $+0.11\pm^{0.08}_{0.09}$ & $+0.08\pm^{0.18}_{0.05}$ & $-0.02\pm^{0.14}_{0.06}$ & $+0.16\pm^{0.09}_{0.09}$ & $+0.09\pm^{0.22}_{0.06}$ & $-0.24\pm^{0.19}_{0.30}$ &  2288 \\ 
	26 & $3899\pm^{  77}_{ 103}$ & $+1.37\pm^{0.21}_{0.33}$ & $-0.47\pm^{0.14}_{0.20}$ & $-0.03\pm^{0.12}_{0.10}$ & $+0.11\pm^{0.11}_{0.10}$ & $+0.14\pm^{0.11}_{0.09}$ & $-0.03\pm^{0.12}_{0.11}$ & $+0.12\pm^{0.11}_{0.11}$ & $+0.18\pm^{0.12}_{0.10}$ & $-0.55\pm^{0.19}_{0.38}$ &  2252 \\ 
	27 & $4243\pm^{ 641}_{ 282}$ & $+4.83\pm^{0.17}_{0.84}$ & $-0.21\pm^{0.21}_{0.20}$ & $+0.03\pm^{0.04}_{0.24}$ & $-0.44\pm^{0.74}_{0.50}$ & $-0.14\pm^{0.11}_{0.20}$ & $+0.02\pm^{0.05}_{0.22}$ & $-0.48\pm^{0.67}_{0.47}$ & $-0.18\pm^{0.14}_{0.20}$ & $-0.19\pm^{0.26}_{0.50}$ &  1431 \\ 
	28 & $4027\pm^{ 119}_{ 165}$ & $+1.47\pm^{0.35}_{0.66}$ & $-0.81\pm^{0.19}_{0.33}$ & $-0.09\pm^{0.15}_{0.30}$ & $+0.09\pm^{0.18}_{0.10}$ & $+0.24\pm^{0.04}_{0.11}$ & $-0.10\pm^{0.16}_{0.31}$ & $+0.10\pm^{0.21}_{0.11}$ & $+0.27\pm^{0.06}_{0.10}$ & $-1.14\pm^{0.42}_{0.73}$ &  1117 \\ 
	29 & $3973\pm^{ 244}_{ 361}$ & $+4.97\pm^{0.03}_{0.79}$ & $-0.40\pm^{0.20}_{0.48}$ & $+0.05\pm^{0.25}_{0.25}$ & $-0.35\pm^{0.79}_{0.59}$ & $-0.16\pm^{0.19}_{0.24}$ & $+0.05\pm^{0.25}_{0.25}$ & $-0.36\pm^{0.76}_{0.51}$ & $-0.19\pm^{0.23}_{0.23}$ & $-0.46\pm^{0.46}_{0.80}$ &  1081 \\ 
	30 & $4202\pm^{1744}_{ 659}$ & $+3.14\pm^{1.16}_{1.05}$ & $-0.54\pm^{0.83}_{0.84}$ & $+0.24\pm^{0.57}_{0.43}$ & $+0.09\pm^{0.78}_{0.56}$ & $+0.02\pm^{0.34}_{0.43}$ & $+0.20\pm^{0.59}_{0.34}$ & $+0.11\pm^{0.76}_{0.59}$ & $+0.06\pm^{0.44}_{0.45}$ & $-0.23\pm^{0.73}_{1.07}$ &   562 \\ 
	31 & $4458\pm^{ 386}_{ 528}$ & $+2.17\pm^{0.83}_{0.73}$ & $-1.34\pm^{0.47}_{0.58}$ & $+0.04\pm^{0.45}_{0.31}$ & $+0.18\pm^{0.53}_{0.32}$ & $+0.20\pm^{0.11}_{0.23}$ & $-0.04\pm^{0.42}_{0.32}$ & $+0.19\pm^{0.49}_{0.35}$ & $+0.29\pm^{0.19}_{0.27}$ & $-1.10\pm^{0.50}_{1.10}$ &   474 \\ 
	\end{tabular}
\end{table}
\end{landscape}
\begin{landscape}
\renewcommand{\arraystretch}{1.35}
\begin{table}
	\centering
	\caption{\label{tab:quantiles1} Median chemical abundances for the 32 most populated classes. Last column contains the number of stars in each class.}
	\begin{tabular}{ccccccccccccc}
Class &  [K/H]  &  [Mg/H]  &  [Mn/H]  &  [Na/H]  &  [Ni/H]  &  [N/H]  &  [O/H]  &  [Si/H]  &  [S/H]  &  [Ti/H]  &  [V/H]  & $N_{\star}$ \\ \hline
	00 & $+0.04\pm^{0.14}_{0.06}$ & $-0.11\pm^{0.14}_{0.13}$ & $+0.20\pm^{0.10}_{0.09}$ & $+0.20\pm^{0.11}_{0.09}$ & $-0.39\pm^{0.11}_{0.13}$ & $+0.03\pm^{0.13}_{0.06}$ & $-0.07\pm^{0.11}_{0.09}$ & $-0.57\pm^{0.42}_{1.12}$ & $-0.35\pm^{0.13}_{0.16}$ & $-0.30\pm^{0.12}_{0.13}$ & $-0.37\pm^{0.11}_{0.13}$ & 15066 \\ 
	01 & $+0.01\pm^{0.13}_{0.06}$ & $-0.04\pm^{0.13}_{0.15}$ & $+0.21\pm^{0.09}_{0.08}$ & $+0.19\pm^{0.10}_{0.08}$ & $-0.31\pm^{0.09}_{0.15}$ & $-0.00\pm^{0.11}_{0.05}$ & $-0.06\pm^{0.10}_{0.07}$ & $-0.43\pm^{0.31}_{0.47}$ & $-0.24\pm^{0.13}_{0.17}$ & $-0.21\pm^{0.11}_{0.15}$ & $-0.26\pm^{0.10}_{0.15}$ & 14177 \\ 
	02 & $-0.02\pm^{0.08}_{0.04}$ & $+0.07\pm^{0.12}_{0.12}$ & $+0.18\pm^{0.07}_{0.07}$ & $+0.16\pm^{0.07}_{0.07}$ & $-0.19\pm^{0.09}_{0.10}$ & $-0.03\pm^{0.08}_{0.04}$ & $-0.06\pm^{0.08}_{0.07}$ & $-0.22\pm^{0.20}_{0.30}$ & $-0.05\pm^{0.12}_{0.13}$ & $-0.06\pm^{0.10}_{0.11}$ & $-0.11\pm^{0.10}_{0.11}$ & 12482 \\ 
	03 & $+0.77\pm^{0.23}_{1.26}$ & $-2.16\pm^{0.64}_{0.34}$ & $+0.26\pm^{0.37}_{0.51}$ & $+0.46\pm^{0.42}_{0.49}$ & $-2.50\pm^{2.02}_{0.00}$ & $-0.31\pm^{1.31}_{0.69}$ & $-0.35\pm^{1.35}_{0.65}$ & $+0.14\pm^{0.36}_{0.71}$ & $-0.60\pm^{0.43}_{1.03}$ & $-1.21\pm^{0.34}_{0.38}$ & $-2.50\pm^{1.30}_{0.00}$ & 10628 \\ 
	04 & $-0.02\pm^{0.07}_{0.04}$ & $+0.19\pm^{0.14}_{0.13}$ & $+0.16\pm^{0.07}_{0.07}$ & $+0.13\pm^{0.06}_{0.07}$ & $-0.08\pm^{0.10}_{0.10}$ & $-0.05\pm^{0.07}_{0.03}$ & $-0.02\pm^{0.08}_{0.06}$ & $-0.06\pm^{0.19}_{0.20}$ & $+0.15\pm^{0.12}_{0.14}$ & $+0.08\pm^{0.10}_{0.11}$ & $+0.04\pm^{0.11}_{0.12}$ & 10253 \\ 
	05 & $+0.12\pm^{0.13}_{0.11}$ & $-0.22\pm^{0.22}_{0.18}$ & $+0.20\pm^{0.14}_{0.16}$ & $+0.26\pm^{0.14}_{0.15}$ & $-0.49\pm^{0.22}_{0.18}$ & $+0.08\pm^{0.13}_{0.10}$ & $-0.02\pm^{0.14}_{0.16}$ & $-0.87\pm^{0.69}_{1.63}$ & $-0.53\pm^{0.32}_{0.24}$ & $-0.45\pm^{0.29}_{0.20}$ & $-0.52\pm^{0.28}_{0.19}$ &  9144 \\ 
	06 & $-0.01\pm^{0.06}_{0.04}$ & $+0.28\pm^{0.14}_{0.14}$ & $+0.16\pm^{0.07}_{0.07}$ & $+0.12\pm^{0.06}_{0.06}$ & $+0.01\pm^{0.11}_{0.12}$ & $-0.07\pm^{0.06}_{0.03}$ & $+0.02\pm^{0.08}_{0.06}$ & $+0.06\pm^{0.21}_{0.18}$ & $+0.30\pm^{0.13}_{0.15}$ & $+0.17\pm^{0.11}_{0.12}$ & $+0.15\pm^{0.12}_{0.13}$ &  8096 \\ 
	07 & $+0.05\pm^{0.13}_{0.05}$ & $-0.12\pm^{0.15}_{0.17}$ & $+0.23\pm^{0.09}_{0.07}$ & $+0.16\pm^{0.12}_{0.06}$ & $-0.37\pm^{0.14}_{0.17}$ & $-0.02\pm^{0.10}_{0.04}$ & $+0.02\pm^{0.11}_{0.07}$ & $-0.43\pm^{0.16}_{0.18}$ & $-0.26\pm^{0.17}_{0.18}$ & $-0.29\pm^{0.14}_{0.15}$ & $-0.31\pm^{0.14}_{0.14}$ &  5325 \\ 
	08 & $+0.03\pm^{0.10}_{0.05}$ & $+0.13\pm^{0.19}_{0.14}$ & $+0.13\pm^{0.07}_{0.07}$ & $+0.13\pm^{0.08}_{0.09}$ & $-0.16\pm^{0.12}_{0.10}$ & $-0.01\pm^{0.15}_{0.05}$ & $-0.07\pm^{0.11}_{0.10}$ & $-0.13\pm^{0.24}_{0.53}$ & $-0.02\pm^{0.17}_{0.14}$ & $-0.02\pm^{0.15}_{0.11}$ & $-0.08\pm^{0.16}_{0.11}$ &  5054 \\ 
	09 & $+0.01\pm^{0.06}_{0.04}$ & $+0.41\pm^{0.09}_{0.15}$ & $+0.16\pm^{0.07}_{0.07}$ & $+0.11\pm^{0.05}_{0.06}$ & $+0.13\pm^{0.11}_{0.14}$ & $-0.09\pm^{0.06}_{0.04}$ & $+0.08\pm^{0.07}_{0.07}$ & $+0.26\pm^{0.18}_{0.19}$ & $+0.49\pm^{0.01}_{0.16}$ & $+0.30\pm^{0.09}_{0.12}$ & $+0.31\pm^{0.11}_{0.14}$ &  4820 \\ 
	10 & $+0.10\pm^{0.11}_{0.07}$ & $-0.04\pm^{0.22}_{0.19}$ & $+0.06\pm^{0.09}_{0.09}$ & $+0.11\pm^{0.13}_{0.10}$ & $-0.28\pm^{0.12}_{0.14}$ & $+0.01\pm^{0.12}_{0.06}$ & $+0.01\pm^{0.12}_{0.16}$ & $-0.31\pm^{0.31}_{0.50}$ & $-0.21\pm^{0.17}_{0.20}$ & $-0.16\pm^{0.16}_{0.18}$ & $-0.24\pm^{0.15}_{0.19}$ &  4771 \\ 
	11 & $+0.06\pm^{0.15}_{0.07}$ & $-0.37\pm^{0.24}_{0.22}$ & $+0.00\pm^{0.15}_{0.14}$ & $+0.20\pm^{0.19}_{0.19}$ & $-0.37\pm^{0.23}_{0.29}$ & $+0.03\pm^{0.11}_{0.07}$ & $+0.13\pm^{0.32}_{0.31}$ & $-0.42\pm^{0.33}_{1.99}$ & $-0.28\pm^{0.28}_{0.40}$ & $-0.28\pm^{0.25}_{0.36}$ & $-0.31\pm^{0.31}_{0.38}$ &  4271 \\ 
	12 & $+0.07\pm^{0.05}_{0.04}$ & $+0.25\pm^{0.22}_{0.19}$ & $+0.04\pm^{0.06}_{0.06}$ & $+0.05\pm^{0.08}_{0.06}$ & $-0.04\pm^{0.15}_{0.12}$ & $-0.02\pm^{0.09}_{0.04}$ & $+0.07\pm^{0.11}_{0.13}$ & $+0.03\pm^{0.26}_{0.27}$ & $+0.14\pm^{0.18}_{0.15}$ & $+0.13\pm^{0.14}_{0.12}$ & $+0.05\pm^{0.17}_{0.14}$ &  4051 \\ 
	13 & $+0.22\pm^{0.25}_{0.21}$ & $-1.19\pm^{0.48}_{0.81}$ & $-0.10\pm^{0.44}_{0.30}$ & $+0.19\pm^{0.42}_{0.28}$ & $-1.22\pm^{0.74}_{1.23}$ & $+0.05\pm^{0.37}_{0.42}$ & $+0.55\pm^{0.45}_{0.67}$ & $-1.65\pm^{1.50}_{0.85}$ & $-0.53\pm^{0.38}_{1.29}$ & $-0.85\pm^{0.45}_{1.03}$ & $-0.85\pm^{0.54}_{1.11}$ &  3696 \\ 
	14 & $+0.06\pm^{0.10}_{0.05}$ & $+0.02\pm^{0.15}_{0.18}$ & $+0.18\pm^{0.07}_{0.07}$ & $+0.10\pm^{0.10}_{0.05}$ & $-0.20\pm^{0.11}_{0.15}$ & $-0.06\pm^{0.06}_{0.04}$ & $+0.08\pm^{0.10}_{0.07}$ & $-0.18\pm^{0.15}_{0.15}$ & $+0.03\pm^{0.17}_{0.18}$ & $-0.10\pm^{0.13}_{0.14}$ & $-0.10\pm^{0.13}_{0.14}$ &  3545 \\ 
	15 & $+0.02\pm^{0.16}_{0.35}$ & $-0.24\pm^{0.17}_{0.17}$ & $-0.02\pm^{0.08}_{0.04}$ & $-0.04\pm^{0.11}_{0.05}$ & $-0.29\pm^{0.12}_{0.13}$ & $-0.02\pm^{0.07}_{0.07}$ & $-0.18\pm^{0.09}_{0.06}$ & $-0.55\pm^{0.23}_{0.36}$ & $-0.19\pm^{0.18}_{0.19}$ & $-0.20\pm^{0.13}_{0.15}$ & $-0.28\pm^{0.15}_{0.18}$ &  3450 \\ 
	16 & $+0.01\pm^{0.12}_{0.04}$ & $+0.18\pm^{0.20}_{0.19}$ & $+0.00\pm^{0.11}_{0.04}$ & $-0.02\pm^{0.05}_{0.04}$ & $+0.11\pm^{0.17}_{0.20}$ & $-0.24\pm^{0.10}_{0.03}$ & $-0.12\pm^{0.20}_{0.05}$ & $+0.28\pm^{0.22}_{0.23}$ & $+0.50\pm^{0.00}_{0.18}$ & $+0.13\pm^{0.13}_{0.14}$ & $+0.29\pm^{0.15}_{0.19}$ &  3157 \\ 
	17 & $+0.10\pm^{0.06}_{0.05}$ & $+0.04\pm^{0.24}_{0.16}$ & $-0.04\pm^{0.06}_{0.04}$ & $-0.01\pm^{0.08}_{0.05}$ & $-0.12\pm^{0.14}_{0.10}$ & $-0.01\pm^{0.07}_{0.04}$ & $-0.09\pm^{0.17}_{0.11}$ & $-0.24\pm^{0.25}_{0.26}$ & $+0.01\pm^{0.18}_{0.14}$ & $-0.02\pm^{0.13}_{0.09}$ & $-0.11\pm^{0.15}_{0.12}$ &  3109 \\ 
	18 & $+0.05\pm^{0.09}_{0.06}$ & $+0.25\pm^{0.19}_{0.21}$ & $+0.11\pm^{0.09}_{0.09}$ & $+0.02\pm^{0.05}_{0.05}$ & $+0.03\pm^{0.18}_{0.18}$ & $-0.16\pm^{0.07}_{0.06}$ & $+0.00\pm^{0.19}_{0.13}$ & $+0.20\pm^{0.24}_{0.24}$ & $+0.46\pm^{0.04}_{0.25}$ & $+0.16\pm^{0.15}_{0.18}$ & $+0.22\pm^{0.17}_{0.20}$ &  2955 \\ 
	19 & $+0.16\pm^{0.09}_{0.11}$ & $-0.27\pm^{0.23}_{0.29}$ & $+0.18\pm^{0.10}_{0.10}$ & $+0.08\pm^{0.15}_{0.05}$ & $-0.40\pm^{0.19}_{0.25}$ & $-0.03\pm^{0.09}_{0.06}$ & $+0.14\pm^{0.11}_{0.18}$ & $-0.35\pm^{0.19}_{0.25}$ & $-0.24\pm^{0.24}_{0.30}$ & $-0.35\pm^{0.16}_{0.24}$ & $-0.33\pm^{0.18}_{0.26}$ &  2874 \\ 
	20 & $-0.27\pm^{0.09}_{0.19}$ & $-0.39\pm^{0.19}_{0.23}$ & $-0.05\pm^{0.08}_{0.11}$ & $-0.05\pm^{0.09}_{0.18}$ & $-0.31\pm^{0.15}_{0.18}$ & $-0.06\pm^{0.08}_{0.22}$ & $-0.19\pm^{0.04}_{0.20}$ & $-0.62\pm^{0.26}_{0.37}$ & $-0.19\pm^{0.22}_{0.22}$ & $-0.26\pm^{0.19}_{0.17}$ & $-0.30\pm^{0.21}_{0.21}$ &  2771 \\ 
	21 & $+0.13\pm^{0.07}_{0.13}$ & $-0.07\pm^{0.24}_{0.30}$ & $-0.02\pm^{0.15}_{0.04}$ & $+0.24\pm^{0.10}_{0.18}$ & $-0.14\pm^{0.26}_{0.35}$ & $-0.04\pm^{0.11}_{0.06}$ & $+0.23\pm^{0.18}_{0.18}$ & $+0.29\pm^{0.21}_{0.25}$ & $+0.02\pm^{0.31}_{0.39}$ & $-0.25\pm^{0.20}_{0.27}$ & $+0.03\pm^{0.25}_{0.25}$ &  2556 \\ 
	22 & $+0.04\pm^{0.07}_{0.05}$ & $+0.32\pm^{0.16}_{0.18}$ & $+0.16\pm^{0.08}_{0.07}$ & $+0.07\pm^{0.06}_{0.06}$ & $+0.06\pm^{0.15}_{0.16}$ & $-0.12\pm^{0.06}_{0.04}$ & $+0.10\pm^{0.09}_{0.09}$ & $+0.21\pm^{0.20}_{0.20}$ & $+0.47\pm^{0.03}_{0.18}$ & $+0.22\pm^{0.13}_{0.14}$ & $+0.25\pm^{0.15}_{0.16}$ &  2494 \\ 
	23 & $-0.23\pm^{0.24}_{0.11}$ & $-0.84\pm^{0.33}_{0.44}$ & $-0.04\pm^{0.24}_{0.21}$ & $+0.02\pm^{0.37}_{0.10}$ & $-0.42\pm^{0.24}_{0.27}$ & $-0.01\pm^{0.35}_{0.15}$ & $-0.18\pm^{0.28}_{0.11}$ & $-0.76\pm^{0.32}_{0.48}$ & $-0.38\pm^{0.42}_{0.65}$ & $-0.81\pm^{0.42}_{0.78}$ & $-0.67\pm^{0.34}_{0.60}$ &  2425 \\ 
	24 & $+0.14\pm^{0.13}_{0.12}$ & $-1.19\pm^{0.55}_{0.39}$ & $+0.35\pm^{0.09}_{0.09}$ & $+0.41\pm^{0.14}_{0.14}$ & $-1.19\pm^{0.31}_{0.39}$ & $+0.17\pm^{0.13}_{0.11}$ & $+0.00\pm^{0.14}_{0.12}$ & $-1.90\pm^{0.55}_{0.60}$ & $-1.40\pm^{0.30}_{0.23}$ & $-1.21\pm^{0.24}_{0.22}$ & $-1.30\pm^{0.24}_{0.26}$ &  2388 \\ 
	25 & $+0.17\pm^{0.08}_{0.12}$ & $-0.26\pm^{0.19}_{0.28}$ & $+0.09\pm^{0.13}_{0.08}$ & $+0.07\pm^{0.20}_{0.05}$ & $-0.33\pm^{0.17}_{0.27}$ & $-0.06\pm^{0.11}_{0.09}$ & $+0.07\pm^{0.17}_{0.18}$ & $-0.20\pm^{0.17}_{0.26}$ & $-0.15\pm^{0.26}_{0.36}$ & $-0.34\pm^{0.17}_{0.27}$ & $-0.25\pm^{0.18}_{0.29}$ &  2288 \\ 
	26 & $+0.18\pm^{0.07}_{0.14}$ & $-0.34\pm^{0.20}_{0.28}$ & $+0.27\pm^{0.10}_{0.11}$ & $+0.16\pm^{0.13}_{0.10}$ & $-0.53\pm^{0.17}_{0.21}$ & $+0.02\pm^{0.10}_{0.08}$ & $+0.13\pm^{0.11}_{0.13}$ & $-0.54\pm^{0.18}_{0.24}$ & $-0.44\pm^{0.18}_{0.27}$ & $-0.47\pm^{0.14}_{0.20}$ & $-0.48\pm^{0.15}_{0.22}$ &  2252 \\ 
	27 & $-0.27\pm^{0.42}_{0.26}$ & $-0.76\pm^{0.38}_{0.42}$ & $-0.11\pm^{0.12}_{0.17}$ & $-0.11\pm^{0.28}_{0.20}$ & $-0.49\pm^{0.17}_{0.25}$ & $-0.18\pm^{0.20}_{0.16}$ & $-0.30\pm^{0.19}_{0.21}$ & $-0.82\pm^{0.78}_{1.32}$ & $-0.24\pm^{0.22}_{0.22}$ & $-0.35\pm^{0.19}_{0.22}$ & $-0.31\pm^{0.22}_{0.23}$ &  1431 \\ 
	28 & $+0.18\pm^{0.06}_{0.15}$ & $-0.69\pm^{0.33}_{0.59}$ & $+0.38\pm^{0.09}_{0.10}$ & $+0.28\pm^{0.08}_{0.13}$ & $-0.90\pm^{0.28}_{0.34}$ & $+0.12\pm^{0.07}_{0.08}$ & $+0.02\pm^{0.10}_{0.11}$ & $-1.00\pm^{0.28}_{0.43}$ & $-0.93\pm^{0.24}_{0.38}$ & $-0.81\pm^{0.19}_{0.33}$ & $-0.87\pm^{0.24}_{0.36}$ &  1117 \\ 
	29 & $-0.32\pm^{0.22}_{0.28}$ & $-1.27\pm^{0.56}_{0.35}$ & $-0.15\pm^{0.21}_{0.22}$ & $-0.13\pm^{0.27}_{0.24}$ & $-0.72\pm^{0.25}_{0.44}$ & $-0.20\pm^{0.30}_{0.19}$ & $-0.29\pm^{0.21}_{0.35}$ & $-1.12\pm^{0.70}_{1.37}$ & $-0.31\pm^{0.25}_{0.50}$ & $-0.61\pm^{0.24}_{0.72}$ & $-0.49\pm^{0.31}_{0.45}$ &  1081 \\ 
	30 & $+0.01\pm^{0.54}_{0.51}$ & $-0.63\pm^{1.13}_{1.21}$ & $+0.04\pm^{0.37}_{0.58}$ & $+0.14\pm^{0.61}_{0.44}$ & $-0.74\pm^{0.78}_{0.90}$ & $+0.11\pm^{0.70}_{0.45}$ & $-0.00\pm^{0.71}_{0.42}$ & $-0.99\pm^{1.23}_{1.51}$ & $-0.55\pm^{0.79}_{0.81}$ & $-0.63\pm^{0.72}_{0.76}$ & $-0.68\pm^{0.77}_{0.98}$ &   562 \\ 
	31 & $+0.09\pm^{0.23}_{0.34}$ & $-1.27\pm^{0.51}_{0.82}$ & $+0.35\pm^{0.12}_{0.36}$ & $+0.31\pm^{0.20}_{0.28}$ & $-1.37\pm^{0.45}_{0.77}$ & $+0.16\pm^{0.19}_{0.18}$ & $-0.00\pm^{0.27}_{0.25}$ & $-1.90\pm^{1.03}_{0.60}$ & $-1.37\pm^{0.50}_{0.49}$ & $-1.36\pm^{0.46}_{0.50}$ & $-1.51\pm^{0.54}_{0.47}$ &   474 \\  
	\end{tabular}
\end{table}
\end{landscape}

\appendix
\section{Hint to repeatability index interpretation}
\label{sec:app0}

We define the centroid of class $i$ as
\begin{equation}
\vec{\mu}_i = \frac{1}{n_i} \sum_{\iota \epsilon \omega_i} \vec{x}_\iota\,,
\end{equation}

where $\omega_i$ is the set of spectra $x_{\iota}$ assigned to class i, and $n_i$ is the number of spectra in the class. So the mean difference between the classes in a particular classification $c$ compared with the chosen classification is given by

\begin{equation}
 \vec{\sigma}_{c}  =
       \sqrt{
            \frac{
                 \sum_{i=0}^{49}
                 ||\vec{\mu}_{i,c} -
                 \vec{\mu}_{i,chosen}||^2
                 }
                 {50}
             }\,.
\end{equation}

Therefore, when we refer to mean difference between the matching classes over the 100 classifications we mean

\begin{equation}
  \langle \vec{\sigma}_{compare} \rangle = \frac{1}{100}\sum_{c=0}^{99}\vec{\sigma}_{c}\,.
\end{equation}

This is the mean pixel by pixel difference between the 99 classifications as compared with the chosen one. This vector can be compared with the mean within the cluster dispersion of the chosen classification,

\begin{equation}
  \langle \vec{\sigma}_{within} \rangle = \frac{1}{50}\sum_{i=0}^{49} \sqrt{\frac{\sum_{\iota \epsilon \omega_i}||\vec{x_{\iota}} -\vec{\mu_i}||^2}{n_i}}\,,
\end{equation}

giving the main difference ratio between these quantities over the 4838 pixels of the spectra:

\begin{equation}
  \langle \sigma_{ratio} \rangle = \frac{1}{4838} \sum_{j=0}^{4837} \frac{\sigma_{j, compare}}{\sigma_{j, within}} \approx 0.064.
\end{equation}

The standard deviation of 3.3\% is given by the standard deviation of $\sigma_{j, compare}/\sigma_{j, within}$ over the 4838 pixels.

\section{Classes summary and online material}
\label{sec:app1}

In Table \ref{tab:desc} we present a summary of the 32 classes containing $\approx 99$ per cent of the spectra in the data set. In this table the first column is the group and the second column is a hyper-link for the appendix supplementary figures for each class. The third column gives the main stellar type found in each class. This information was inferred based only on the range of atmospheric parameters covered by each class \citep{Payne} and should be taken just as an idea of what kind of object is dominant in each class. The fourth column gives information about the main spatial distribution of each class. It is also a simple approximation based on their distribution of galactic coordinates and \aM-\MH\, (see \citealt{Bensby2003, Bensby2007}). Finally, the fifth column presents some extra comments about the main features of the class.

%One pdf file with the whole mean spectra for the 32 first classes is also included. It consists of 46 figures like Figure \ref{fig:spectra_pile}, but with a shorter wavelength coverage, $\approx 36$ \AA. The pdf file can be found \href{https://garciadias.github.io/APOGEE/images/Deteiled_Class_Spectra.pdf}{here}.

The complete information about the classification is also available as online material in the form of three tables; Table \ref{tab:assign} presents the classification for each spectra, APOGEE ID, and class; Table \ref{tab:mean_spectra} gives the mean spectra for each class, in the form of normalised fluxes and wavelengths; and \ref{tab:std_spectra} contains the spectral within-class standard deviation for each class, normalised fluxes, and wavelengths. In both Tables \ref{tab:mean_spectra} and \ref{tab:std_spectra} the last column gives the mask applied to the spectra: a binary index, where zero means the wavelength was not considered during classification and one means it was included in the classification procedure.

\begin{table*}
\caption{\label{tab:desc} Summary of the classes and complementary material.}
\begin{minipage}{\textwidth}
\centering
\begin{tabular}{ccccc}
\centering
Group & Class\footnote{Hyper-links to figures as described in Appendix.} & Stellar type\footnote{The stellar types here are inferred simply from the distribution of \Teff\, in the classes.} & Gal. component\footnote{Based on the mean distribution of the class on the galactic plane and on the \aM-\MH\, plane. } & Comment \\ \hline
\multicolumn{5}{l}{\underline{Metal-rich RC and RGB}} \\
\hyperref[subsec:G0]{0}  &   \hyperref[class02]{Class  02} &   K-Giants            &   Thin disk   &   Lowest \MH\, in the group, 31\% RC.\\
\hyperref[subsec:G0]{0}  &   \hyperref[class04]{Class 04} &   K-Giants            &   Thin disk   &   26\% RC\\
\hyperref[subsec:G0]{0}  &   \hyperref[class06]{Class 06} &   K-Giants            &   Thin disk   &   26\% RC\\
\hyperref[subsec:G0]{0}  &   \hyperref[class08]{Class 08} &   Sub Giants, K-Giants&   Thin disk   &   Warmest in the group, 1\% RC.\\
\hyperref[subsec:G0]{0}  &   \hyperref[class09]{Class 09} &   K-Giants            &   Thin disk   &   \MH\, near to grid limits, 21\% RC.\\ \hline
\multicolumn{5}{l}{\underline{Metal-poor cool RGB}} \\
\hyperref[subsec:G1]{1}  &   \hyperref[class07]{Class 07} &   K-Giants            &   Disk        &   Thick disk.\\
\hyperref[subsec:G1]{1}  &   \hyperref[class14]{Class 14} &   K-Giants            &   Disk        &   ---\\
\hyperref[subsec:G1]{1}  &   \hyperref[class19]{Class 19} &   K/M-Dwarfs          &   Disk        &   \Teff\, near to the grid limits.\\
\hyperref[subsec:G1]{1}  &   \hyperref[class25]{Class 25} &   M-Giants            &   Disk        &   \Teff\, near to the grid limits.\\
\hyperref[subsec:G1]{1}  &   \hyperref[class26]{Class 26} &   K-Giants            &   Disk        &   High $[\alpha/M]$ blob.\\
\hyperref[subsec:G1]{1}  &   \hyperref[class28]{Class 28} &   K-Giants            &   Bulge/centre&   Most metal-poor stars.\\ \hline
\multicolumn{5}{l}{\underline{Warm stars}} \\
\hyperref[subsec:G2]{2}  &   \hyperref[class03]{Class 03} &   Blue stars          &   Disk        &   Warmest telluric standards\\
\hyperref[subsec:G2]{2}  &   \hyperref[class11]{Class 11} &   F/G-Dwarfs          &   High g. latitude&   Warm, telluric standards.\\
\hyperref[subsec:G2]{2}  &   \hyperref[class13]{Class 13} &   Blue stars          &   ---         &   Warm fast rotation stars. Telluric standards.\\ \hline
\multicolumn{5}{l}{\underline{Fast rotators}} \\
\hyperref[subsec:G3]{3}  &   \hyperref[class27]{Class 27} &   K/M-Dwarfs          &   ---         &   Fast rotators.\\
\hyperref[subsec:G3]{3}  &   \hyperref[class29]{Class 29} &   M-Dwarfs            &   ---         &   Fast rotators.\\ \hline
\multicolumn{5}{l}{\underline{Metal-rich cool RGB}} \\
\hyperref[subsec:G4]{4}  &   \hyperref[class16]{Class 16} &   K/M-Giants          &   Disk        &   \Teff\, near to the grid limits.\\
\hyperref[subsec:G4]{4}  &   \hyperref[class18]{Class 18} &   K-Giants            &   Disk        &   --- \\
\hyperref[subsec:G4]{4}  &   \hyperref[class22]{Class 22} &   K-Giants            &   Thin disk   &   \MH\, near to the grid limits.\\ \hline
\multicolumn{5}{l}{\underline{Metal-poor RC and RGB}} \\
\hyperref[subsec:G5]{5}  &   \hyperref[class00]{Class 00} &   K-Giants            &   Disk        &   Broad in atmospheric parameters.\\
\hyperref[subsec:G5]{5}  &   \hyperref[class01]{Class 01} &   K-Giants            &   Disk        &   Whole RGB\\
\hyperref[subsec:G5]{5}  &   \hyperref[class05]{Class 05} &   Sub Giants, G/K-Giants& Disk        &   Broad in atmospheric parameters.\\ \hline
\multicolumn{5}{l}{\underline{Dwarf stars}} \\
\hyperref[subsec:G6]{6}  &   \hyperref[class10]{Class 10} &   G/K-Dwarfs          &   Thin disk   &   ---\\
\hyperref[subsec:G6]{6}  &   \hyperref[class12]{Class 12} &   K-Dwarfs            &   Thin disk   &   ---\\
\hyperref[subsec:G6]{6}  &   \hyperref[class15]{Class 15} &   K-Dwarfs            &   High g. latitude&   ---\\
\hyperref[subsec:G6]{6}  &   \hyperref[class17]{Class 17} &   K-Dwarfs            &   Thin disk   &   ---\\
\hyperref[subsec:G6]{6}  &   \hyperref[class20]{Class 20} &   M-Dwarfs            &   High g. latitude&   Atmospheric parameter near to the grid limits.\\ \hline
\multicolumn{5}{l}{\underline{Sparse classes}} \\
\hyperref[subsec:G7]{7}  &   \hyperref[class21]{Class 21} &   M-Giants            &   Bulge/Centre/Disk   &   Atmospheric parameter near to the grid limits.\\
\hyperref[subsec:G7]{7}  &   \hyperref[class23]{Class 23} &   M-Dwarfs            &   ---         &   Atmospheric parameter near to the grid limits.\\
\hyperref[subsec:G7]{7}  &   \hyperref[class24]{Class 24} &   Giants              &   Halo        &   High $[\alpha/M]$, metal-poor stars.\\
\hyperref[subsec:G7]{7}  &   \hyperref[class30]{Class 30} &   ---                 &   ---        &   Poor fit, M31 clusters, high  g. latitude.\\
\hyperref[subsec:G7]{7}  &   \hyperref[class31]{Class 31} &   Giants              &   High g. latitude&   metal-poor high \aM. \\ \hline
\end{tabular}
\end{minipage}
\end{table*}

\begin{table}
\caption{\label{tab:assign} Spectral classification. Complete table can be found in online material.}
\centering
\begin{tabular}{cc}
APOGEE ID & Class \\ \hline
2M03183846+7216305  &    11 \\
2M03470204+4125397  &   11      \\
2M04425018+6644089  &    5      \\
2M04575928+3416050  &    5      \\
2M05373344+7441194  &    5      \\
$\vdots$                        &       $\vdots$  \\
\end{tabular}
\end{table}
{\setlength{\tabcolsep}{1mm}
\begin{table}
\caption{\label{tab:mean_spectra} Mean spectra of the 50 classes. Complete table can be found in online material.}
\centering
\begin{tabular}{cccccc}
Class 0 & Class 1 &     $\cdots$        & Class 49 & Wavelength \AA & Mask      \\ \hline
$\vdots$ & $\vdots$ & $\vdots$ & $\vdots$ & $\vdots$ & $\vdots$ \\
0.99505127  &   0.99435151  &   $\cdots$        &   0.16018607  &   16178.34    &   1   \\
0.98787344  &   0.98545332  &   $\cdots$        &   1.02000000  &   16178.57    &   1   \\
0.97224899  &   0.96870929  &   $\cdots$        &   1.02000000  &   16178.79    &   1   \\
0.93429393  &   0.92675939  &   $\cdots$        &   0.97670869  &   16179.01    &   1   \\
0.89091408  &   0.87721260  &   $\cdots$        &   0.09924513  &   16179.24    &   1   \\
$\vdots$ & $\vdots$ & $\vdots$ & $\vdots$ & $\vdots$ & $\vdots$ \\
\end{tabular}
\end{table}

\begin{table}
\caption{\label{tab:std_spectra} Within-class spectral standard deviation for the 50 classes. Complete table can be found in online material.}
\centering
\begin{tabular}{cccccc}
Class 0 & Class 1 & $\cdots$ & Class 49 & Wavelength \AA & Mask \\ \hline
$\vdots$ & $\vdots$ & $\vdots$ & $\vdots$ & $\vdots$ & $\vdots$ \\
0.01448872  &   0.01866457  & $\cdots$      &   0.00000000  &   16178.34    &   1   \\
0.00886448  &   0.02040061  & $\cdots$      &   0.00000000  &   16178.57    &   1   \\
0.01245967  &   0.02015303  & $\cdots$      &   0.00000000  &   16178.79    &   1   \\
0.01557365  &   0.01858692  & $\cdots$      &   0.00000000  &   16179.01    &   1   \\
0.01527482  &   0.01647230  & $\cdots$      &   0.00000000  &   16179.24    &   1   \\
$\vdots$ & $\vdots$ & $\vdots$ & $\vdots$ & $\vdots$ & $\vdots$ \\
\end{tabular}
\end{table}}

\section{Appendix images}

\begin{figure*}
    \centering
    \includegraphics[width=0.95\textwidth]
        {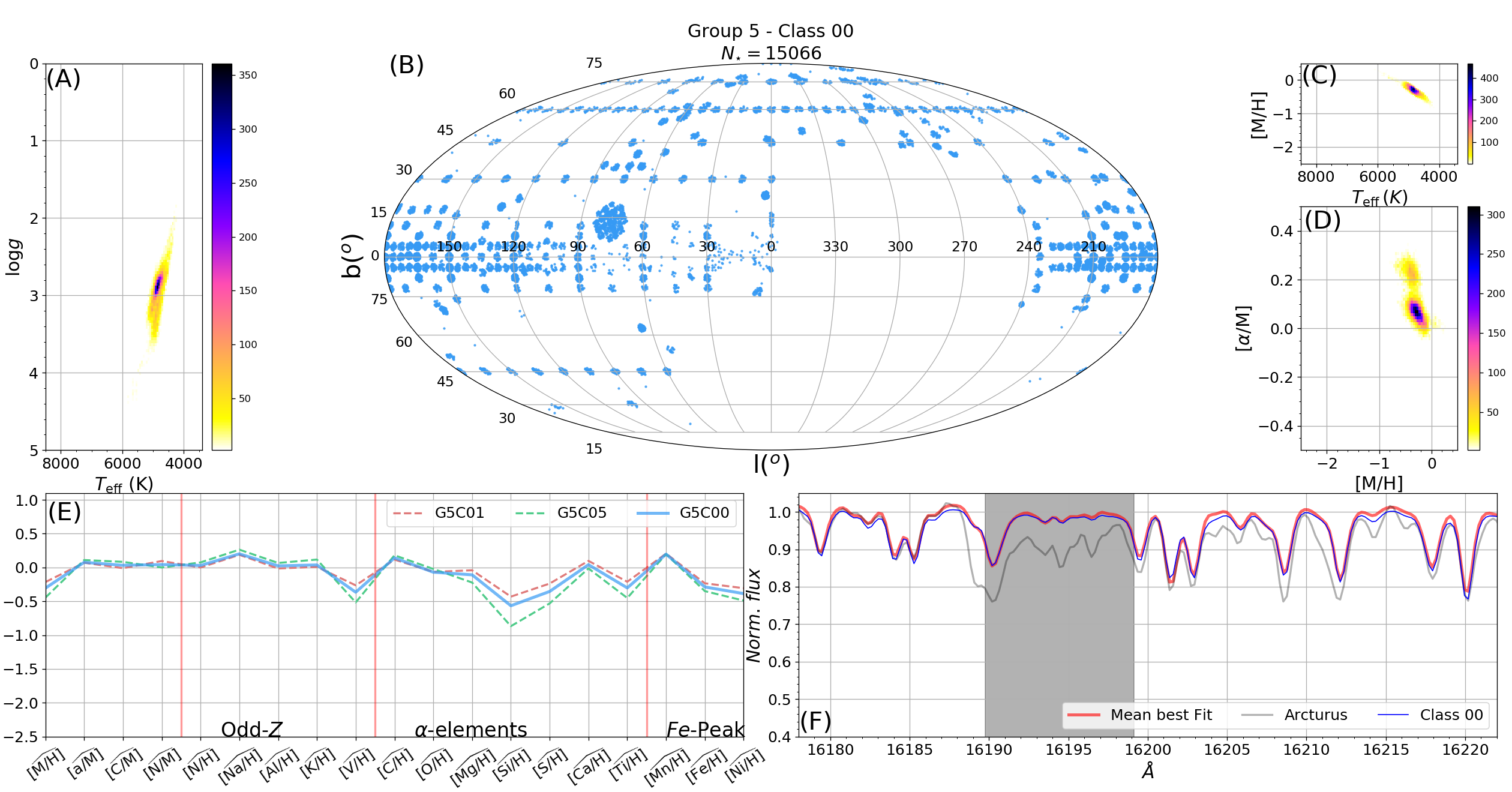}
    \includegraphics[width=0.65\textwidth]
        {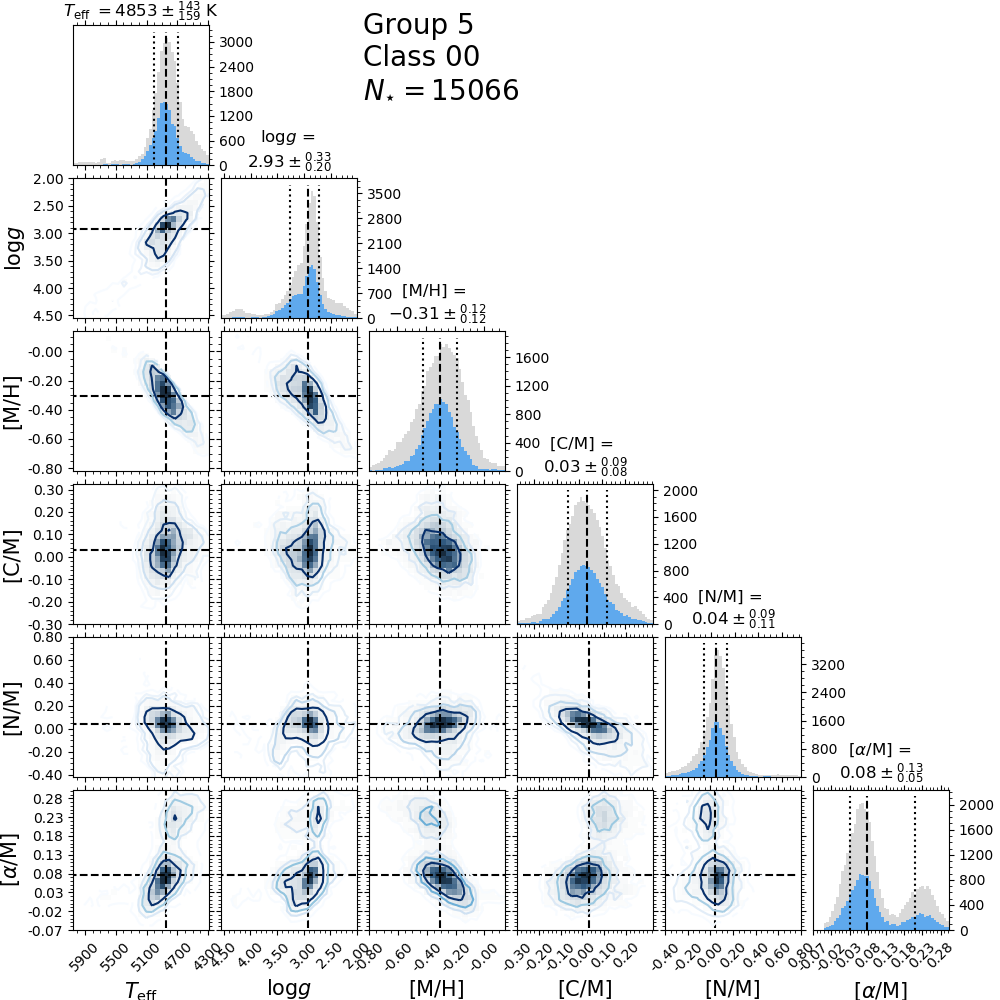}
    \caption{\label{class00} Panel (A) is a 2D histogram in the \Teff-\LOGG\, plane; panel (B) shows the galactic coordinates distribution; panel (C) shows a 2D histogram of the class in the \Teff-\MH\, plane; panel (D) presents a 2D histogram in the \MH-\aM\, plane; panel (E) gives the parallel plot for all the atmospheric parameters and individual chemical elements available in DR12 for all classes in each group, depicting the main class as a solid line and the other classes as dashed lines. The colours used here are the same as those used in Figure \ref{fig:Groups_MH_Teff}; and panel (F) compares the mean spectra of the class (blue line) with the mean best fit model for each spectra in the class (red line) and also shows the Arcturus spectrum (grey line) for comparison. The second figure in the file is a corner plot for \Teff, \LOGG, \MH, \C, \N and \aM. The figure contains 15 panels comparing these quantities with each other in 2D histograms. Four contours mark the levels enclosing 15, 30, 45, and 68.3 per cent of the points in each class. The top panel in each column gives the histogram  comparing the class parameter distribution (using the same colours used in Figure \ref{fig:Groups_MH_Teff}) with the distribution in its corresponding group (gray bars); the median values and the region enclosing 68.3 per cent of the points around the mean are marked by vertical lines and the values are shown above the top panels.}
    \end{figure*}

\begin{figure*}
    \centering
    \includegraphics[width=\textwidth]
        {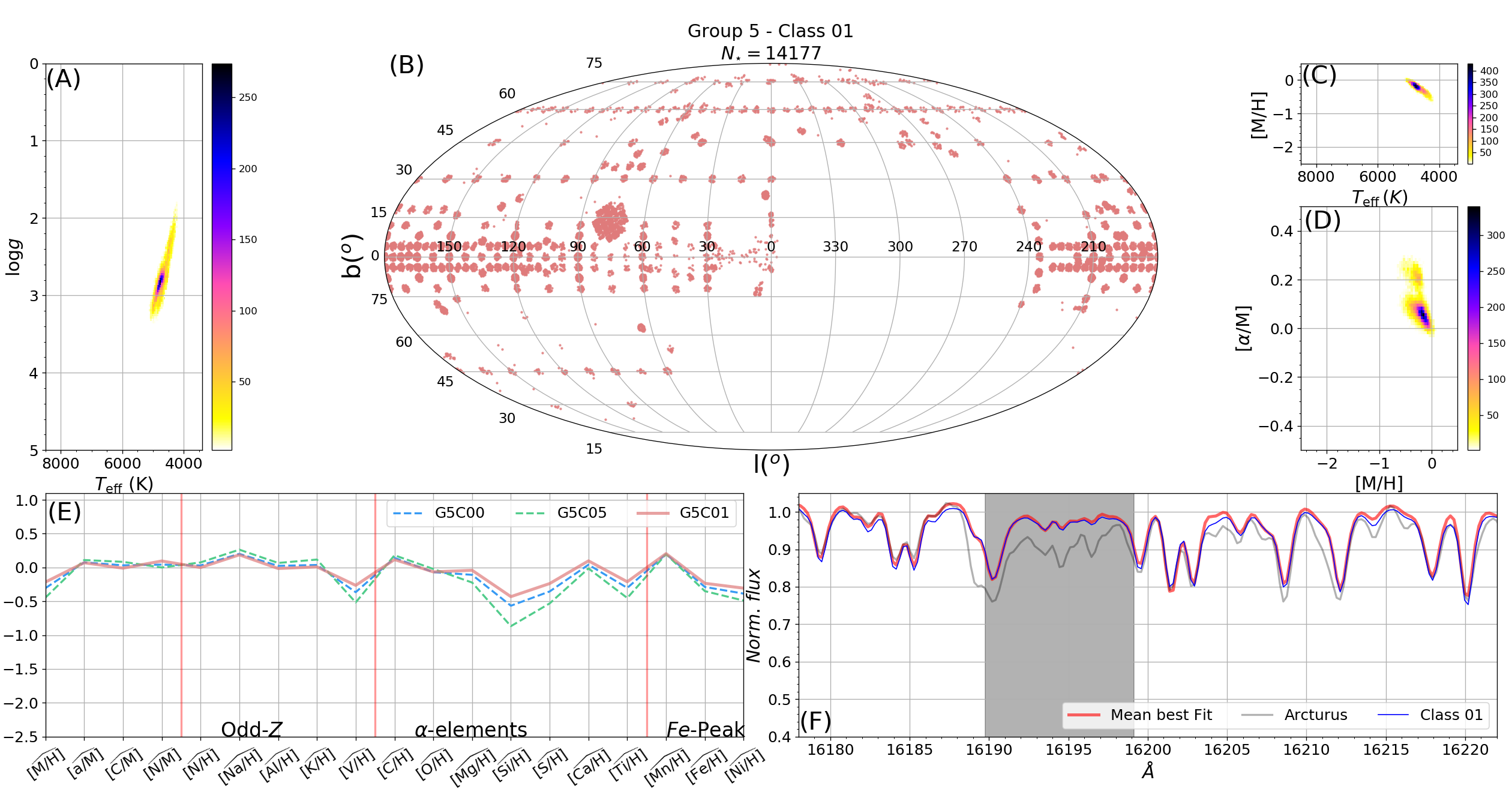}
    \includegraphics[width=0.7\textwidth]
        {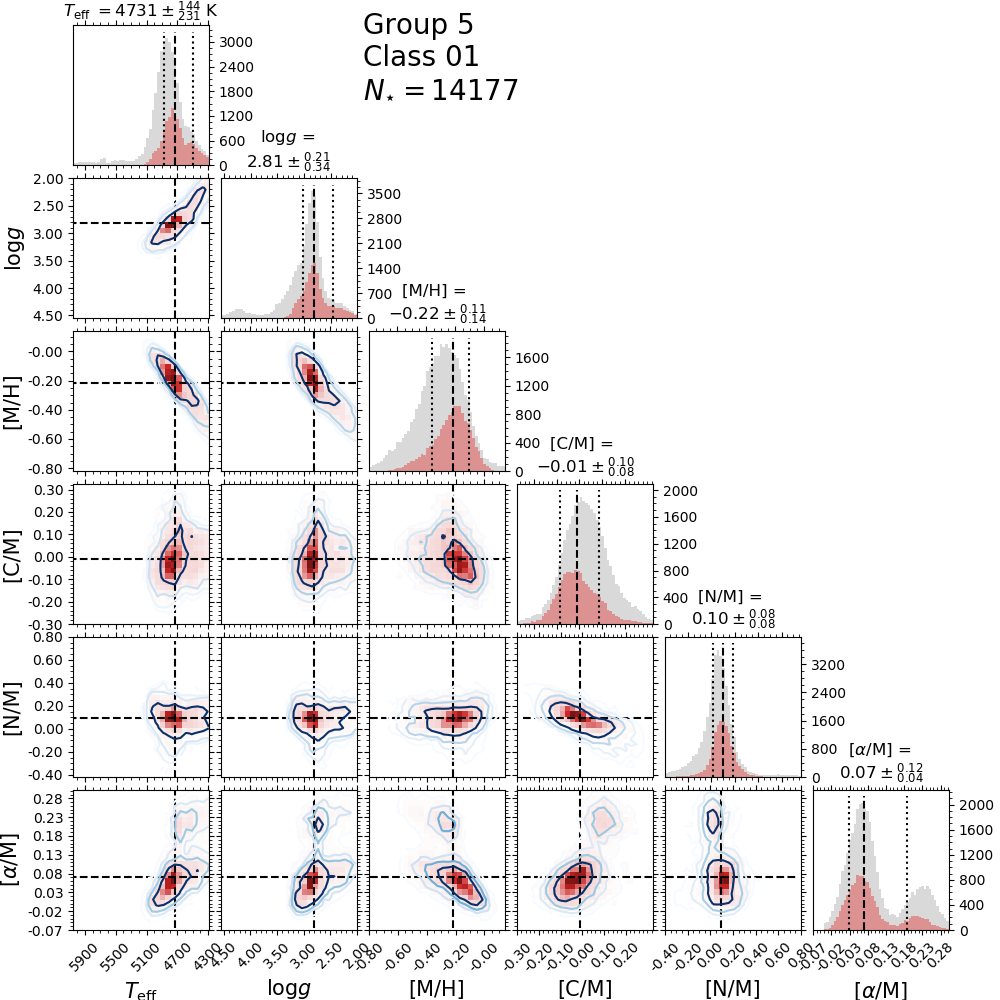}
    \caption{\label{class01} This figure folows the same pattern  from
    figure \ref{class00}. All the classes are described at Table  \ref{tab:desc}.}
    \end{figure*}

\begin{figure*}
    \centering
    \includegraphics[width=\textwidth]
        {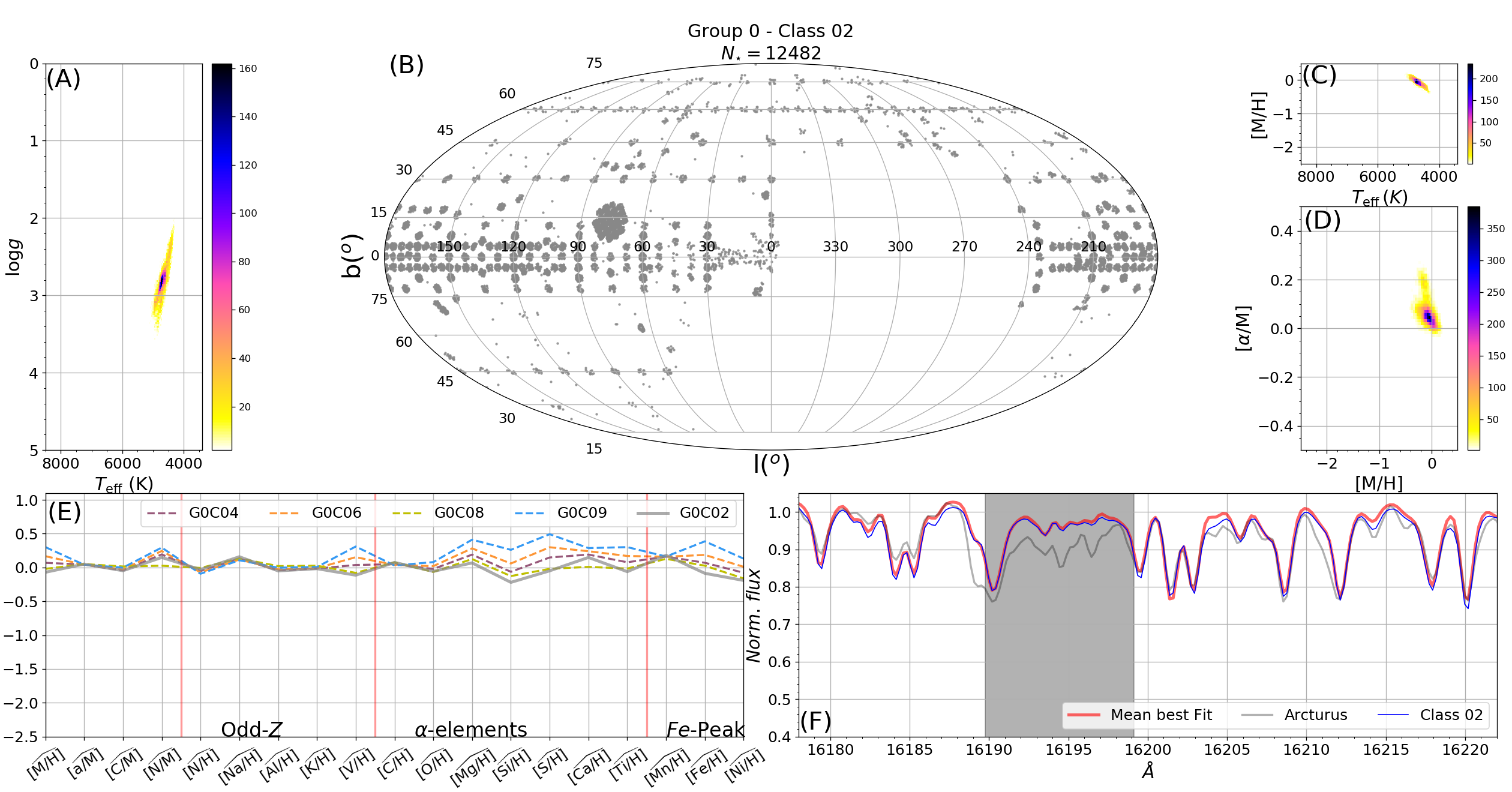}
    \includegraphics[width=0.7\textwidth]
        {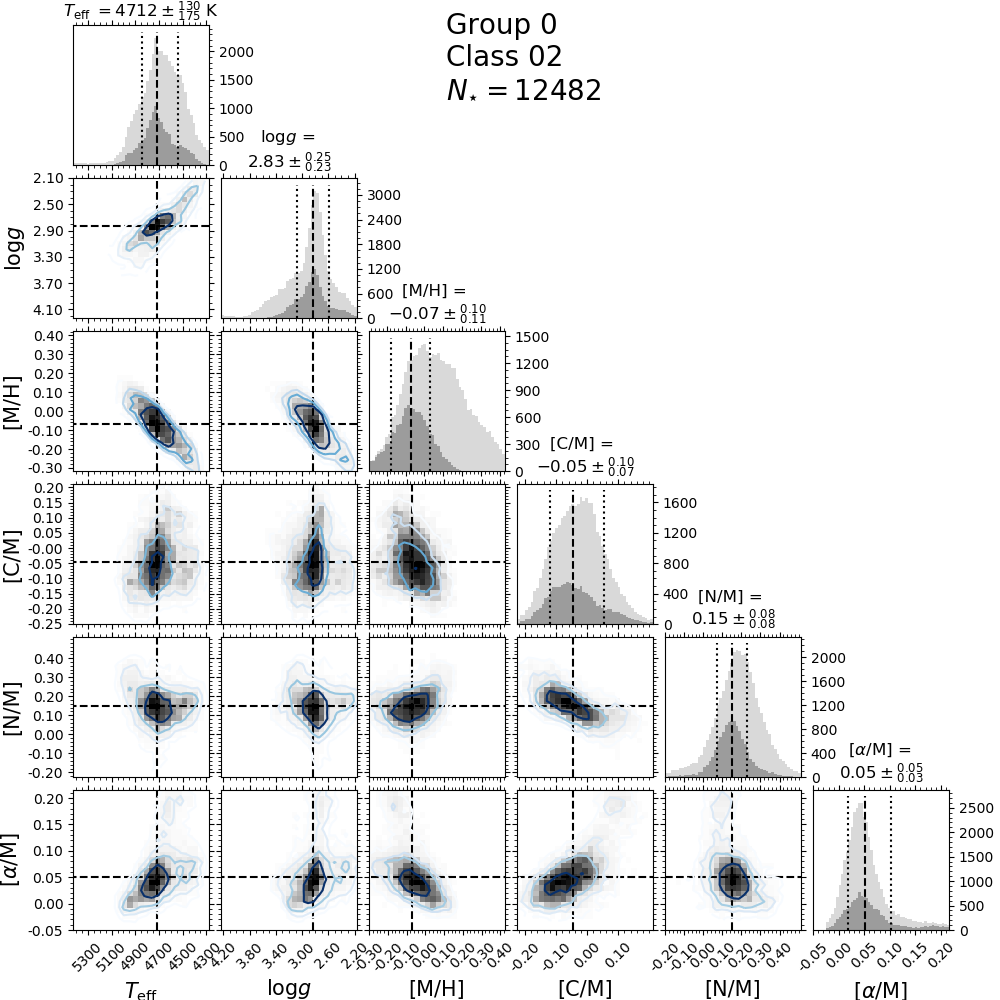}
    \caption{\label{class02} This figure folows the same pattern  from
    figure \ref{class00}. All the classes are described at Table  \ref{tab:desc}.}
    \end{figure*}

\begin{figure*}
    \centering
    \includegraphics[width=\textwidth]
        {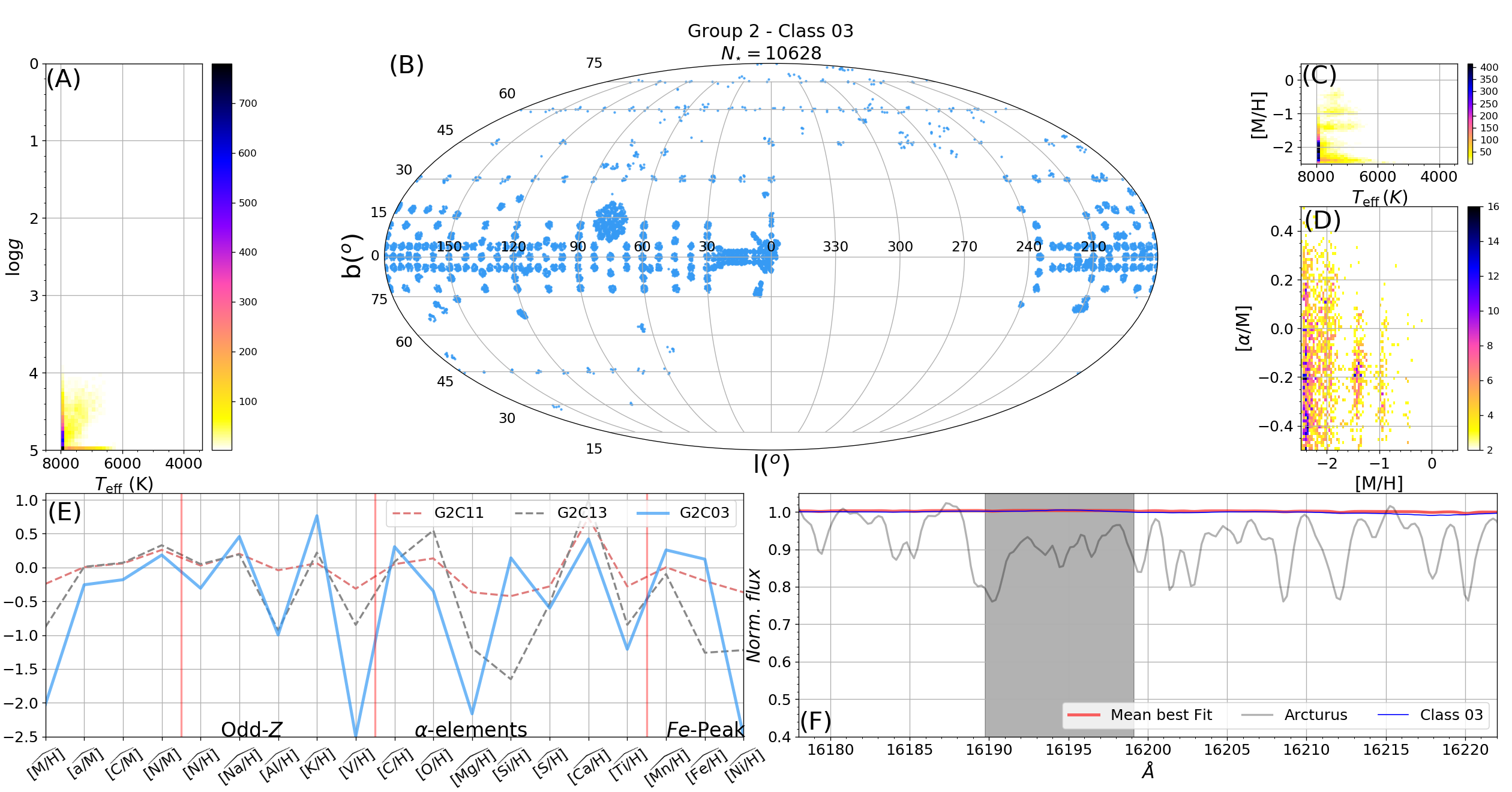}
    \includegraphics[width=0.7\textwidth]
        {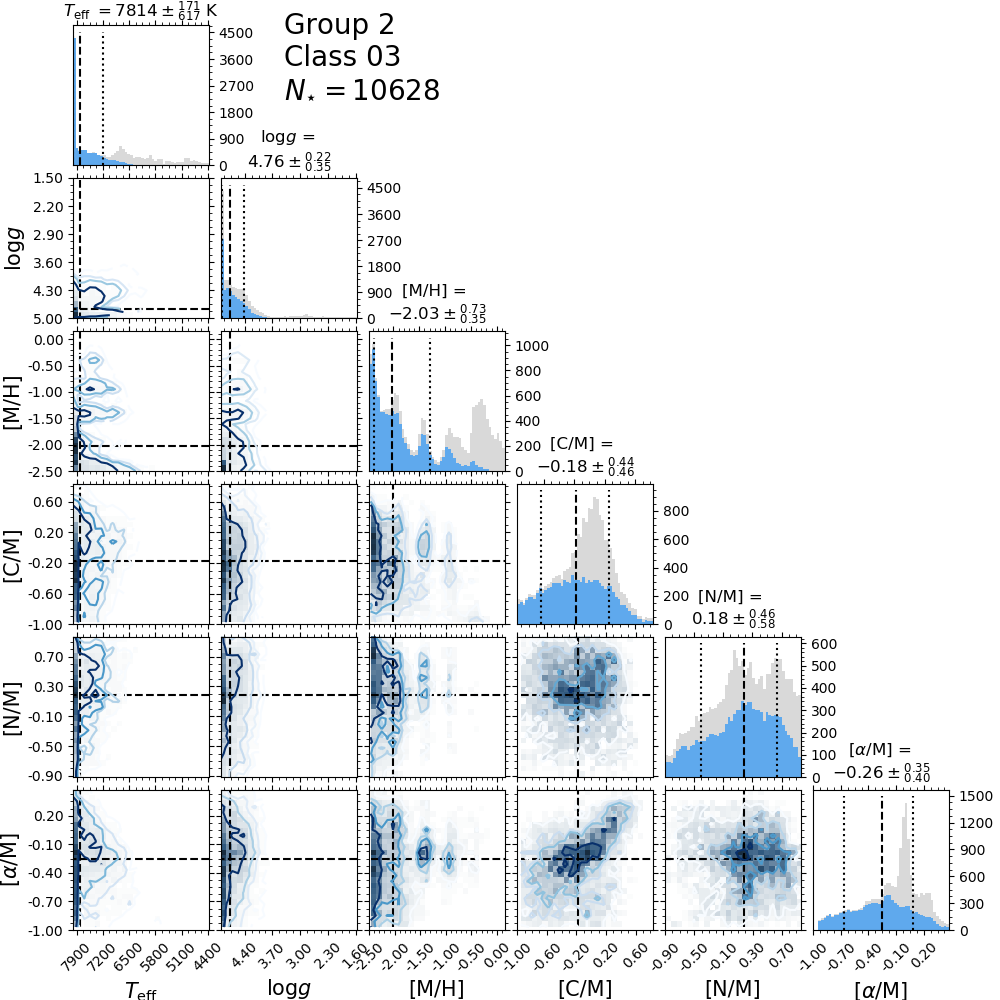}
    \caption{\label{class03} This figure folows the same pattern  from
    figure \ref{class00}. All the classes are described at Table  \ref{tab:desc}.}
    \end{figure*}

\begin{figure*}
    \centering
    \includegraphics[width=\textwidth]
        {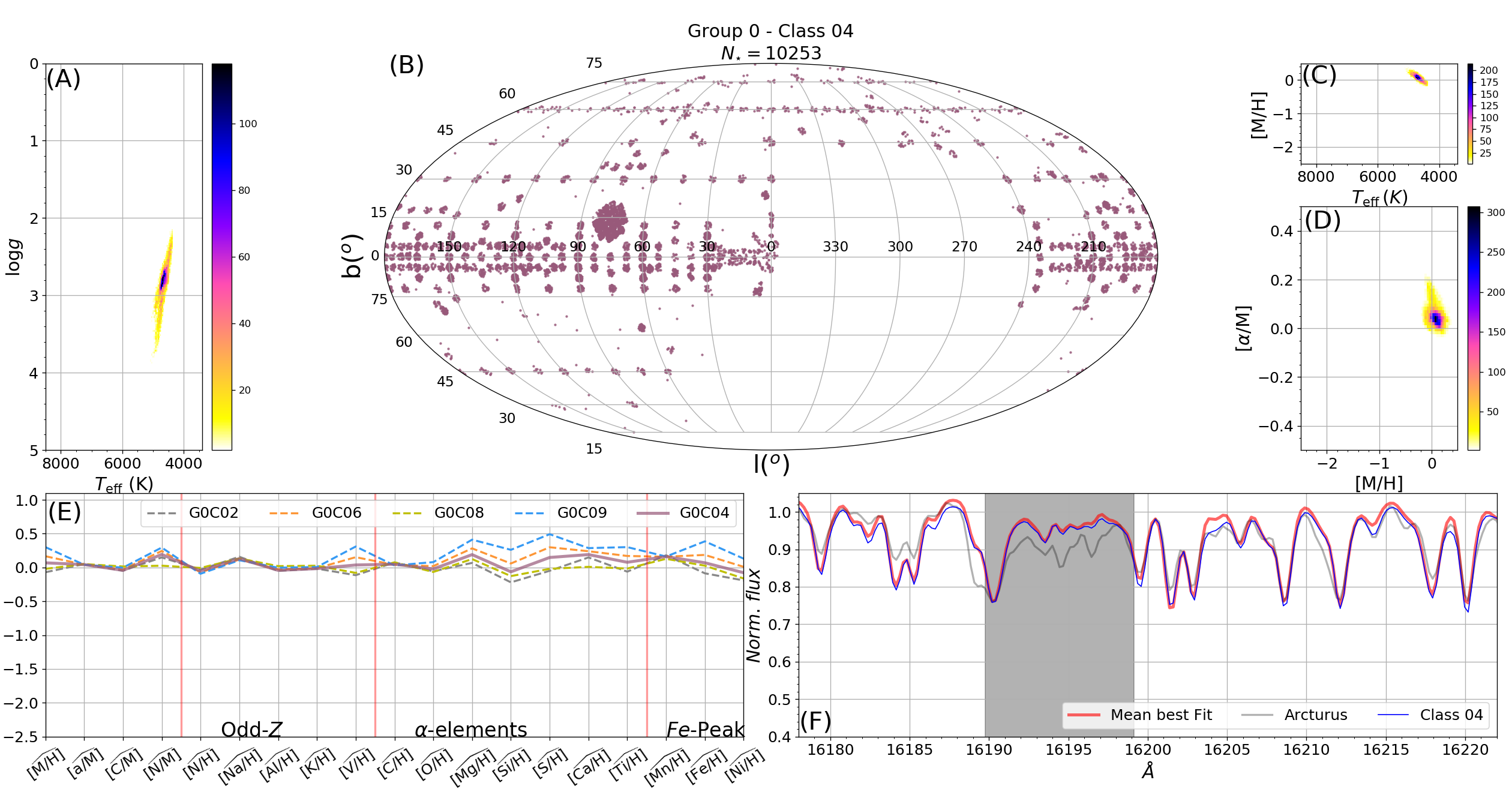}
    \includegraphics[width=0.7\textwidth]
        {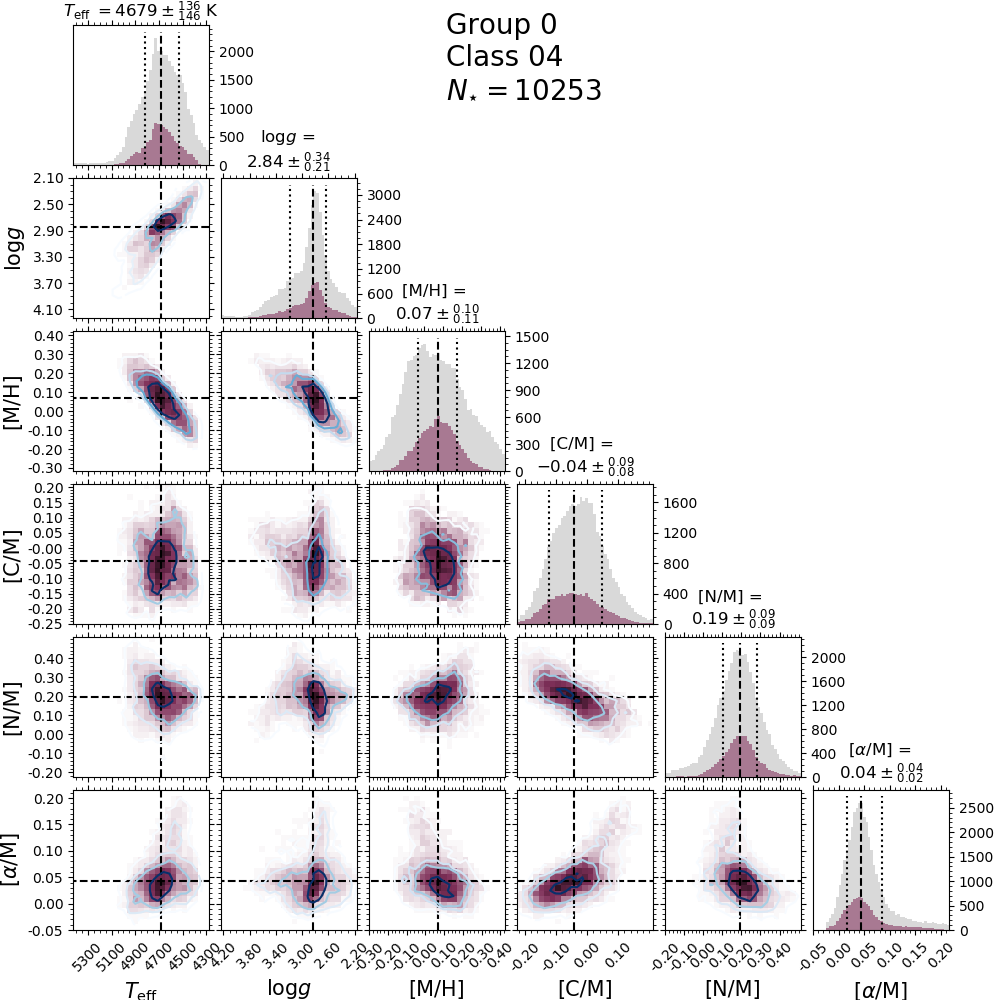}
    \caption{\label{class04} This figure folows the same pattern  from
    figure \ref{class00}. All the classes are described at Table  \ref{tab:desc}.}
    \end{figure*}

\begin{figure*}
    \centering
    \includegraphics[width=\textwidth]
        {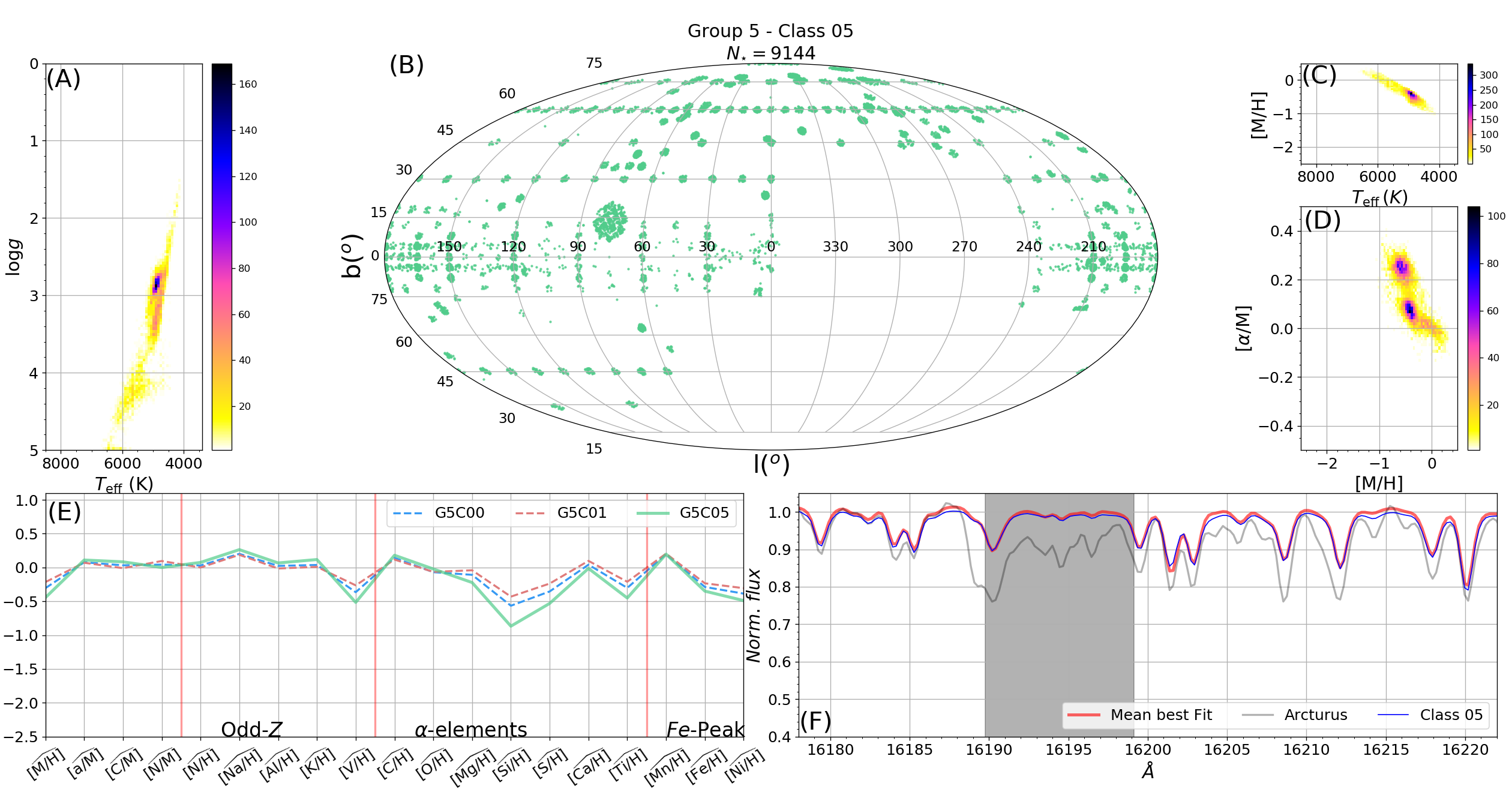}
    \includegraphics[width=0.7\textwidth]
        {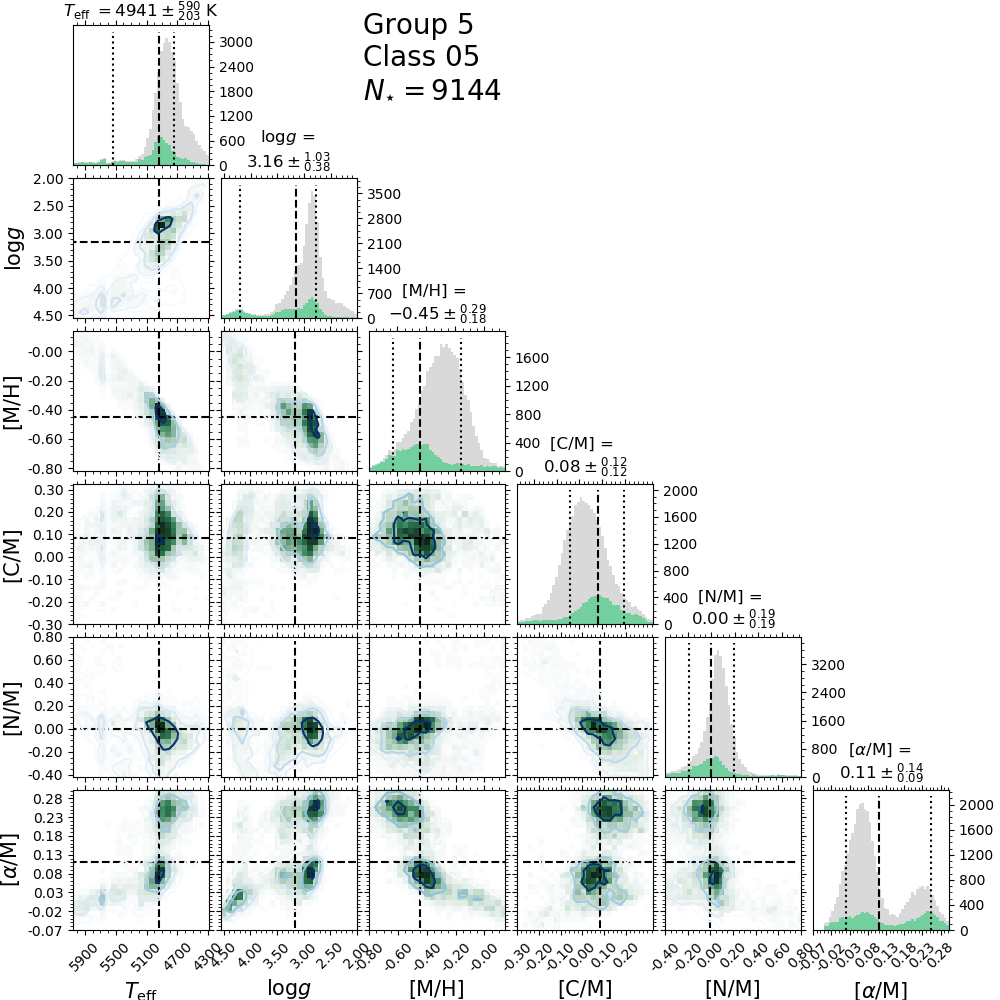}
    \caption{\label{class05} This figure folows the same pattern  from
    figure \ref{class00}. All the classes are described at Table  \ref{tab:desc}.}
    \end{figure*}

\begin{figure*}
    \centering
    \includegraphics[width=\textwidth]
        {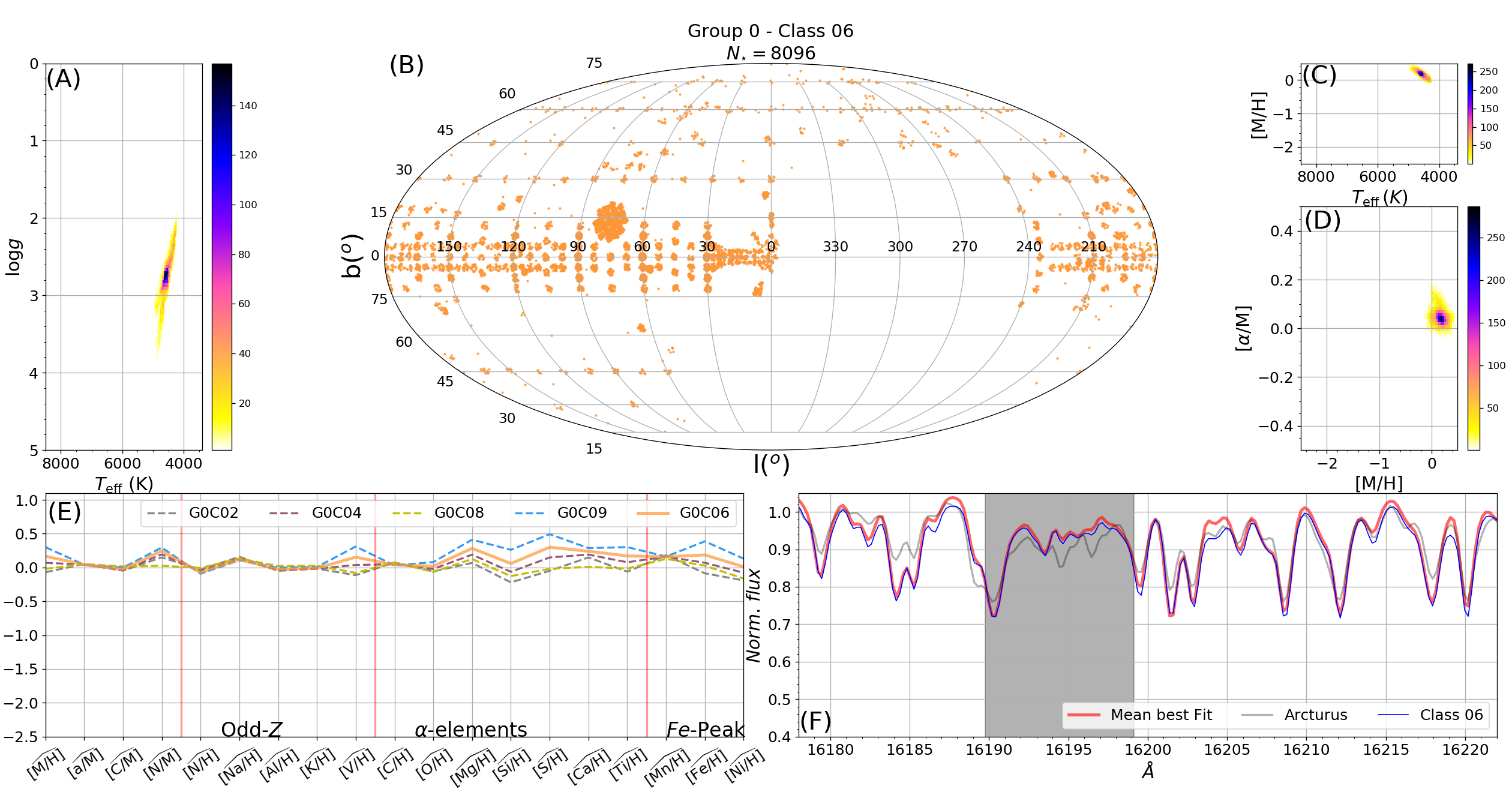}
    \includegraphics[width=0.7\textwidth]
        {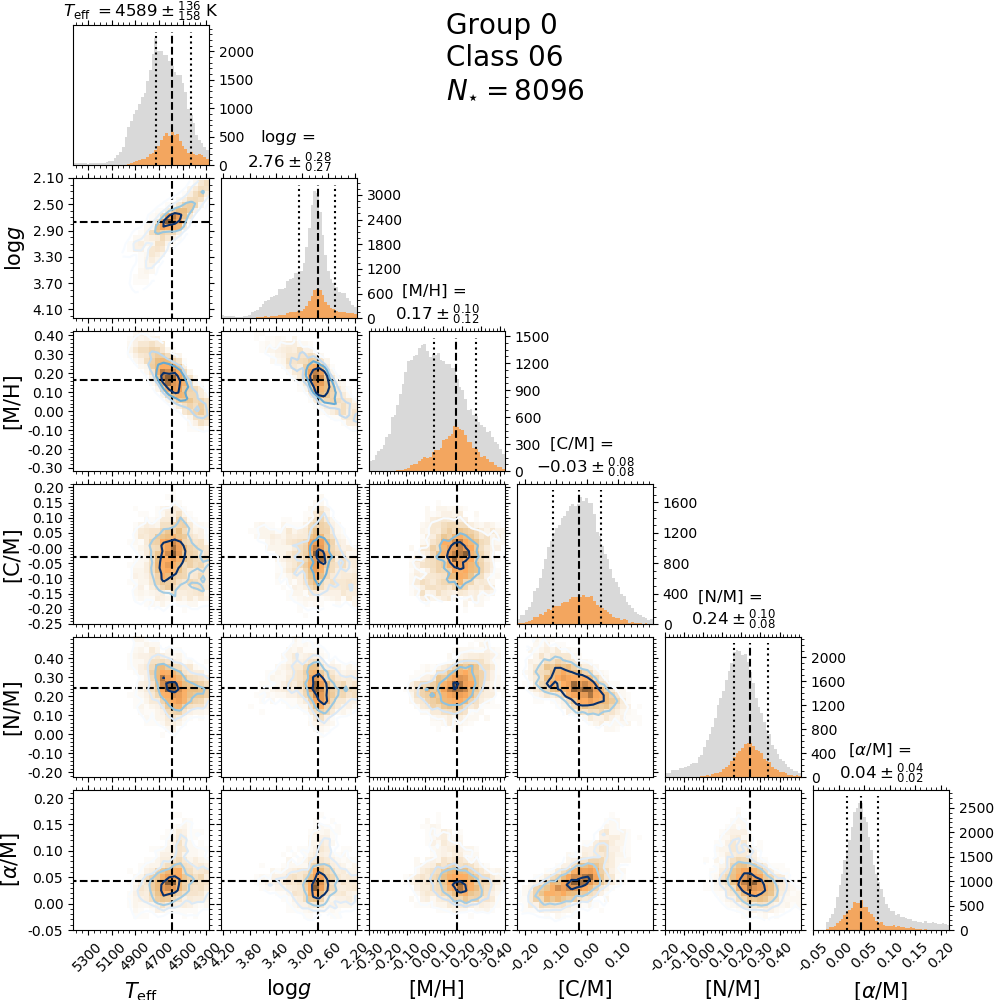}
    \caption{\label{class06} This figure folows the same pattern  from
    figure \ref{class00}. All the classes are described at Table  \ref{tab:desc}.}
    \end{figure*}

\begin{figure*}
    \centering
    \includegraphics[width=\textwidth]
        {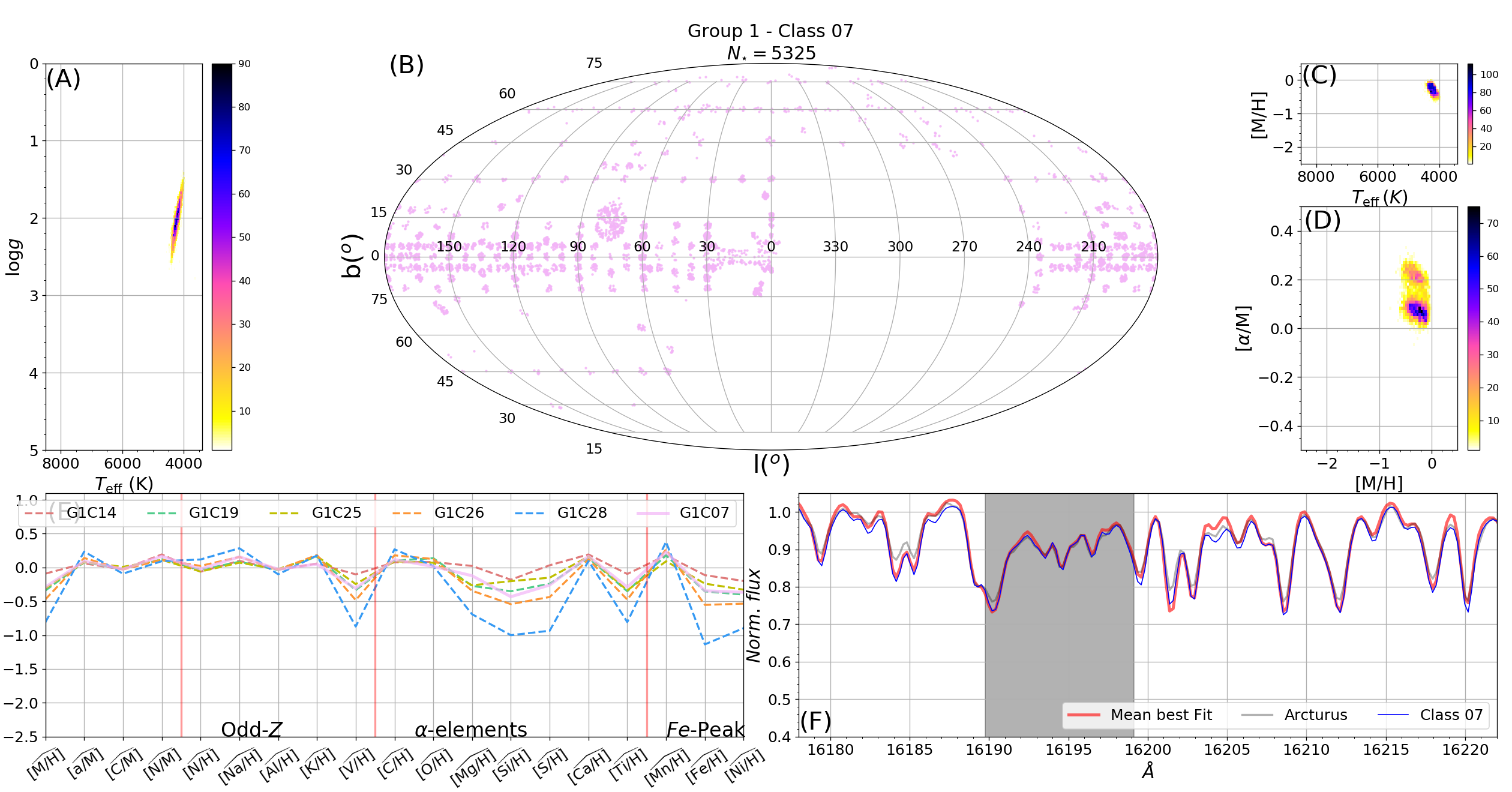}
    \includegraphics[width=0.7\textwidth]
        {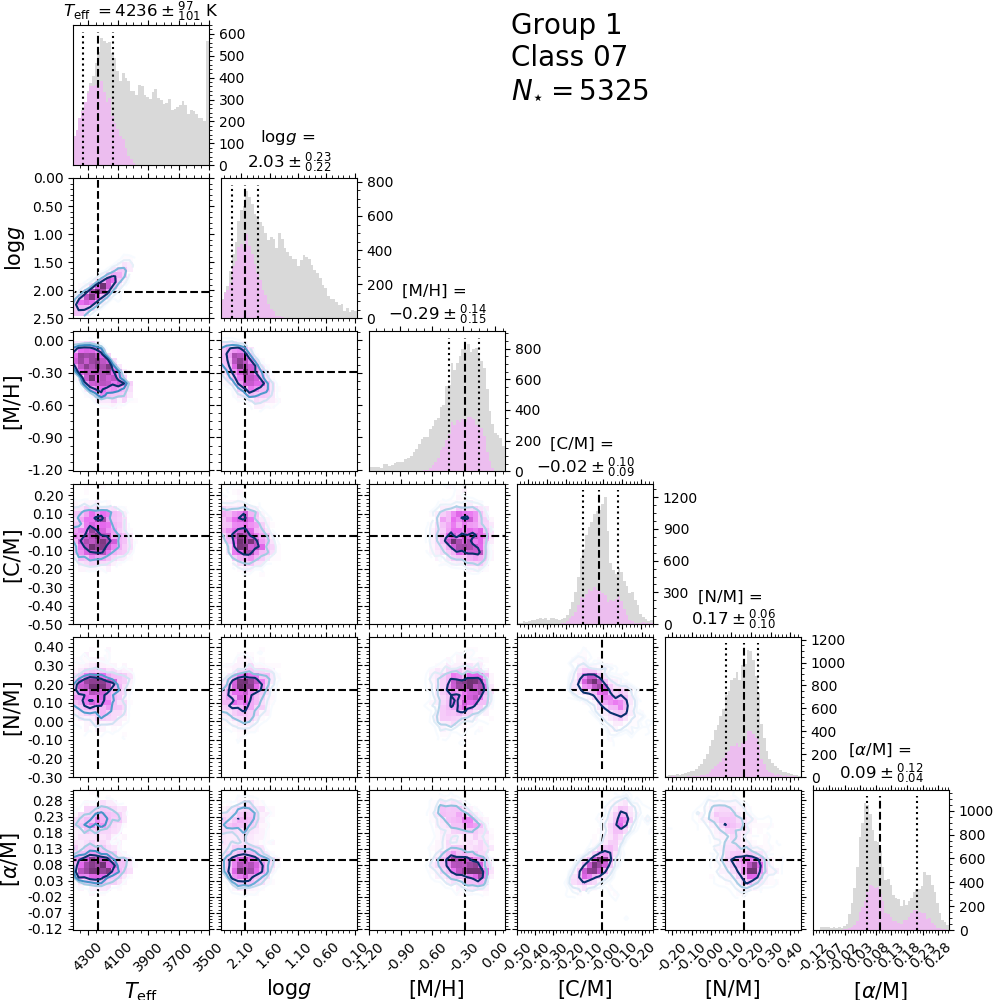}
    \caption{\label{class07} This figure folows the same pattern  from
    figure \ref{class00}. All the classes are described at Table  \ref{tab:desc}.}
    \end{figure*}

\begin{figure*}
    \centering
    \includegraphics[width=\textwidth]
        {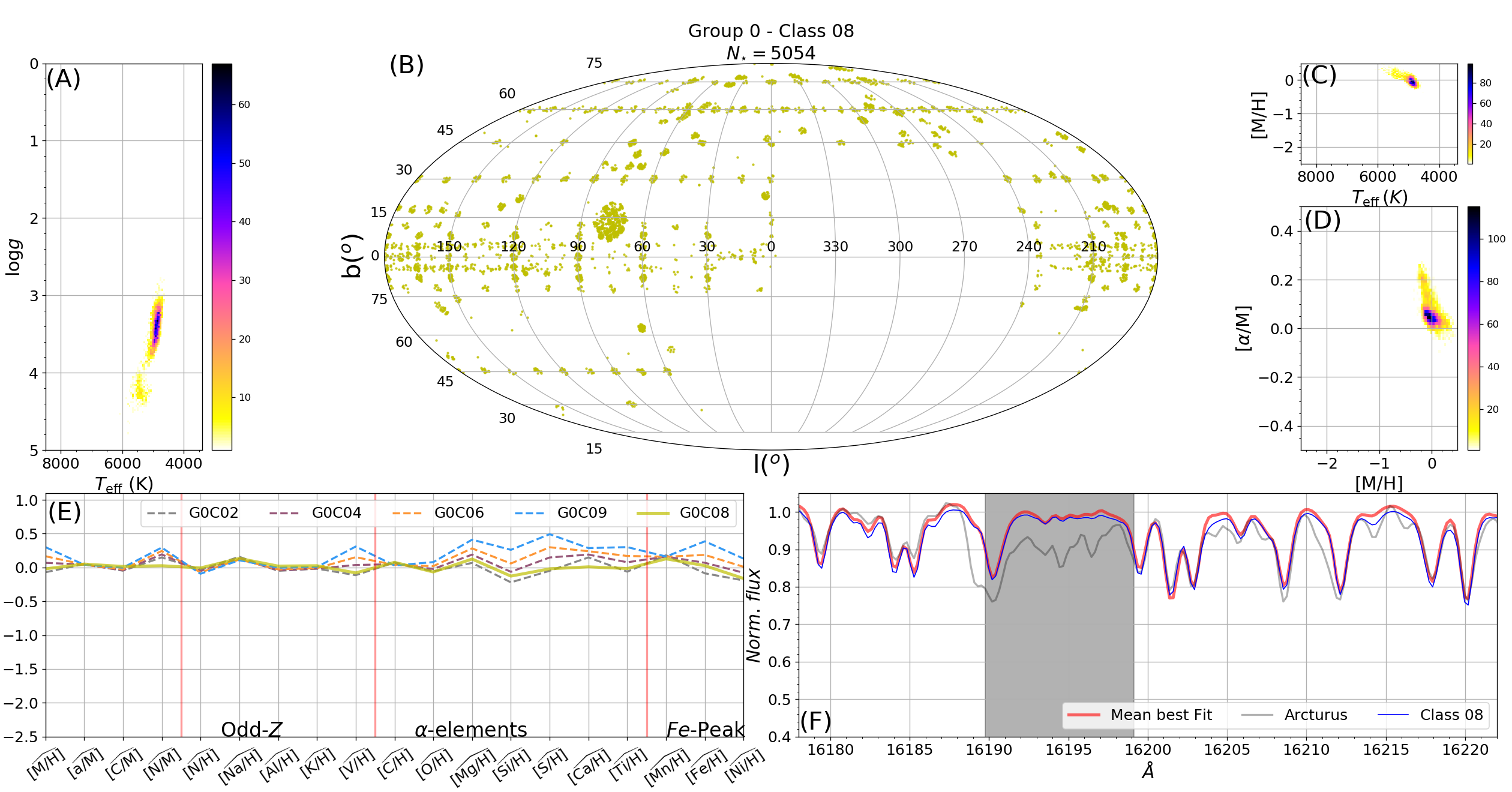}
    \includegraphics[width=0.7\textwidth]
        {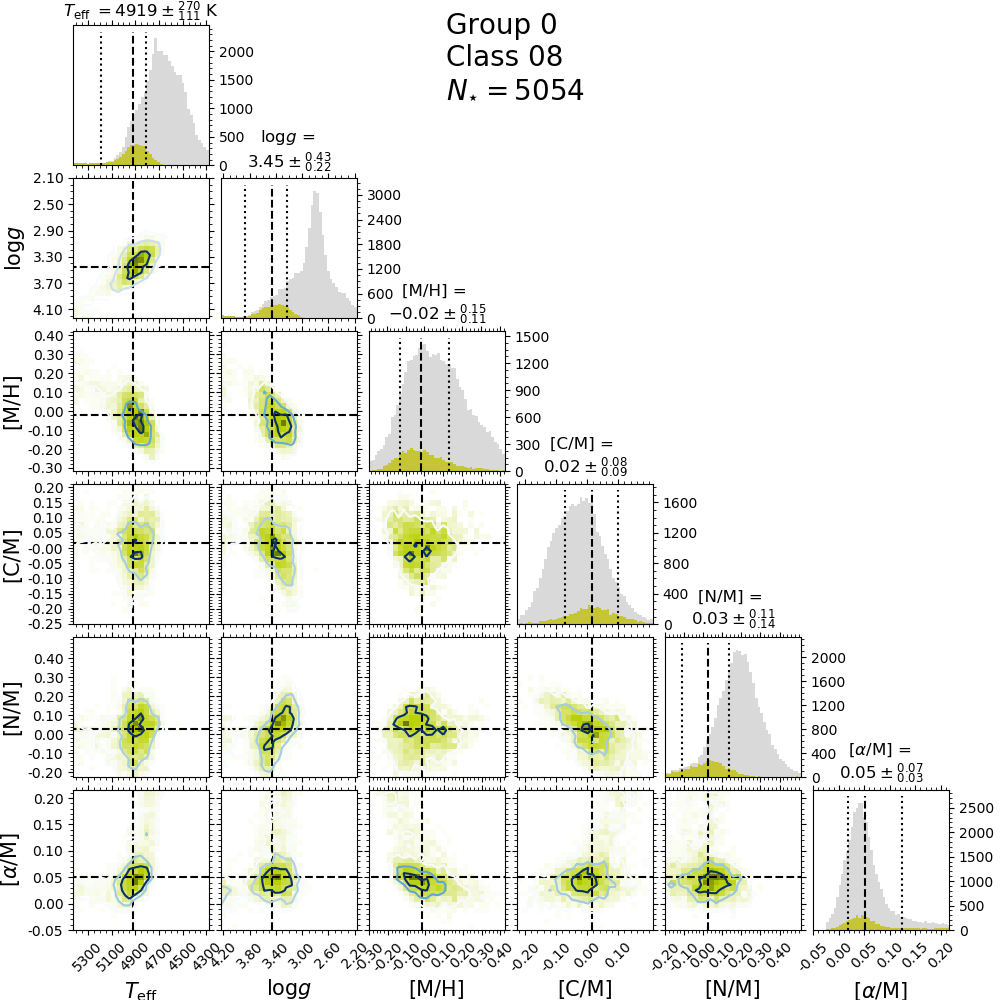}
    \caption{\label{class08} This figure folows the same pattern  from
    figure \ref{class00}. All the classes are described at Table  \ref{tab:desc}.}
    \end{figure*}

\begin{figure*}
    \centering
    \includegraphics[width=\textwidth]
        {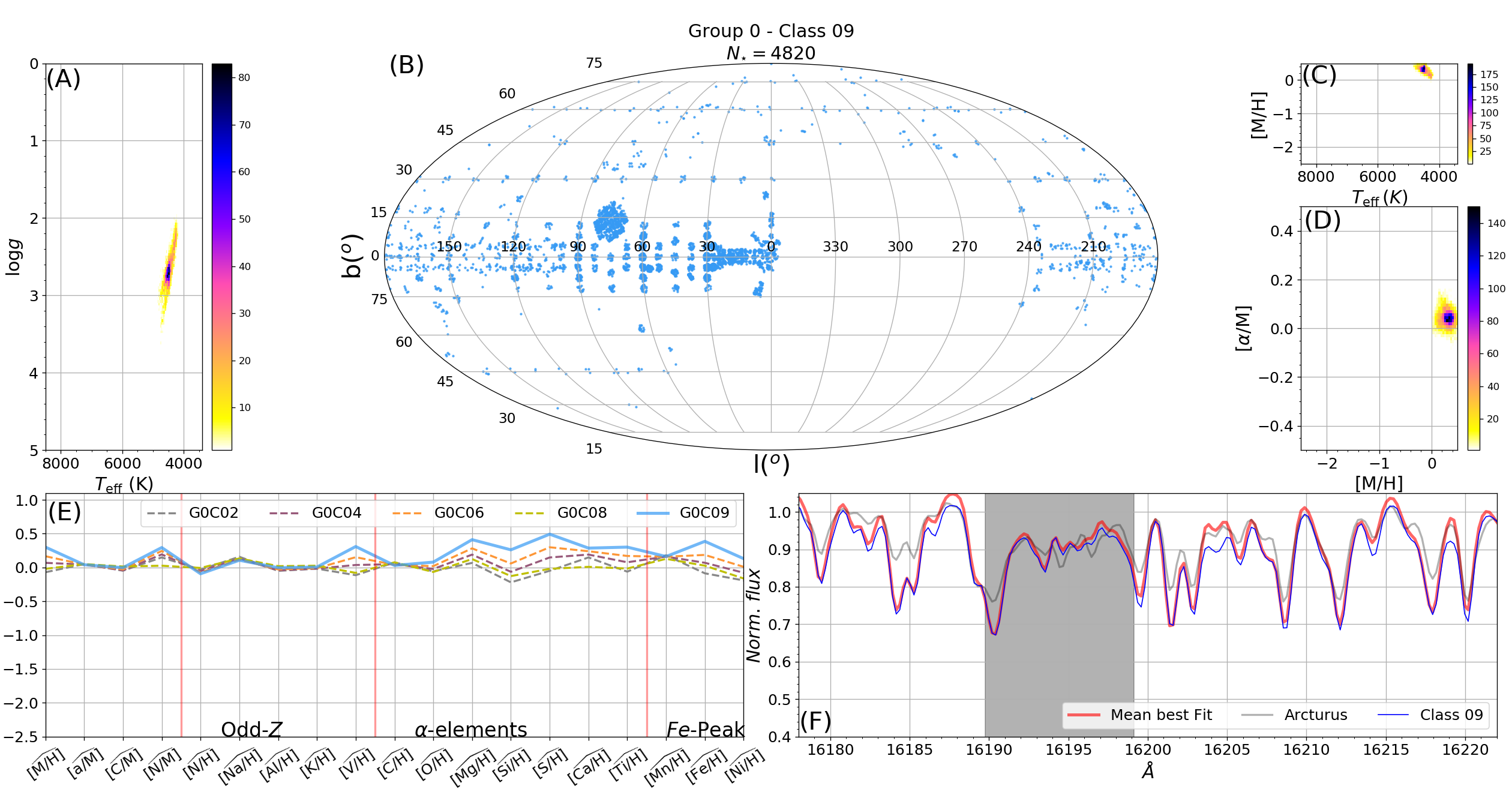}
    \includegraphics[width=0.7\textwidth]
        {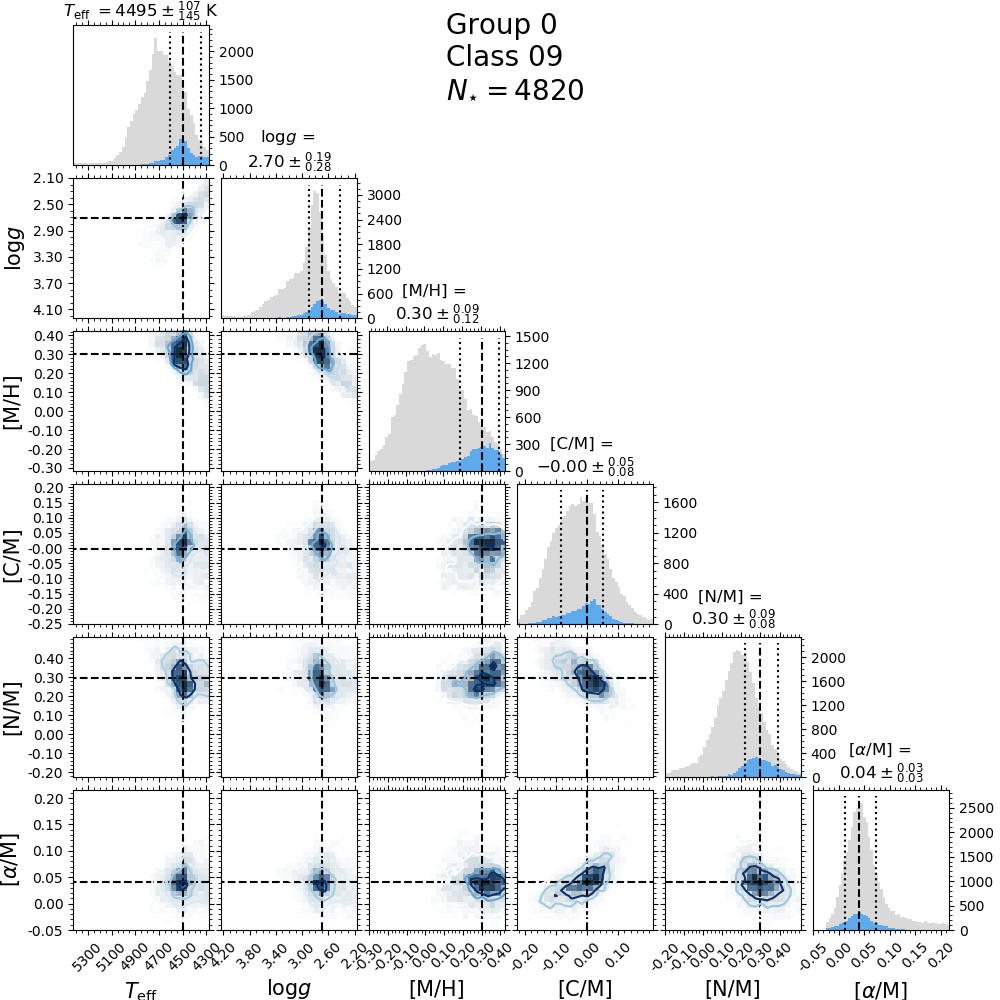}
    \caption{\label{class09} This figure folows the same pattern  from
    figure \ref{class00}. All the classes are described at Table  \ref{tab:desc}.}
    \end{figure*}

\begin{figure*}
    \centering
    \includegraphics[width=\textwidth]
        {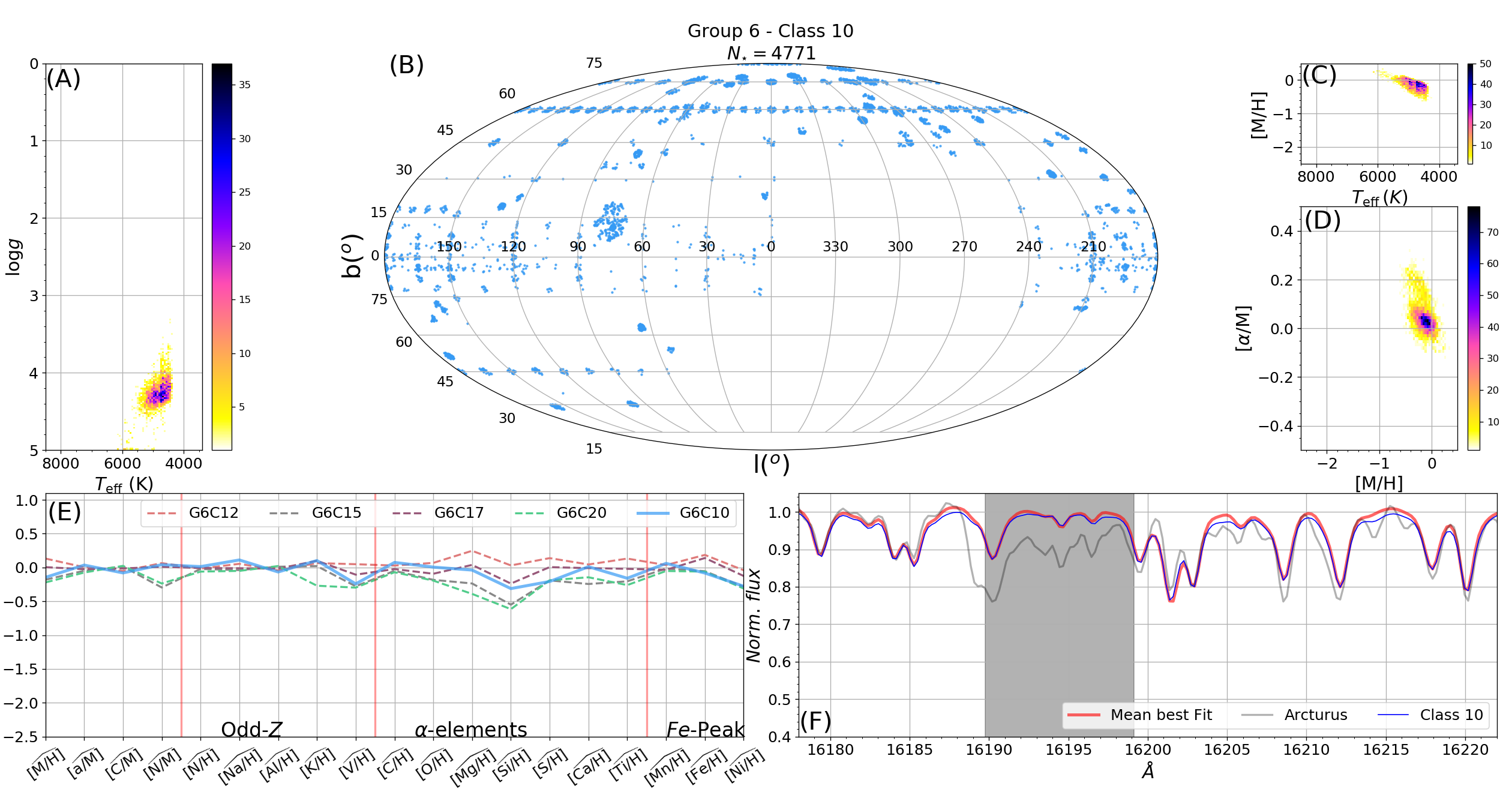}
    \includegraphics[width=0.7\textwidth]
        {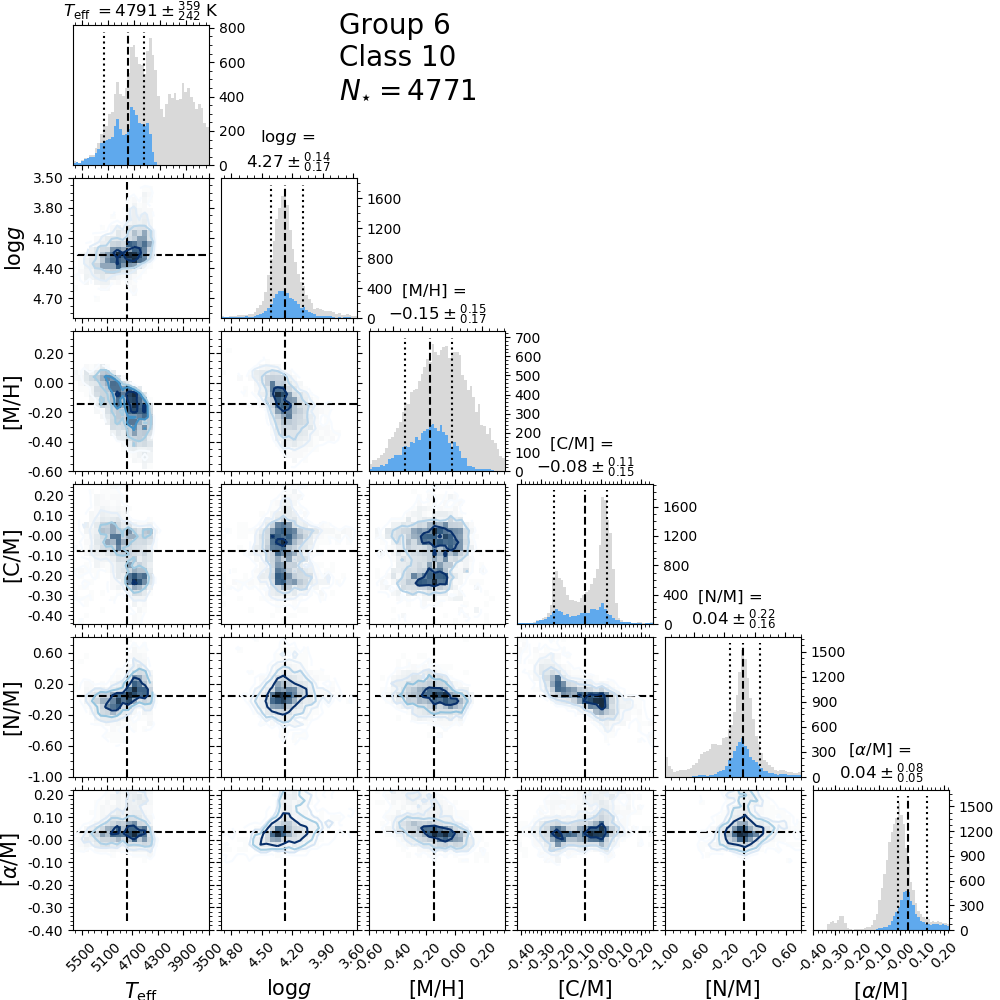}
    \caption{\label{class10} This figure folows the same pattern  from
    figure \ref{class00}. All the classes are described at Table  \ref{tab:desc}.}
    \end{figure*}

\begin{figure*}
    \centering
    \includegraphics[width=\textwidth]
        {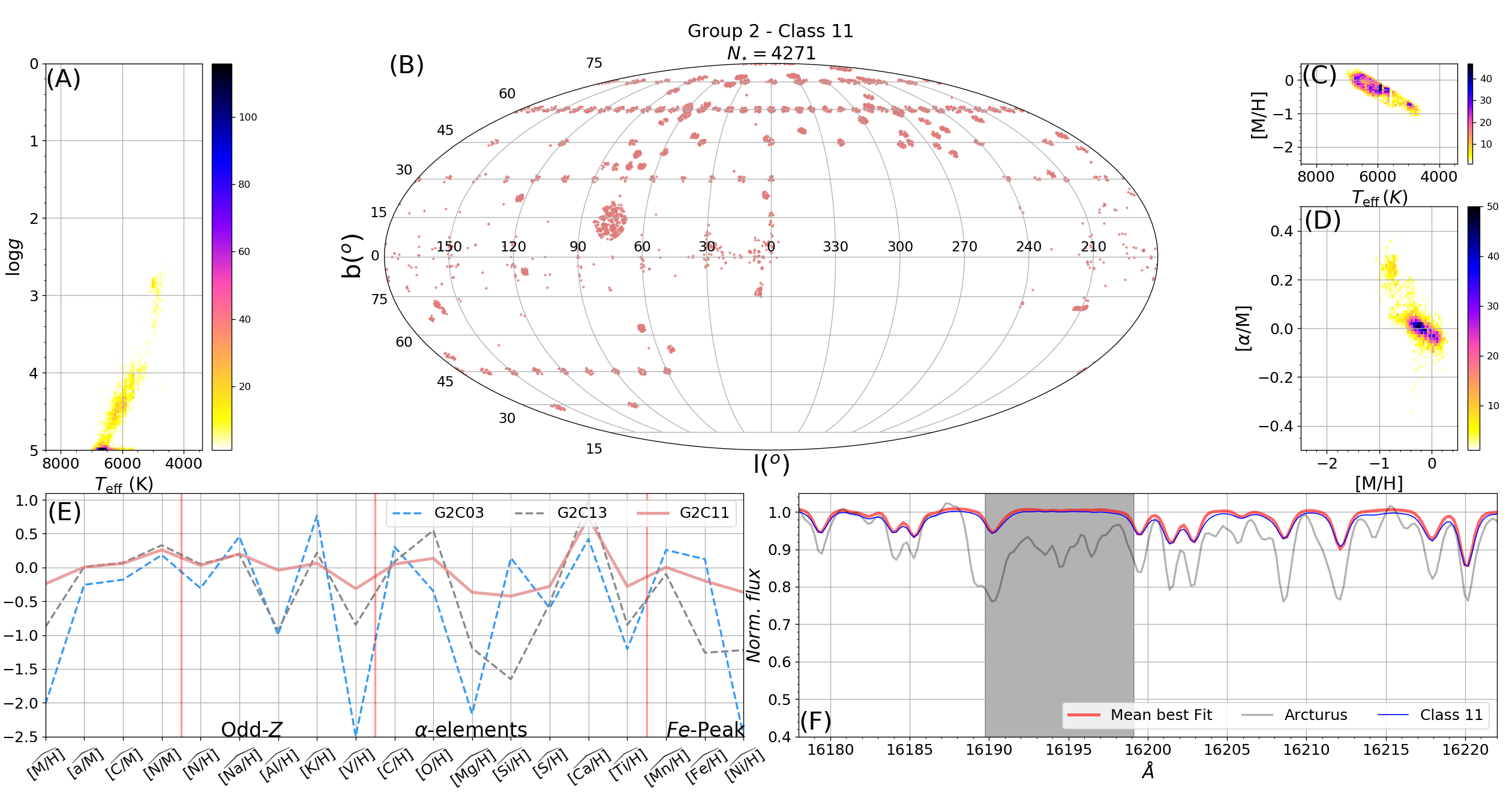}
    \includegraphics[width=0.7\textwidth]
        {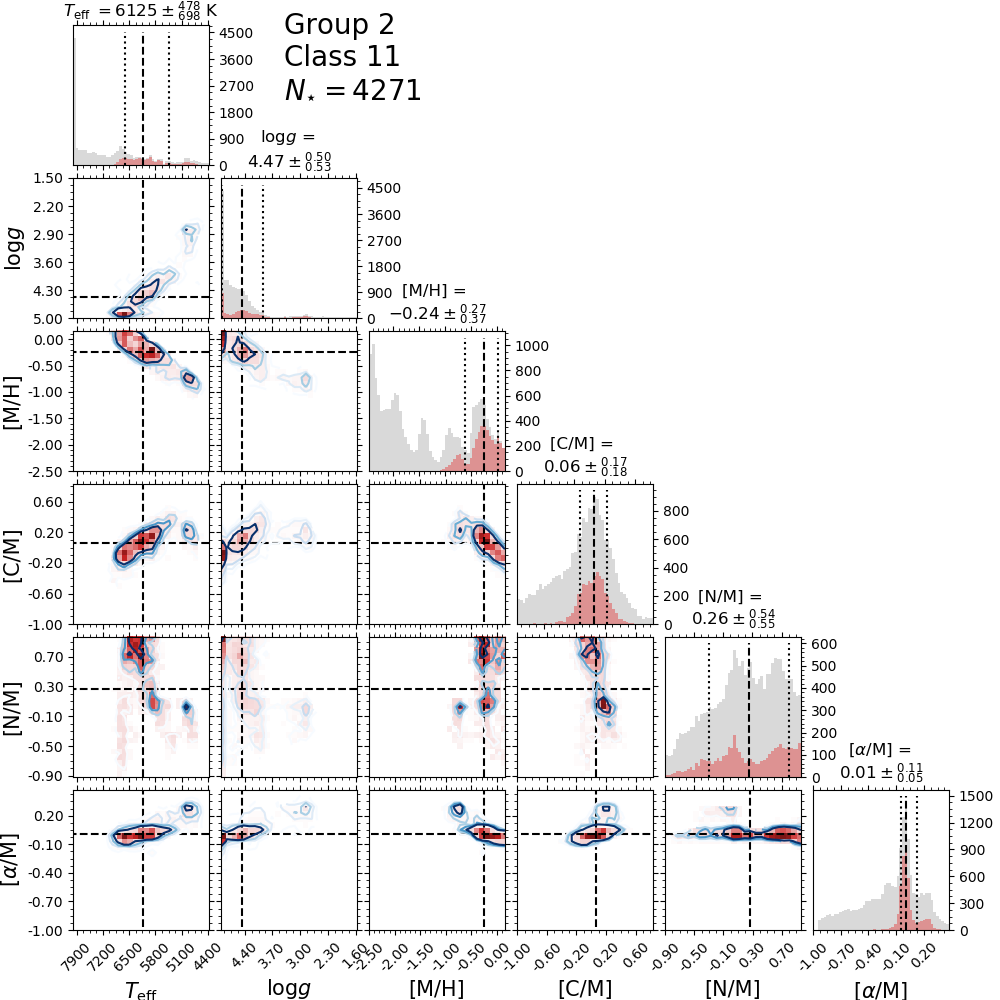}
    \caption{\label{class11} This figure folows the same pattern  from
    figure \ref{class00}. All the classes are described at Table  \ref{tab:desc}.}
    \end{figure*}

\begin{figure*}
    \centering
    \includegraphics[width=\textwidth]
        {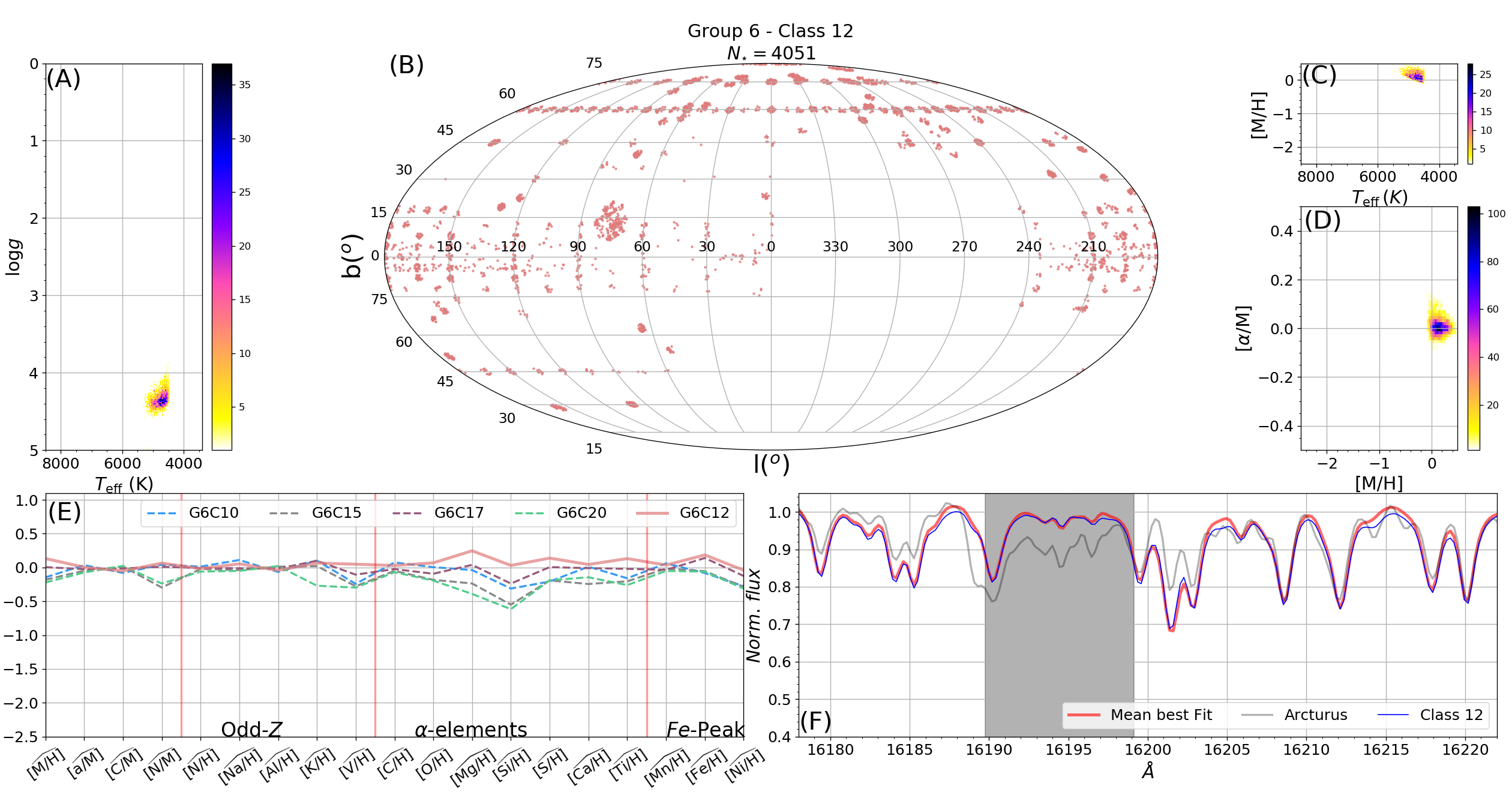}
    \includegraphics[width=0.7\textwidth]
        {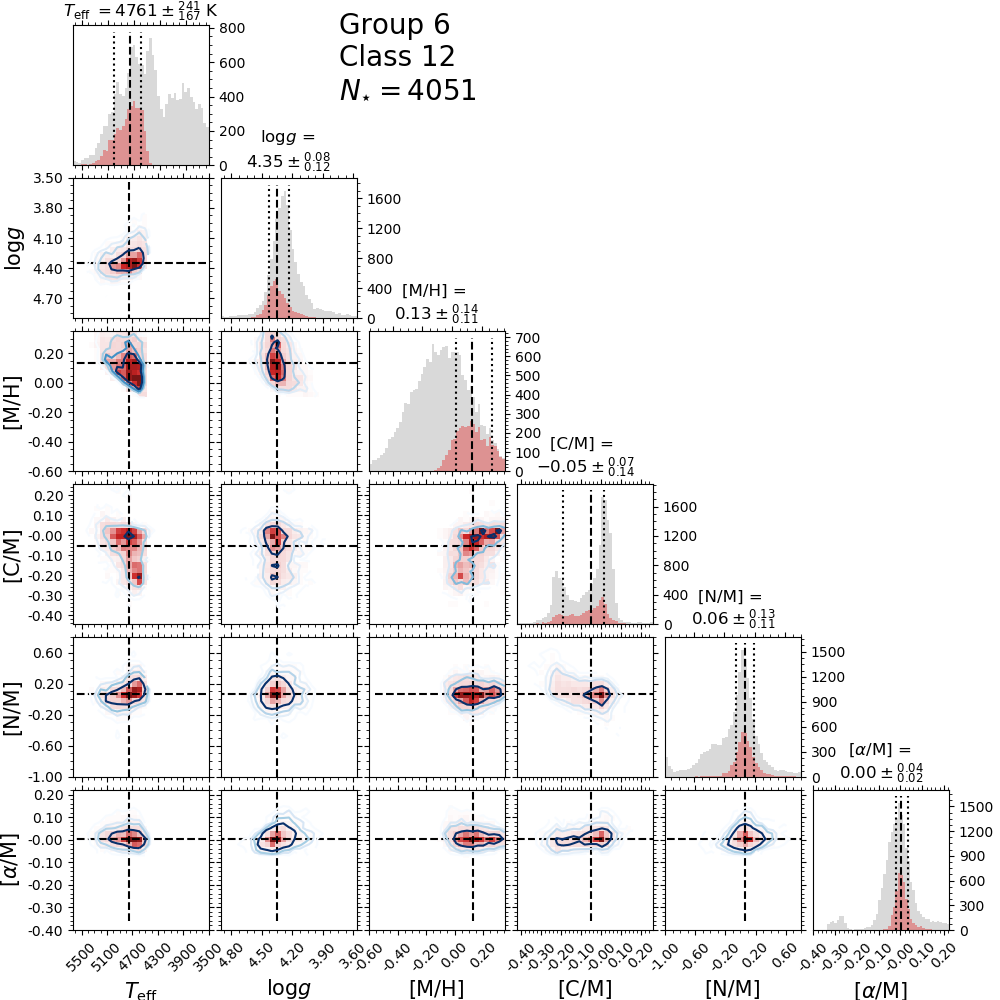}
    \caption{\label{class12} This figure folows the same pattern  from
    figure \ref{class00}. All the classes are described at Table  \ref{tab:desc}.}
    \end{figure*}

\begin{figure*}
    \centering
    \includegraphics[width=\textwidth]
        {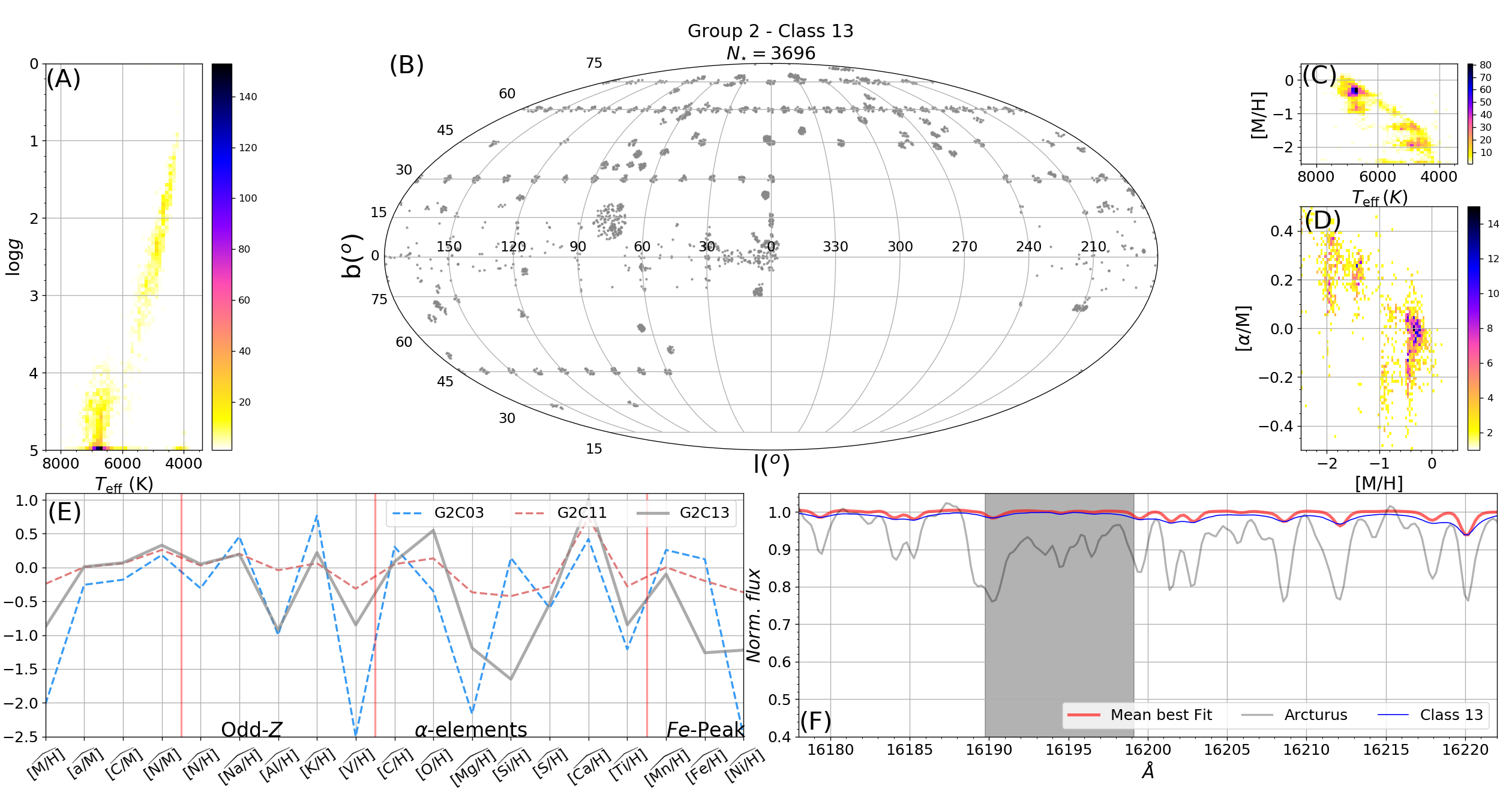}
    \includegraphics[width=0.7\textwidth]
        {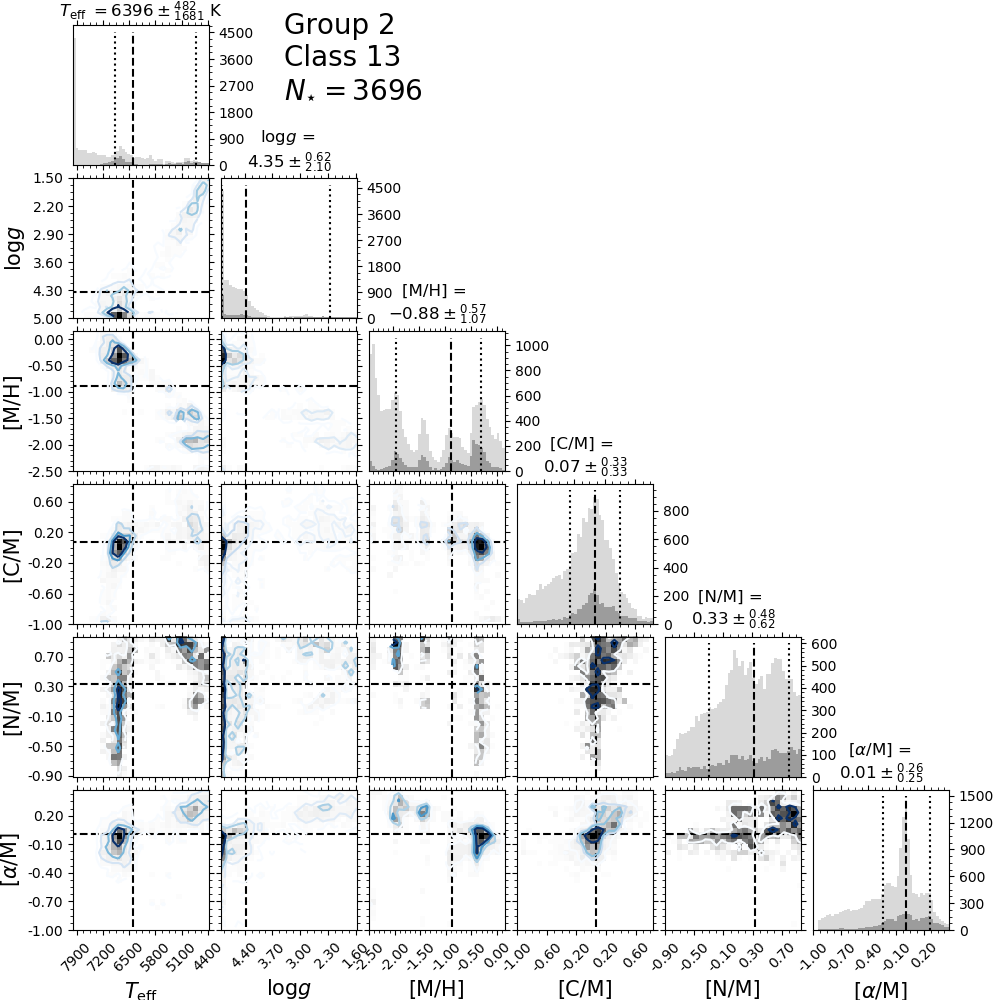}
    \caption{\label{class13} This figure folows the same pattern  from
    figure \ref{class00}. All the classes are described at Table  \ref{tab:desc}.}
    \end{figure*}

\begin{figure*}
    \centering
    \includegraphics[width=\textwidth]
        {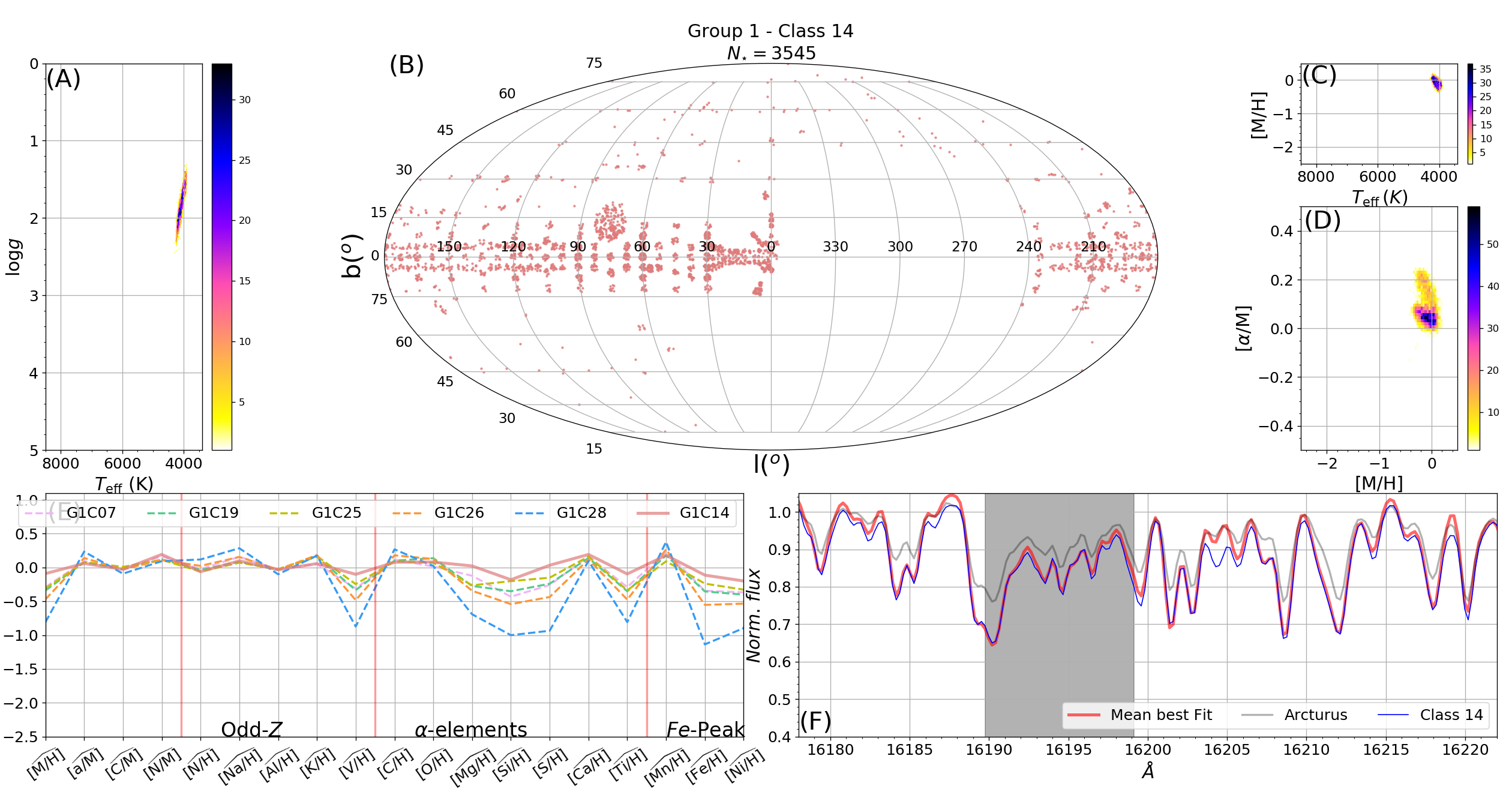}
    \includegraphics[width=0.7\textwidth]
        {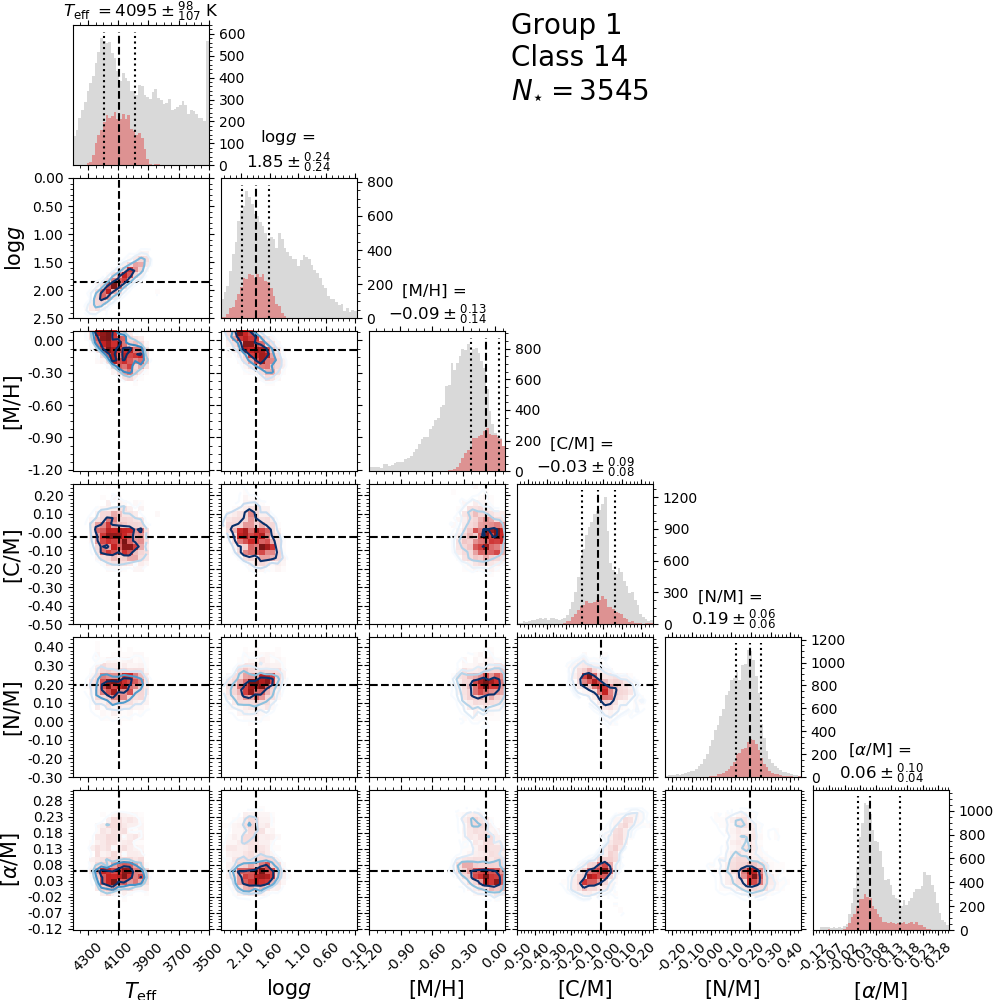}
    \caption{\label{class14} This figure folows the same pattern  from
    figure \ref{class00}. All the classes are described at Table  \ref{tab:desc}.}
    \end{figure*}

\begin{figure*}
    \centering
    \includegraphics[width=\textwidth]
        {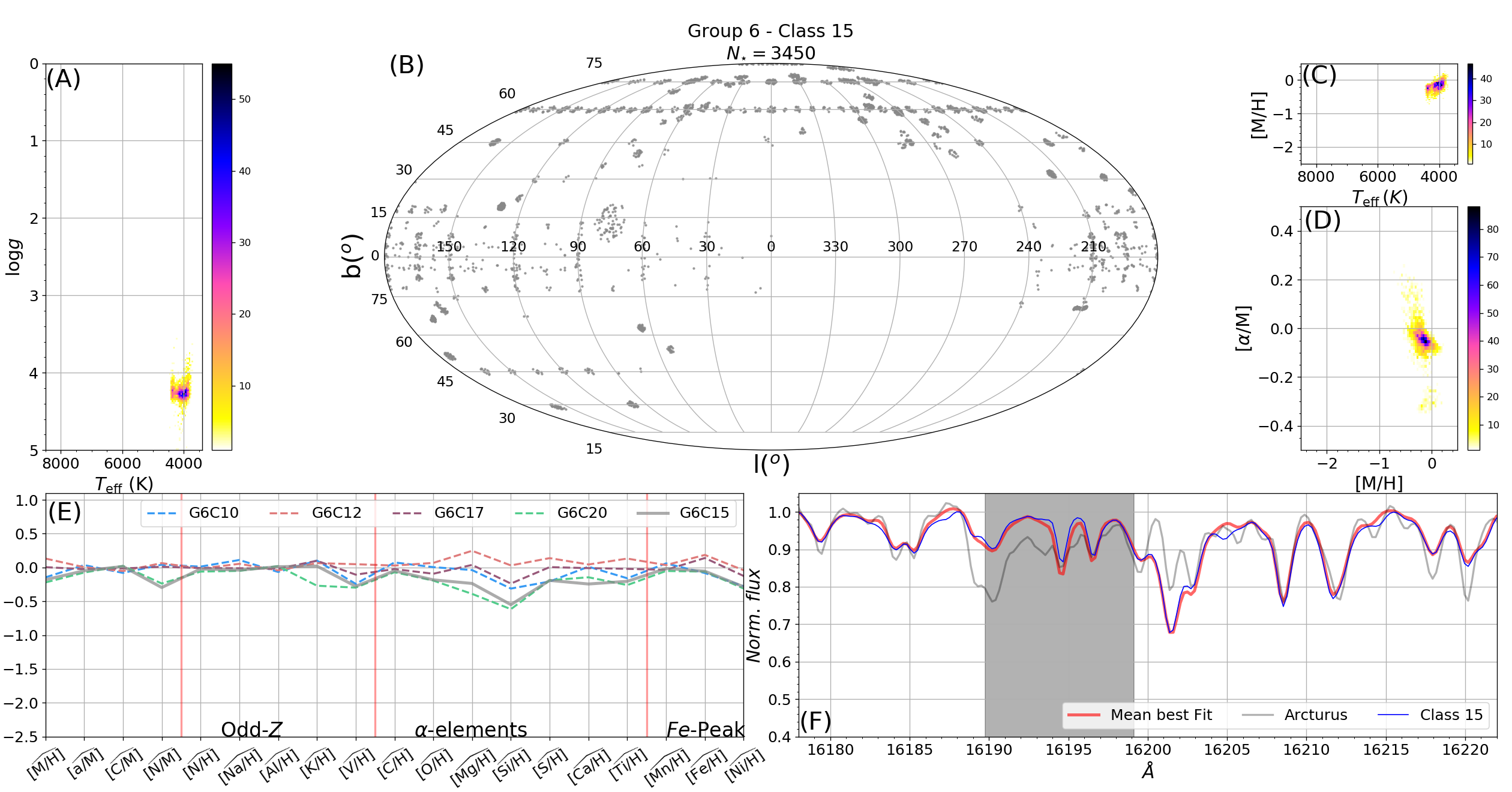}
    \includegraphics[width=0.7\textwidth]
        {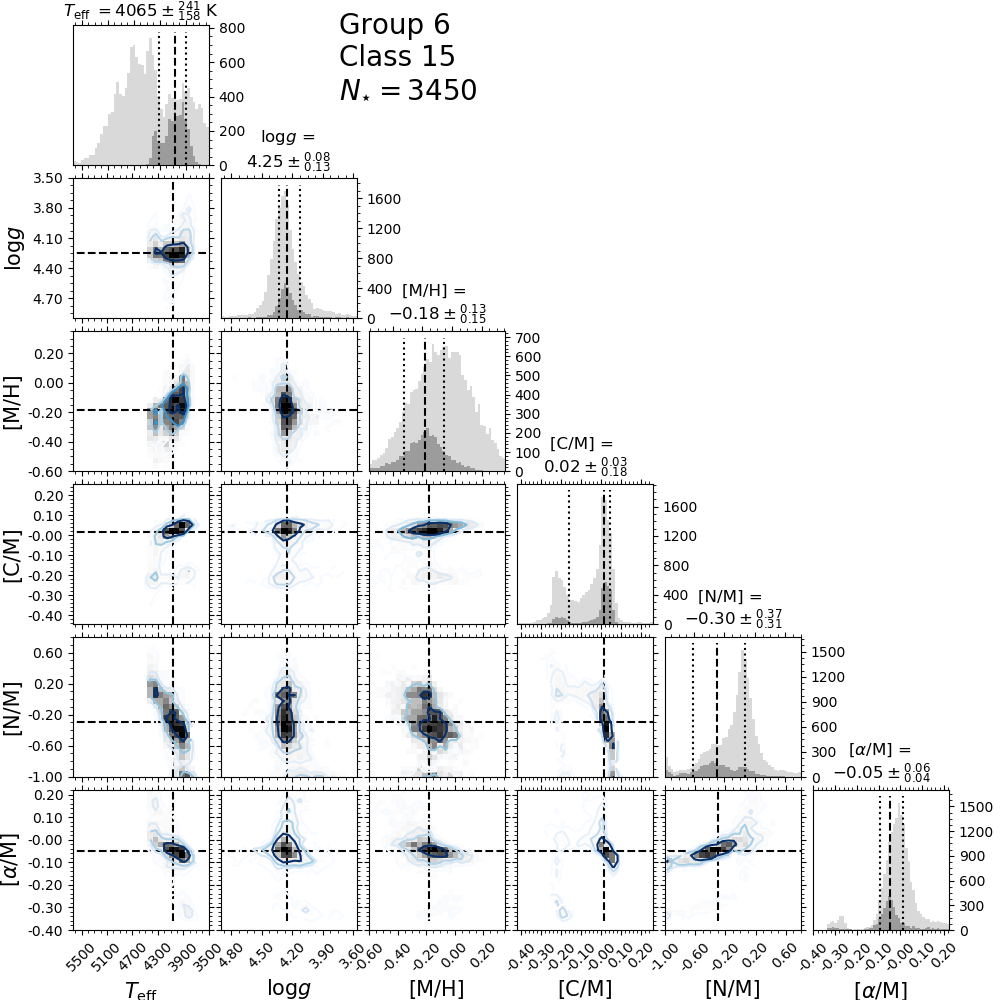}
    \caption{\label{class15} This figure folows the same pattern  from
    figure \ref{class00}. All the classes are described at Table  \ref{tab:desc}.}
    \end{figure*}

\begin{figure*}
    \centering
    \includegraphics[width=\textwidth]
        {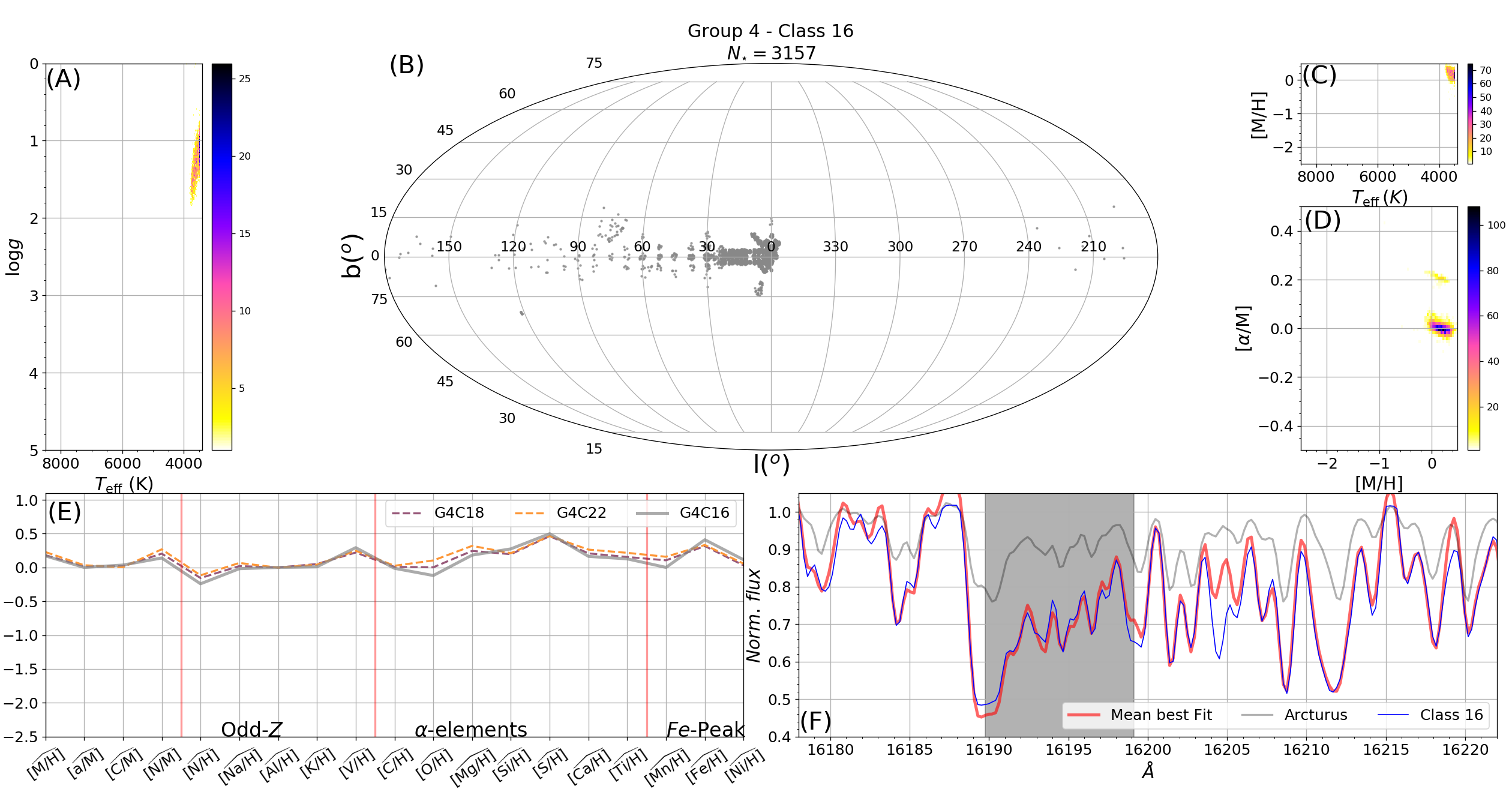}
    \includegraphics[width=0.7\textwidth]
        {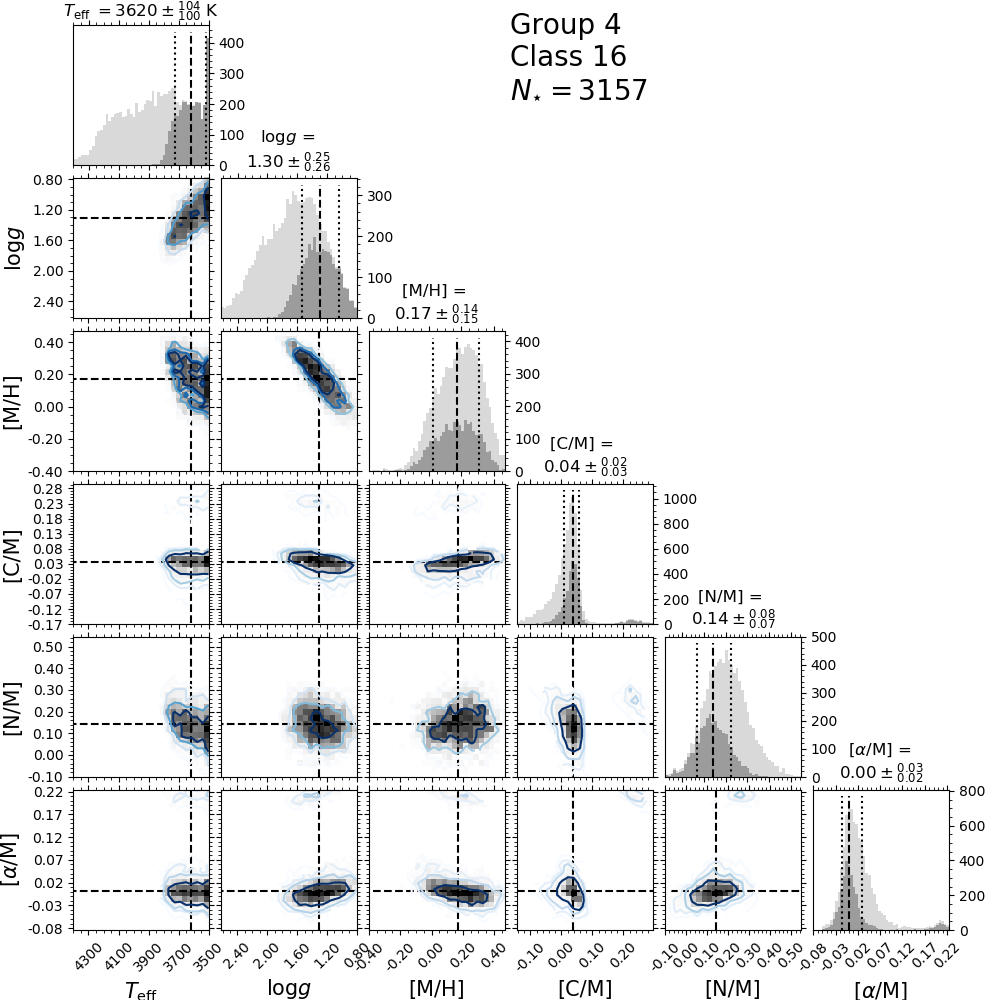}
    \caption{\label{class16} This figure folows the same pattern  from
    figure \ref{class00}. All the classes are described at Table  \ref{tab:desc}.}
    \end{figure*}

\begin{figure*}
    \centering
    \includegraphics[width=\textwidth]
        {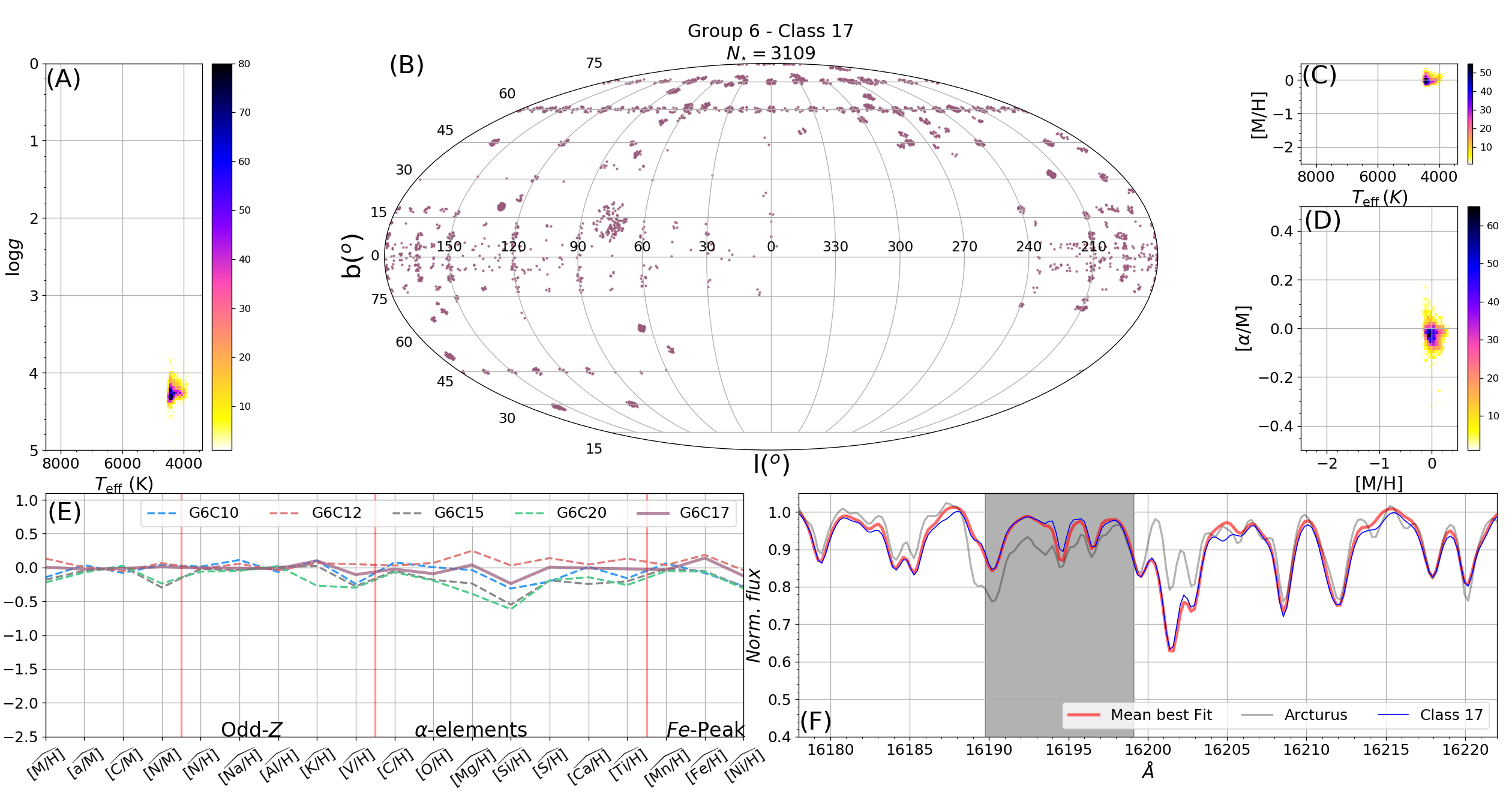}
    \includegraphics[width=0.7\textwidth]
        {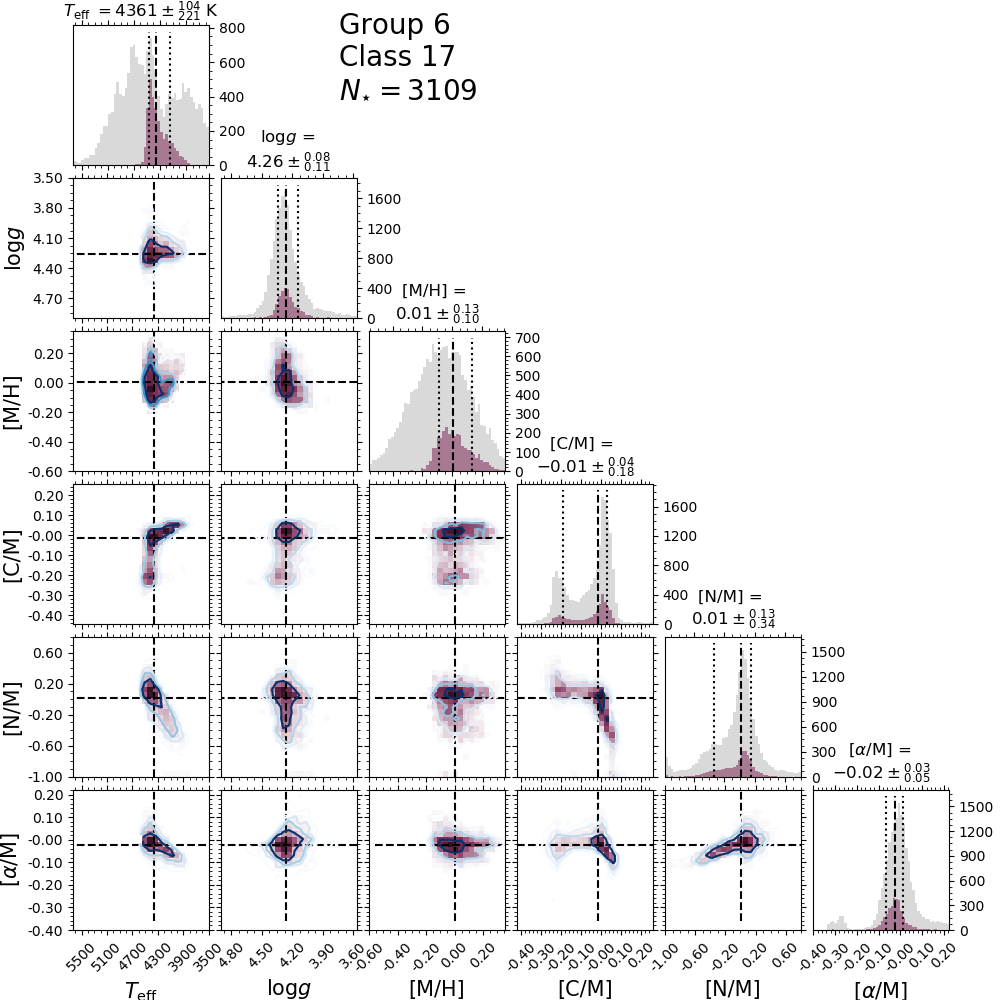}
    \caption{\label{class17} This figure folows the same pattern  from
    figure \ref{class00}. All the classes are described at Table  \ref{tab:desc}.}
    \end{figure*}

\begin{figure*}
    \centering
    \includegraphics[width=\textwidth]
        {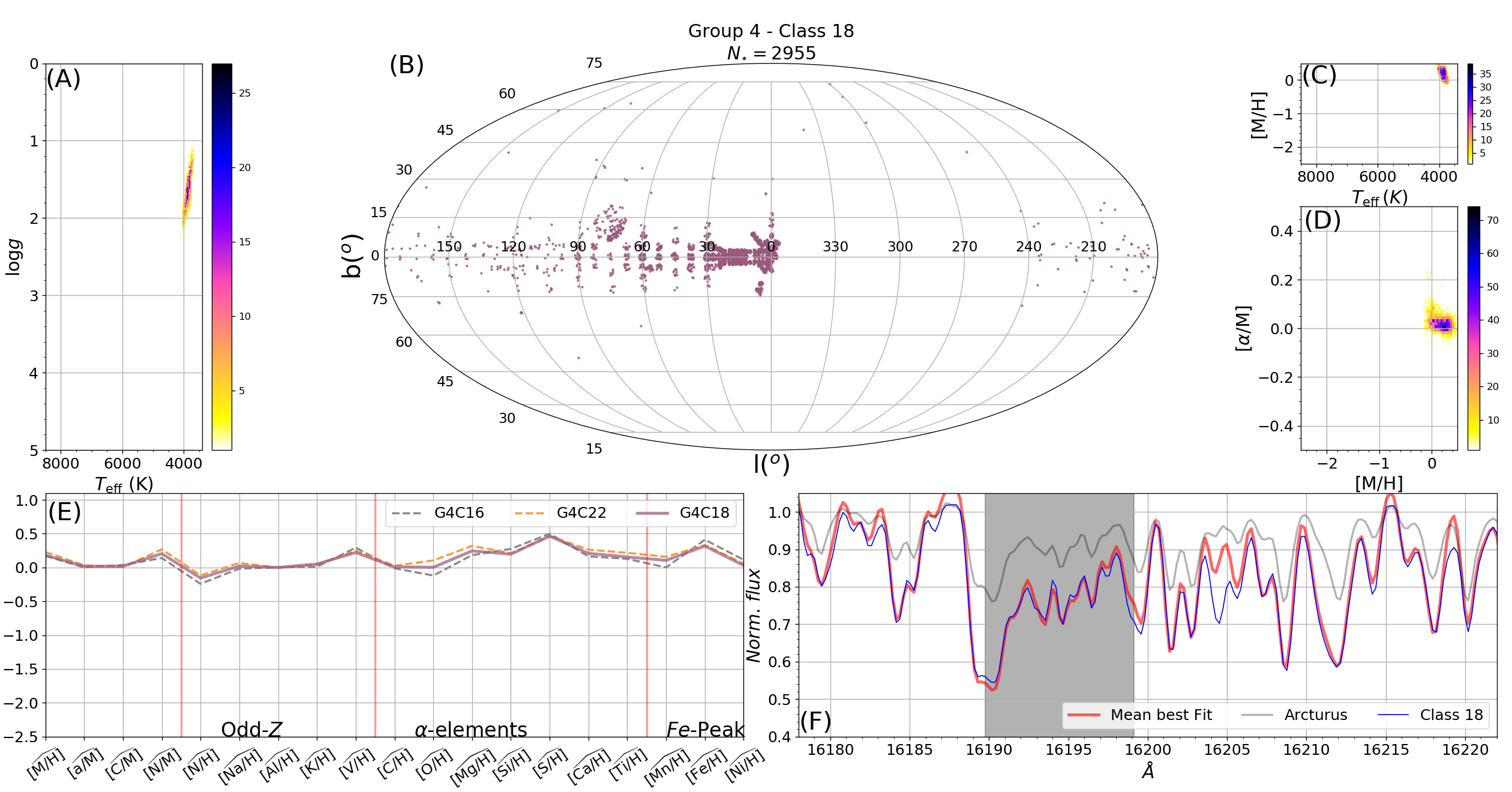}
    \includegraphics[width=0.7\textwidth]
        {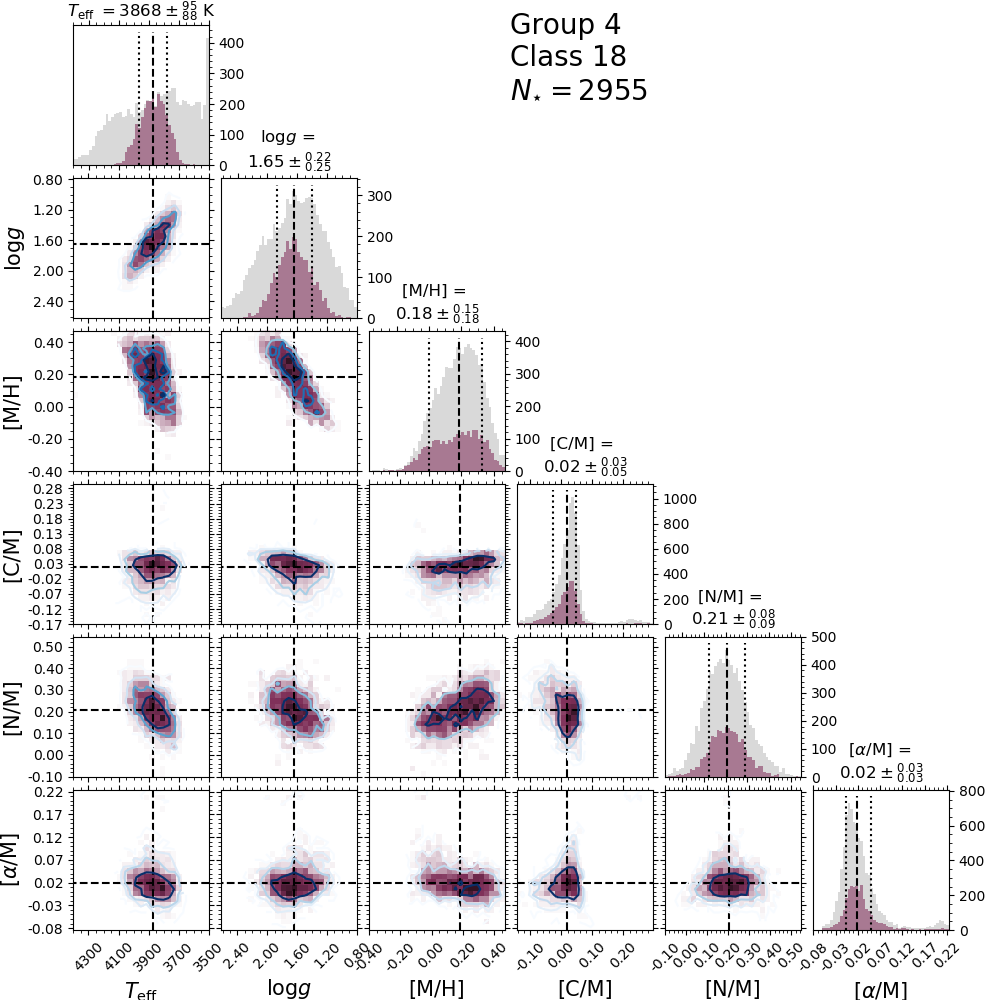}
    \caption{\label{class18} This figure folows the same pattern  from
    figure \ref{class00}. All the classes are described at Table  \ref{tab:desc}.}
    \end{figure*}

\begin{figure*}
    \centering
    \includegraphics[width=\textwidth]
        {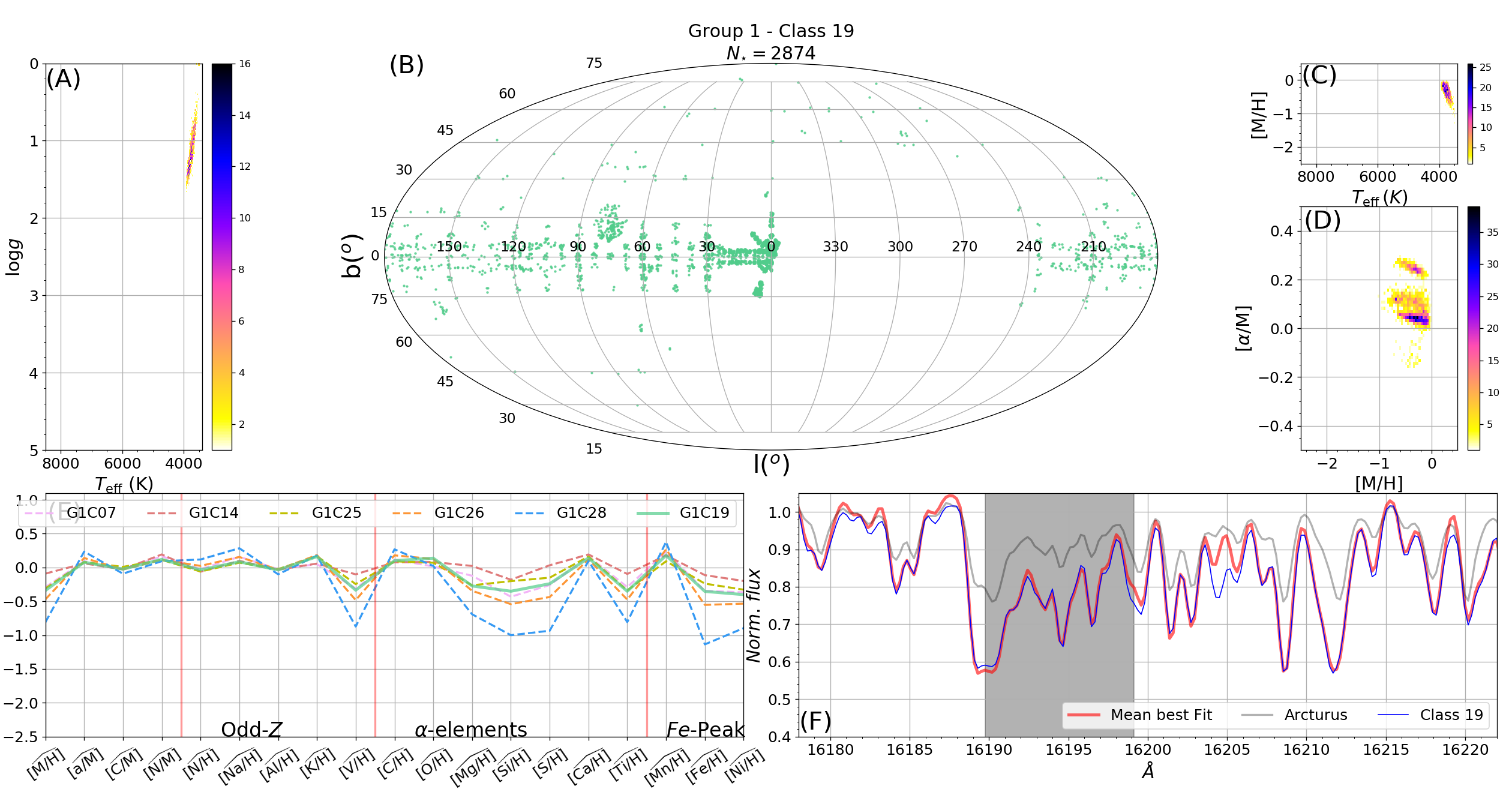}
    \includegraphics[width=0.7\textwidth]
        {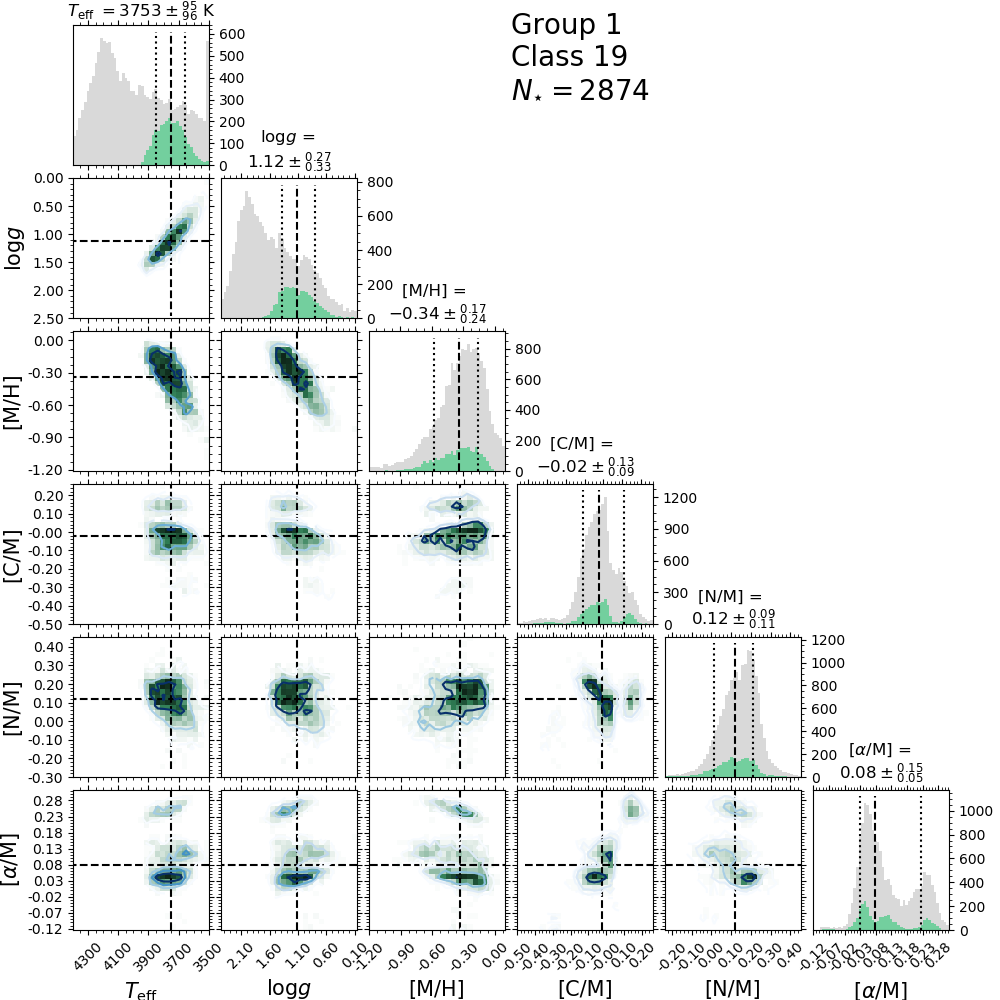}
    \caption{\label{class19} This figure folows the same pattern  from
    figure \ref{class00}. All the classes are described at Table  \ref{tab:desc}.}
    \end{figure*}

\begin{figure*}
    \centering
    \includegraphics[width=\textwidth]
        {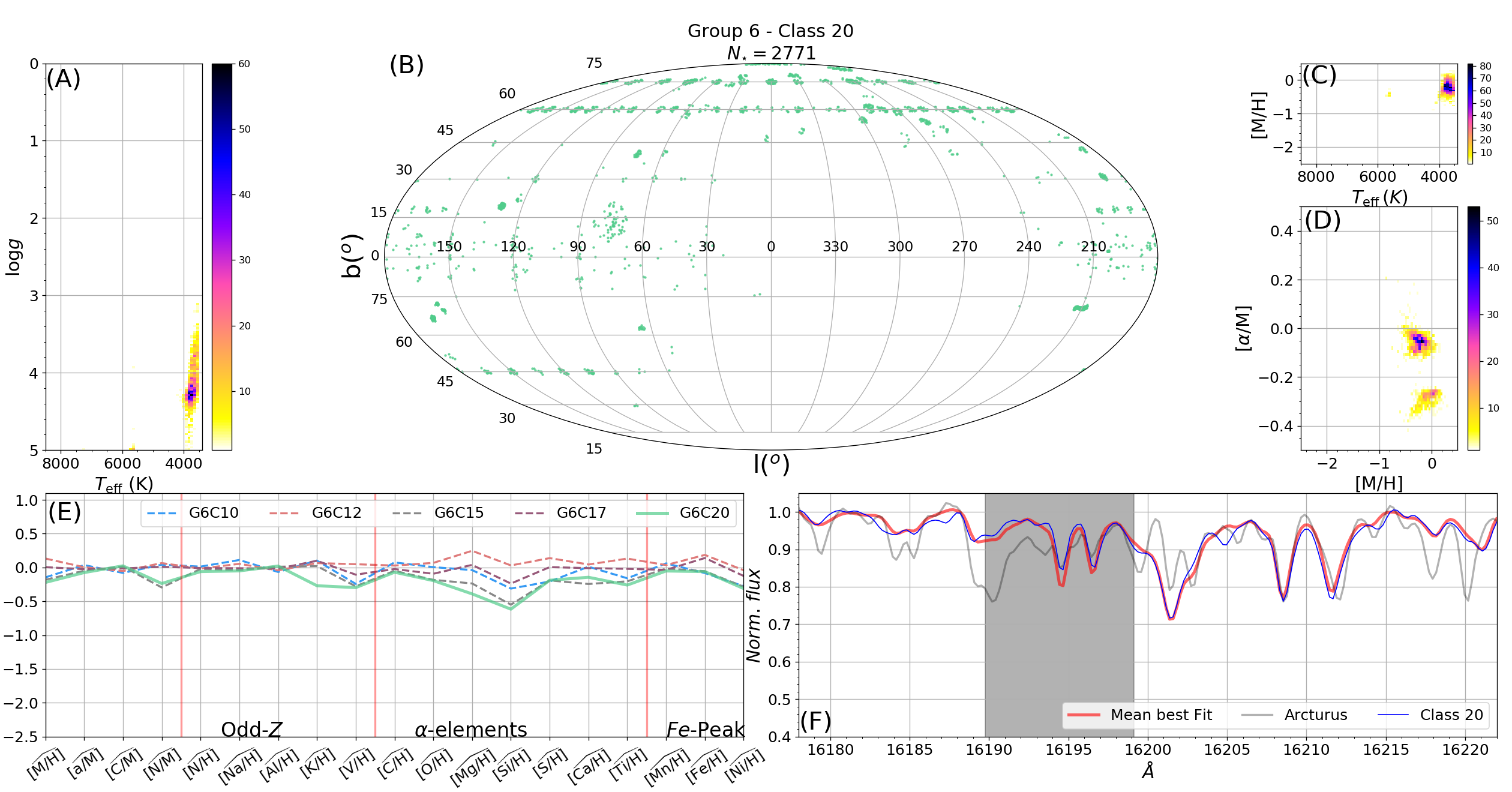}
    \includegraphics[width=0.7\textwidth]
        {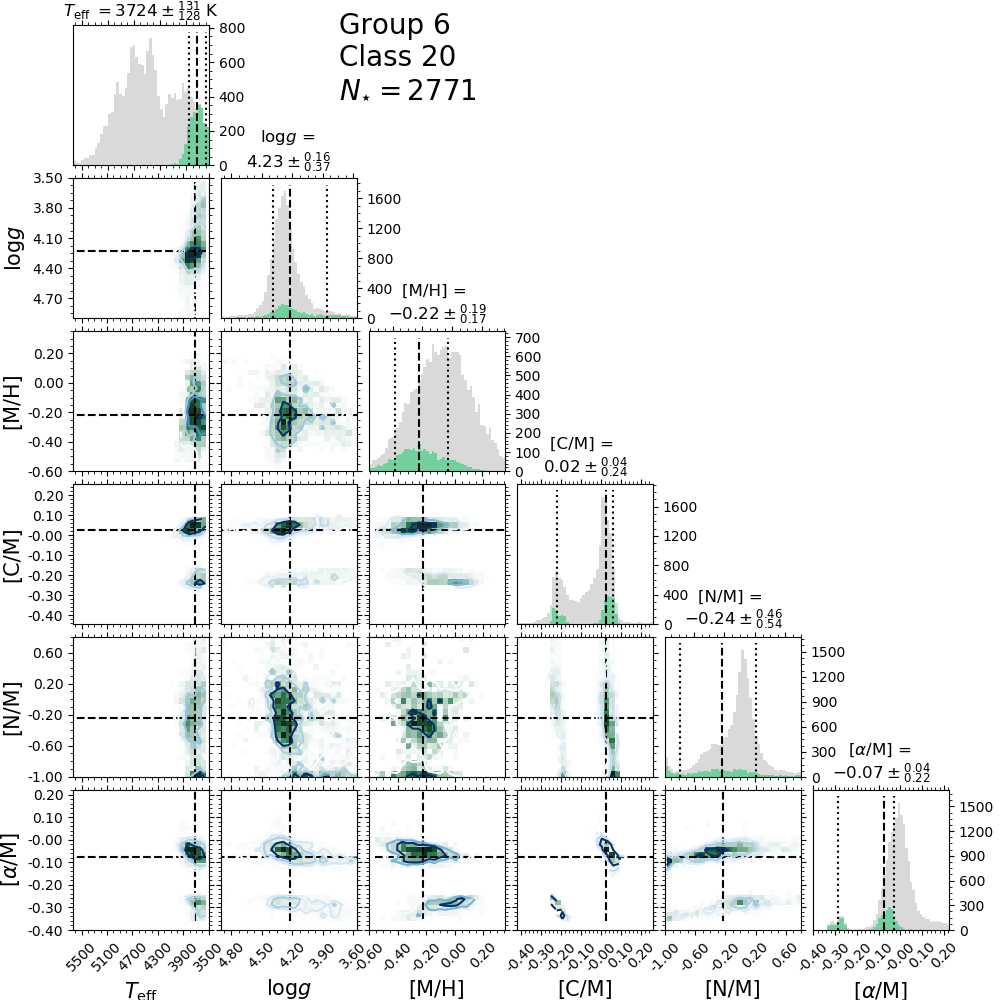}
    \caption{\label{class20} This figure folows the same pattern  from
    figure \ref{class00}. All the classes are described at Table  \ref{tab:desc}.}
    \end{figure*}

\begin{figure*}
    \centering
    \includegraphics[width=\textwidth]
        {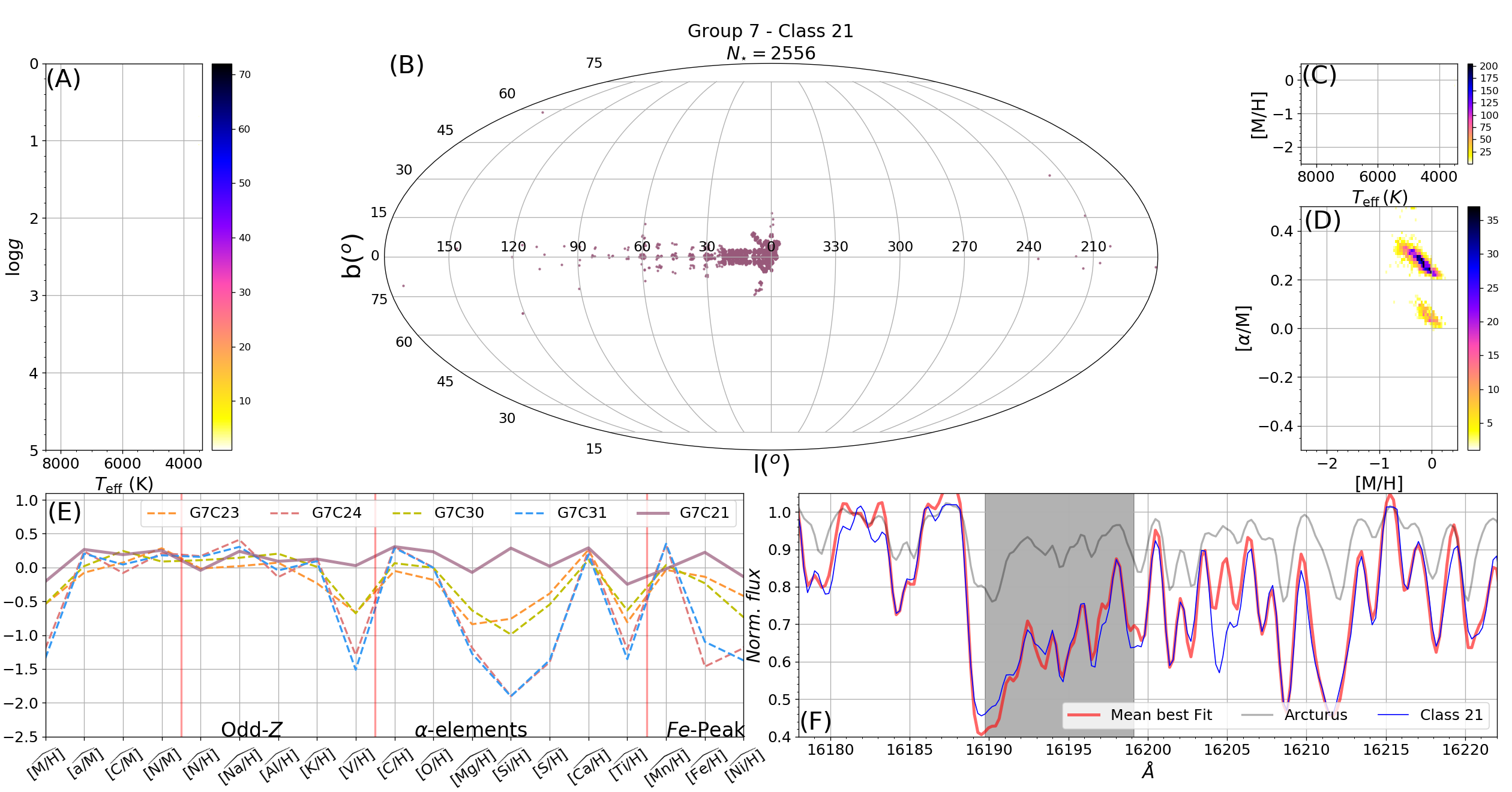}
    \includegraphics[width=0.7\textwidth]
        {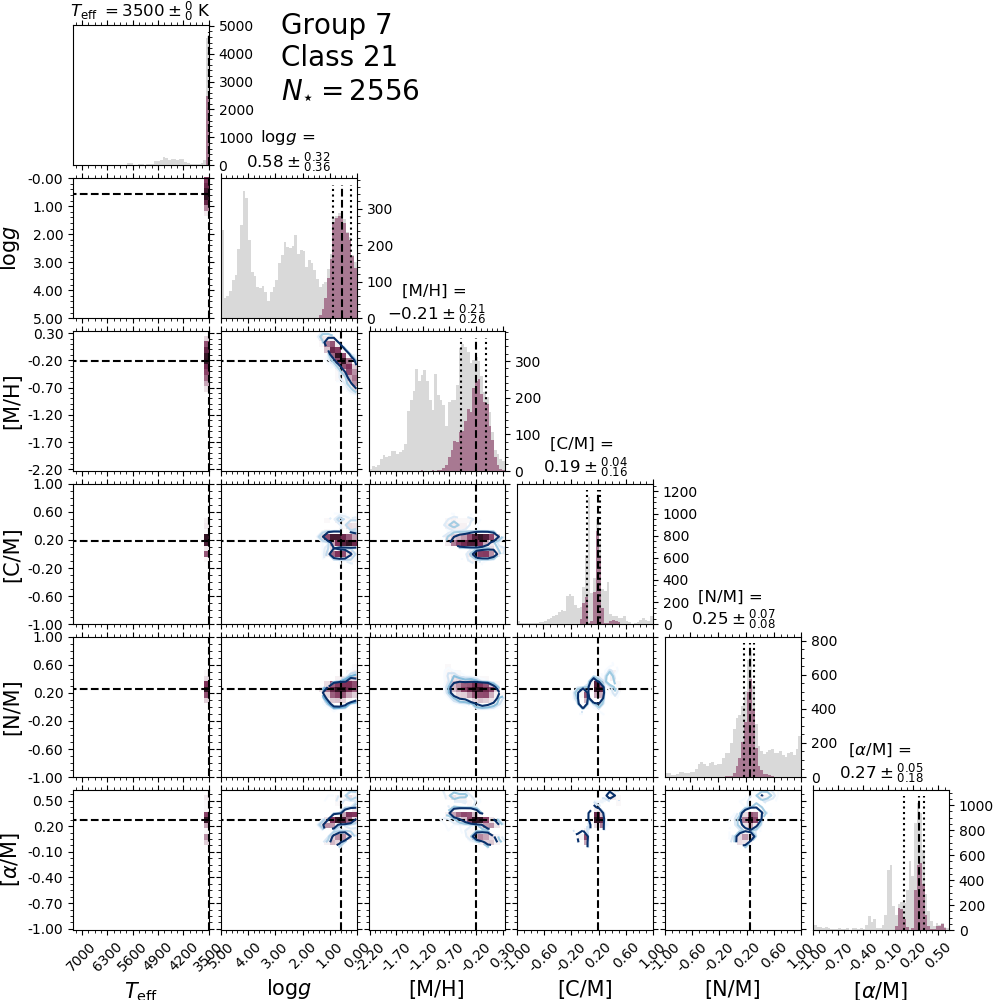}
    \caption{\label{class21} This figure folows the same pattern  from
    figure \ref{class00}. All the classes are described at Table  \ref{tab:desc}.}
    \end{figure*}

\begin{figure*}
    \centering
    \includegraphics[width=\textwidth]
        {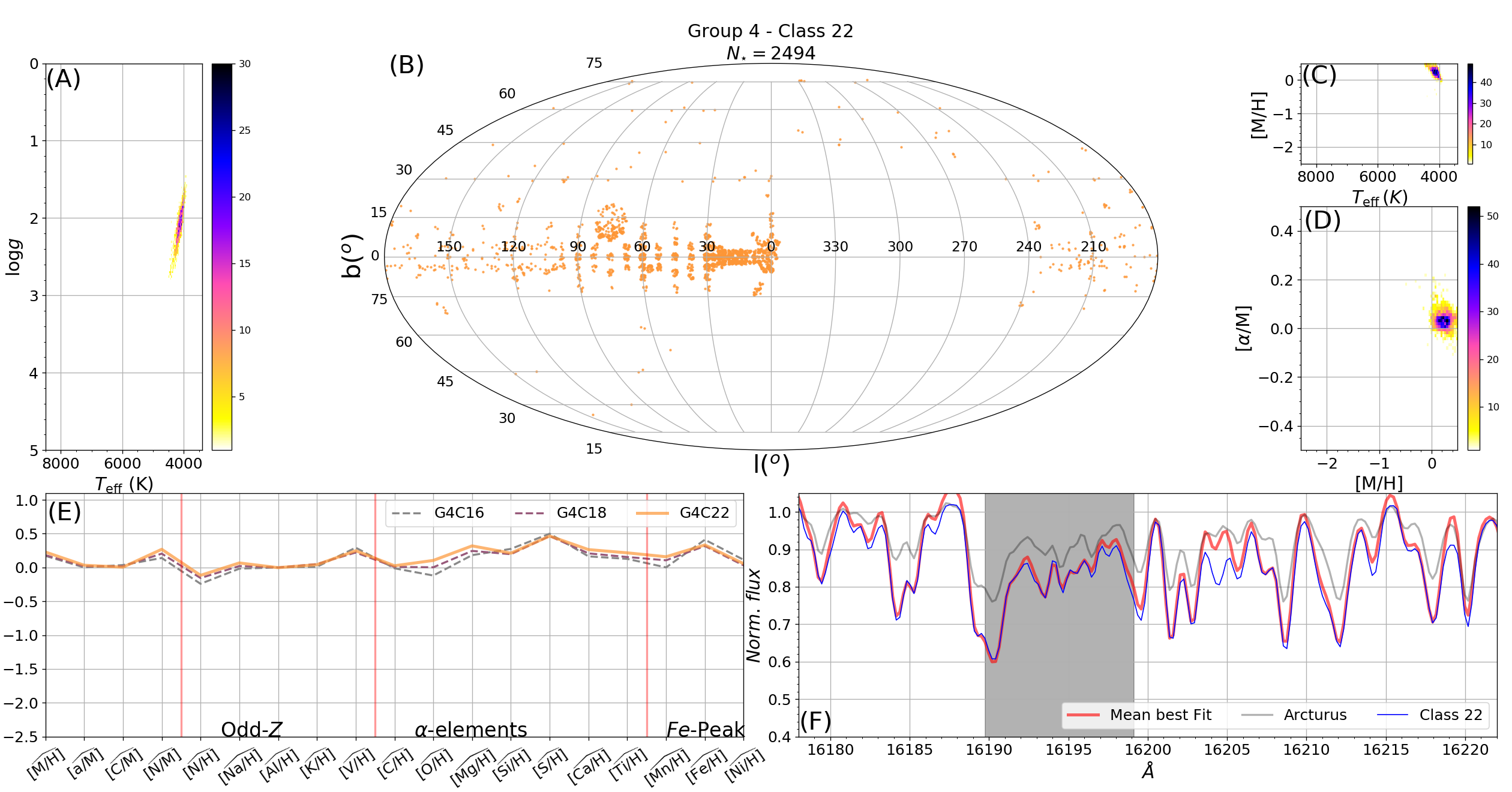}
    \includegraphics[width=0.7\textwidth]
        {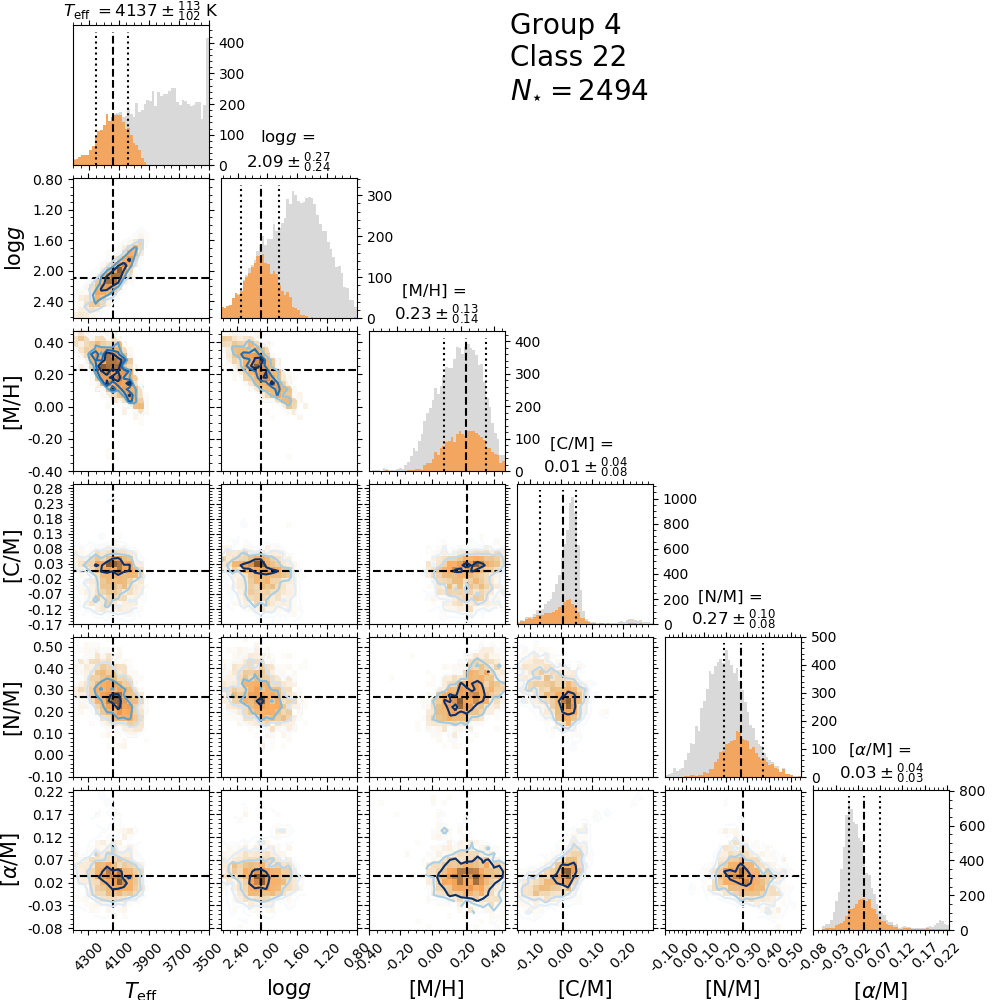}
    \caption{\label{class22} This figure folows the same pattern  from
    figure \ref{class00}. All the classes are described at Table  \ref{tab:desc}.}
    \end{figure*}

\begin{figure*}
    \centering
    \includegraphics[width=\textwidth]
        {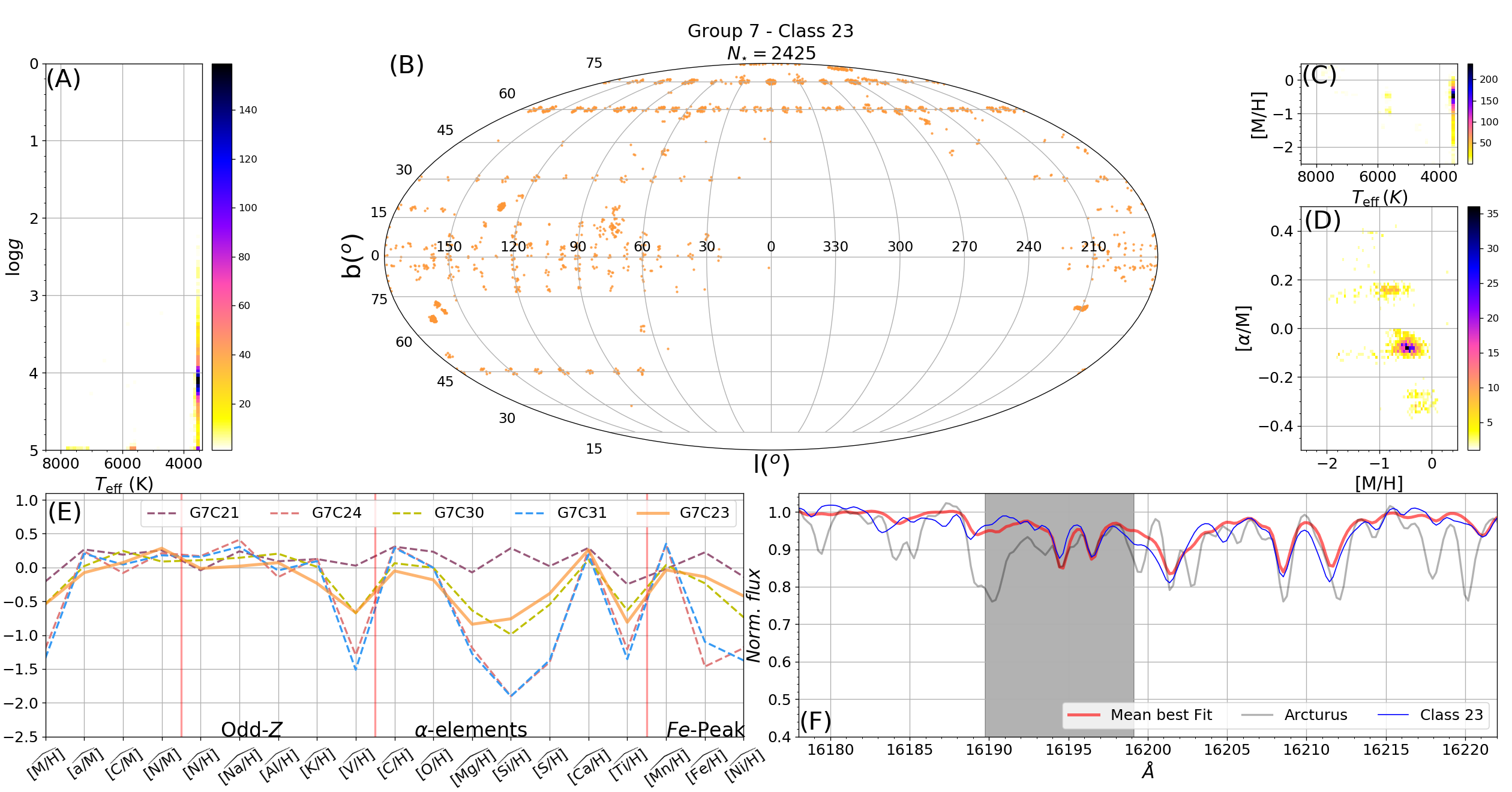}
    \includegraphics[width=0.7\textwidth]
        {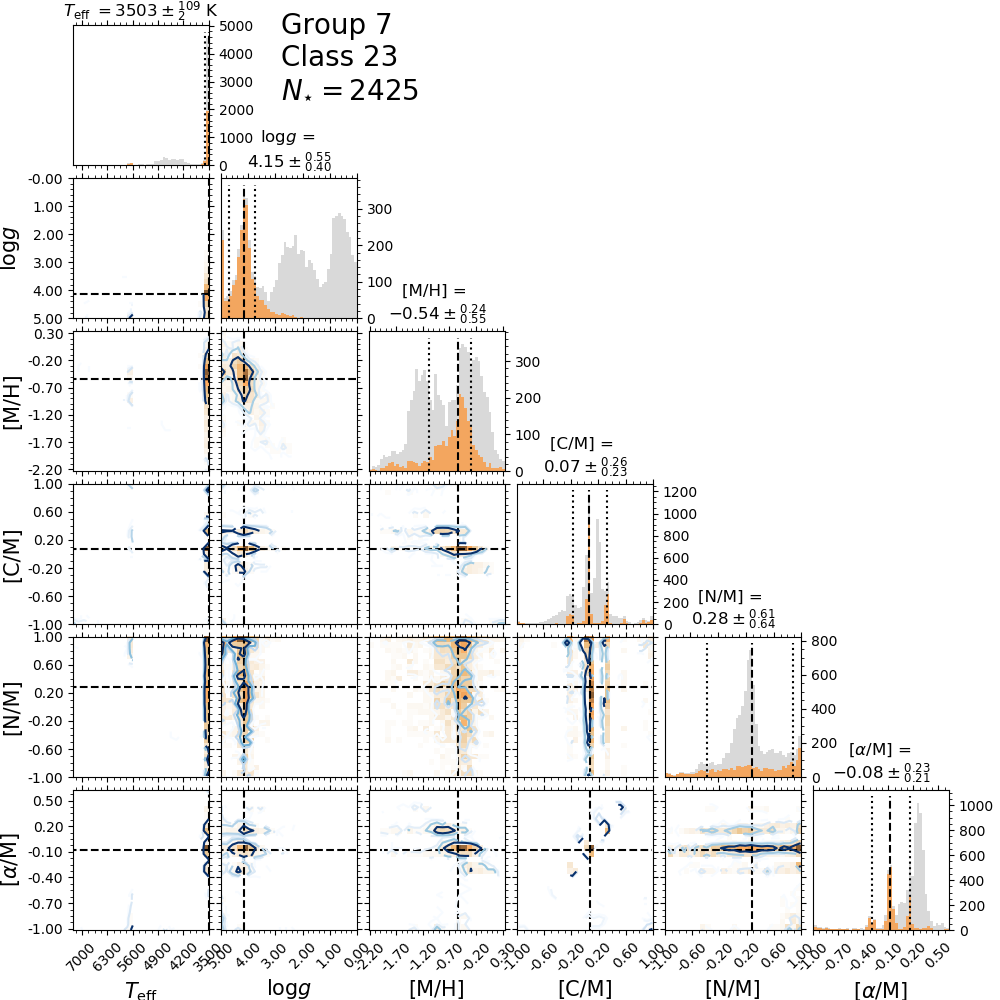}
    \caption{\label{class23} This figure folows the same pattern  from
    figure \ref{class00}. All the classes are described at Table  \ref{tab:desc}.}
    \end{figure*}

\begin{figure*}
    \centering
    \includegraphics[width=\textwidth]
        {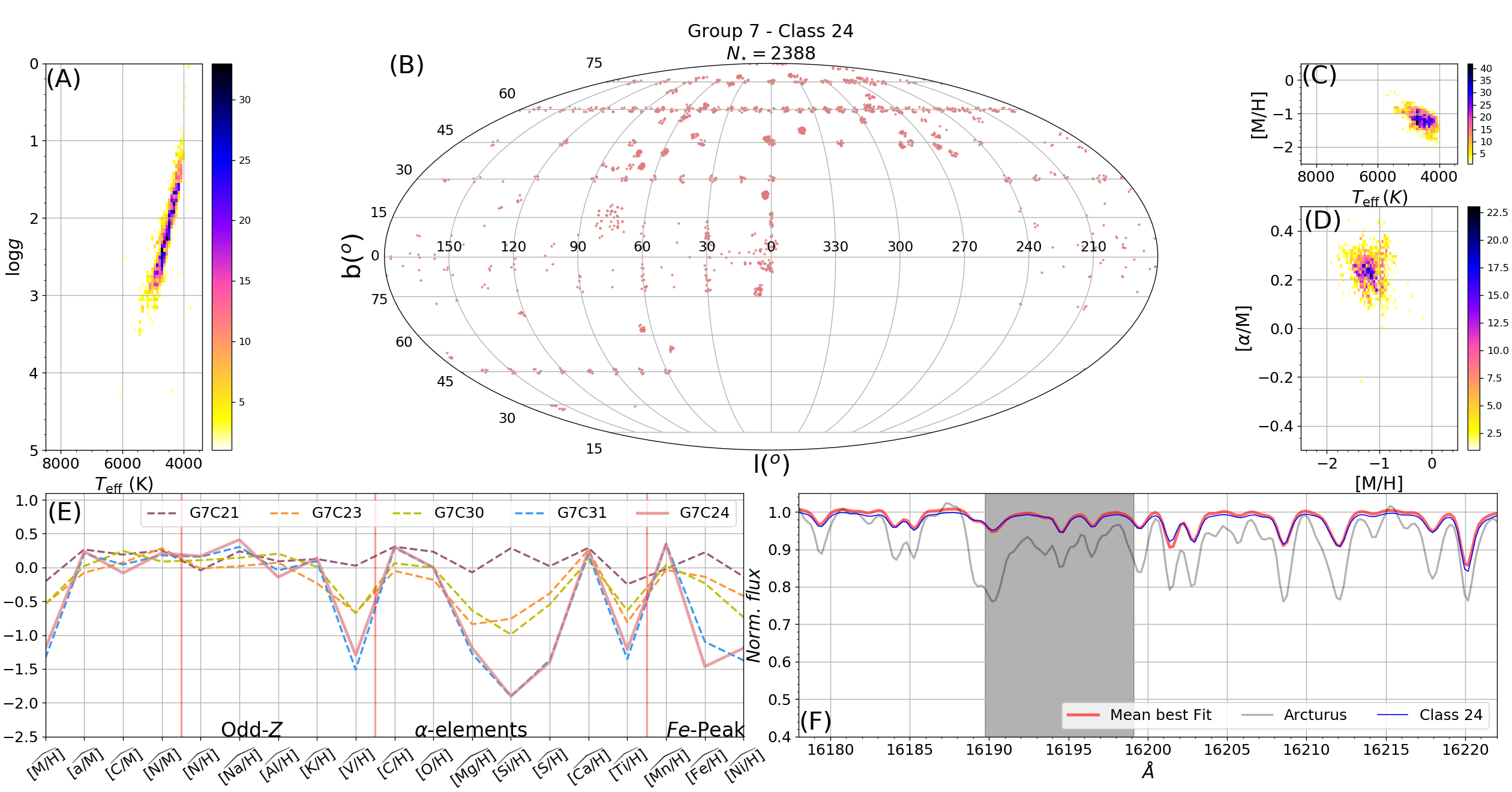}
    \includegraphics[width=0.7\textwidth]
        {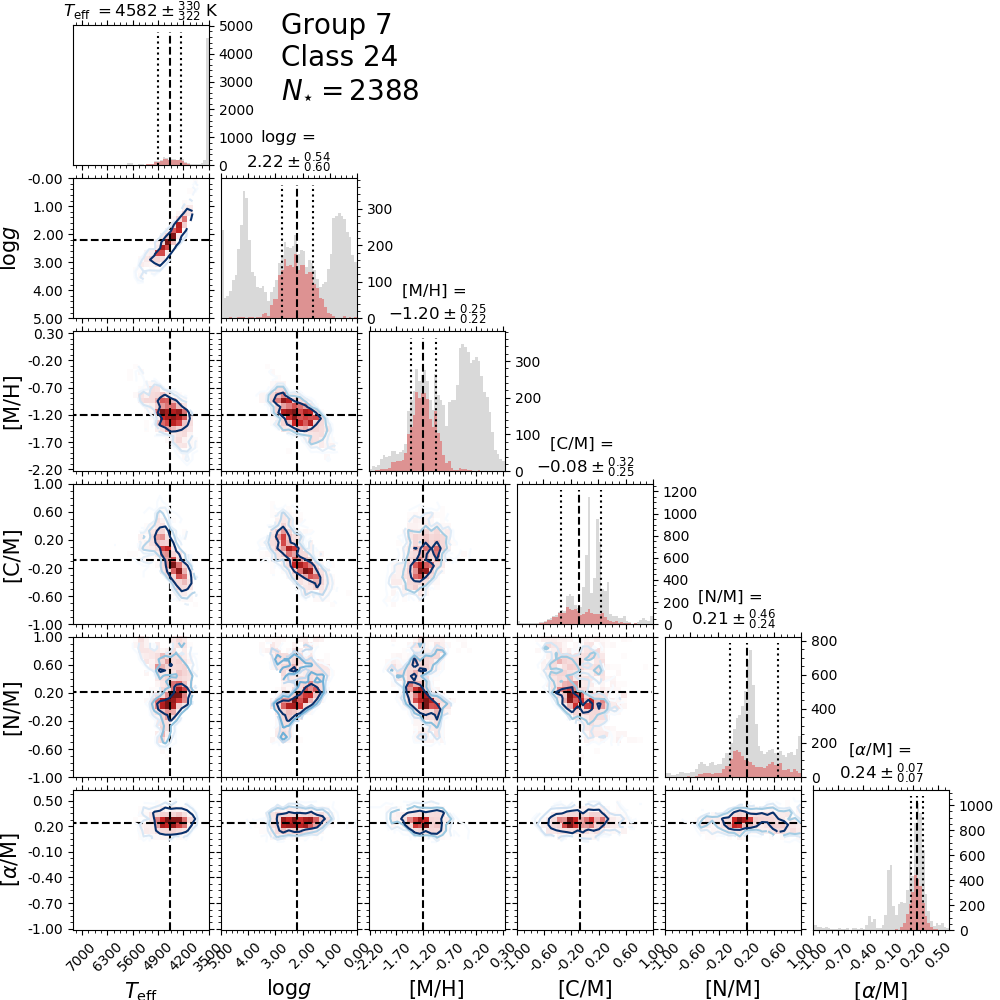}
    \caption{\label{class24} This figure folows the same pattern  from
    figure \ref{class00}. All the classes are described at Table  \ref{tab:desc}.}
    \end{figure*}

\begin{figure*}
    \centering
    \includegraphics[width=\textwidth]
        {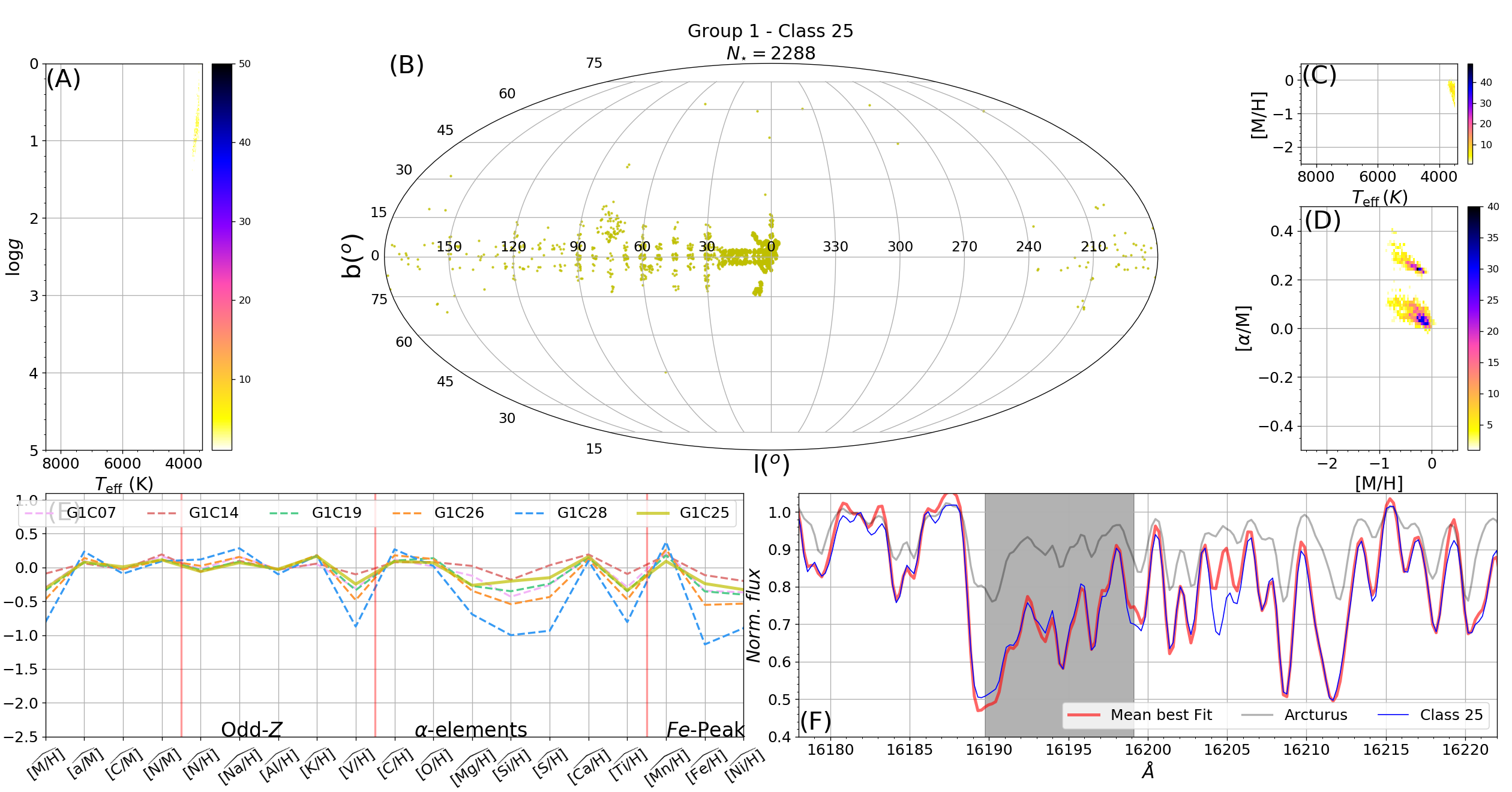}
    \includegraphics[width=0.7\textwidth]
        {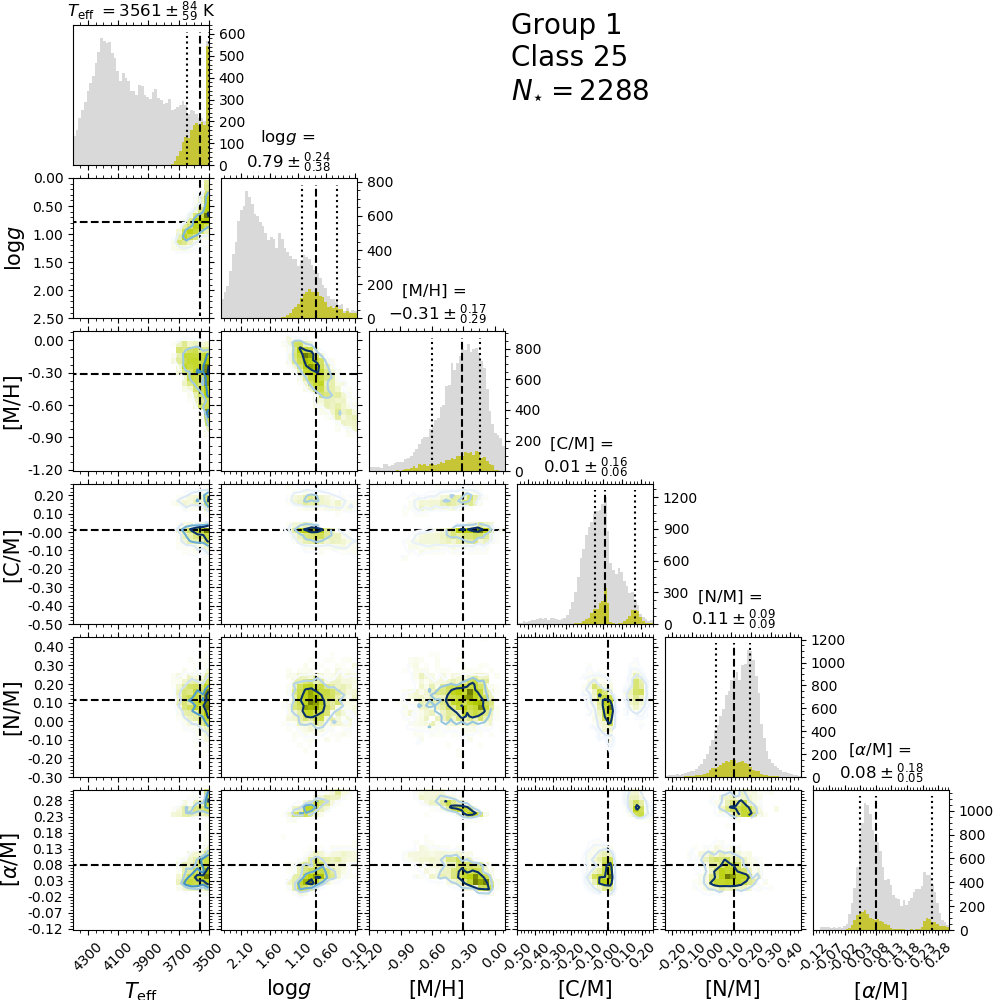}
    \caption{\label{class25} This figure folows the same pattern  from
    figure \ref{class00}. All the classes are described at Table  \ref{tab:desc}.}
    \end{figure*}

\begin{figure*}
    \centering
    \includegraphics[width=\textwidth]
        {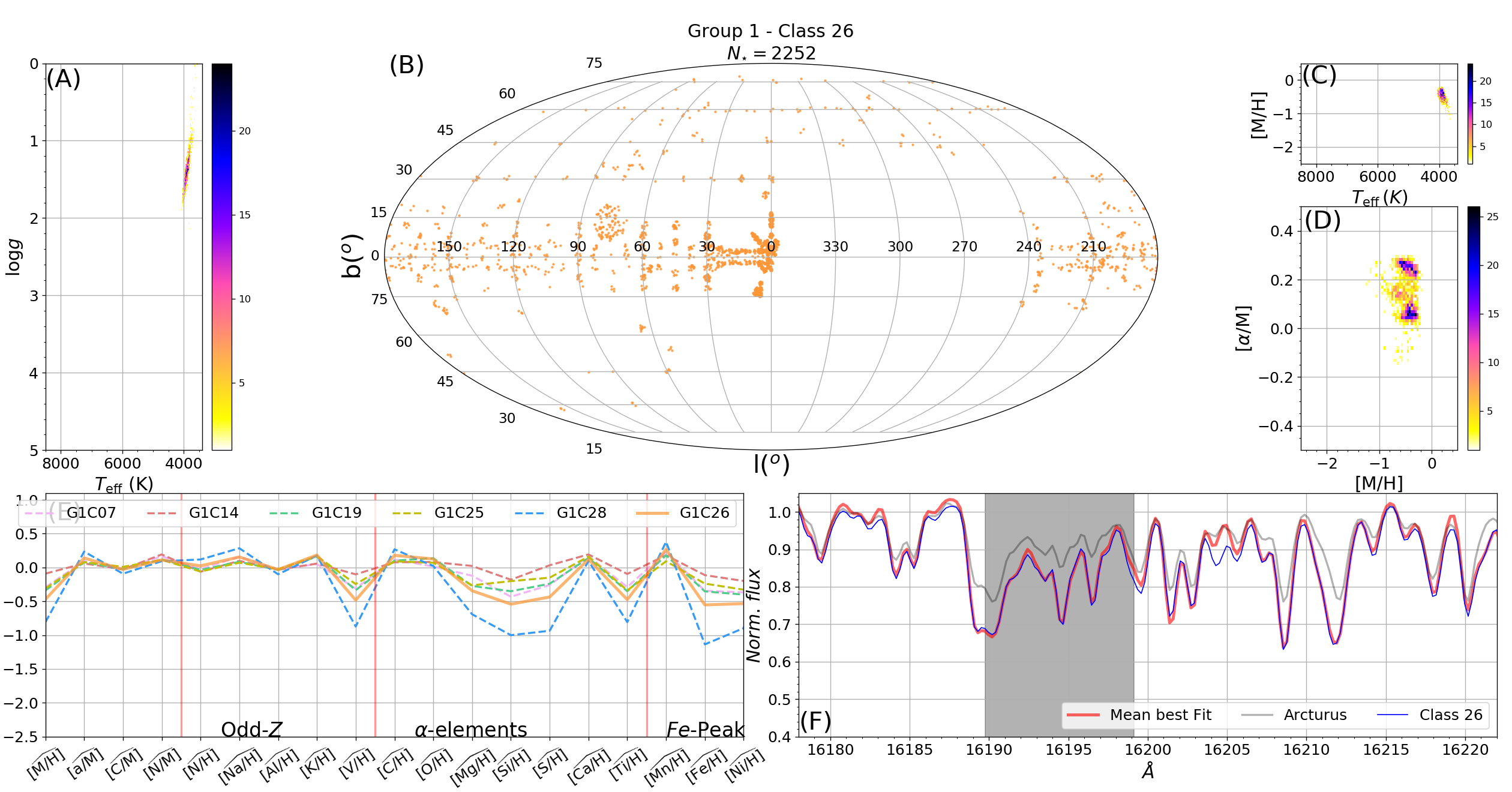}
    \includegraphics[width=0.7\textwidth]
        {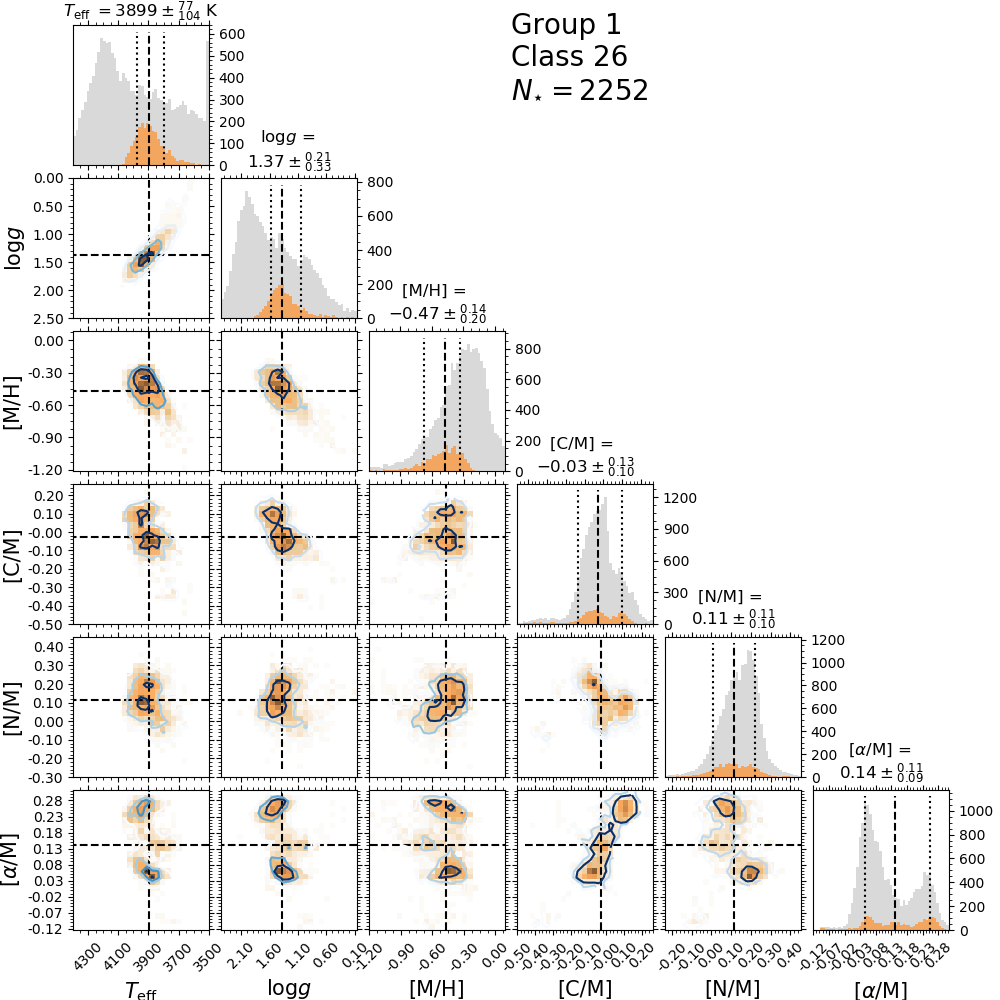}
    \caption{\label{class26} This figure folows the same pattern  from
    figure \ref{class00}. All the classes are described at Table  \ref{tab:desc}.}
    \end{figure*}

\begin{figure*}
    \centering
    \includegraphics[width=\textwidth]
        {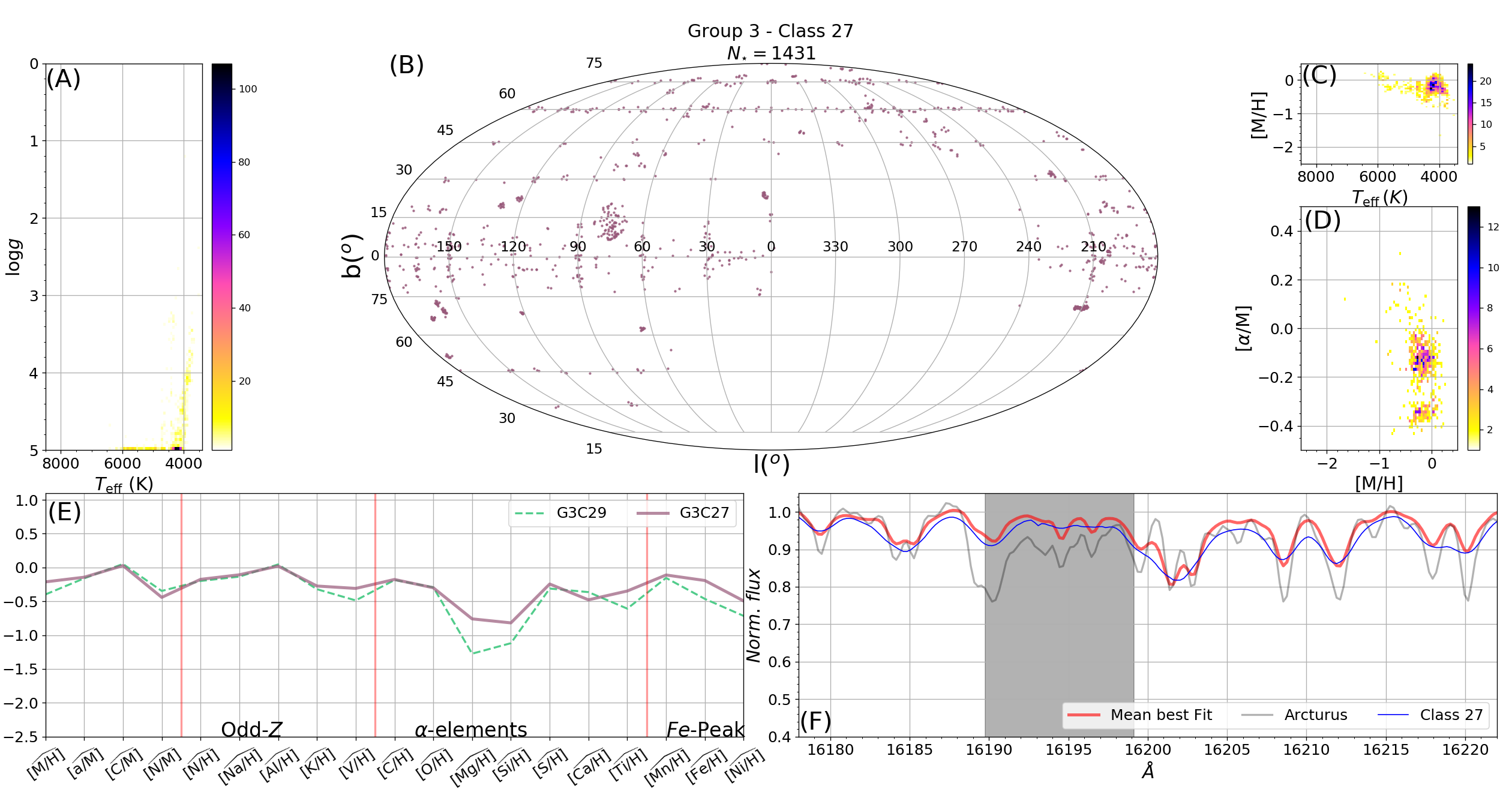}
    \includegraphics[width=0.7\textwidth]
        {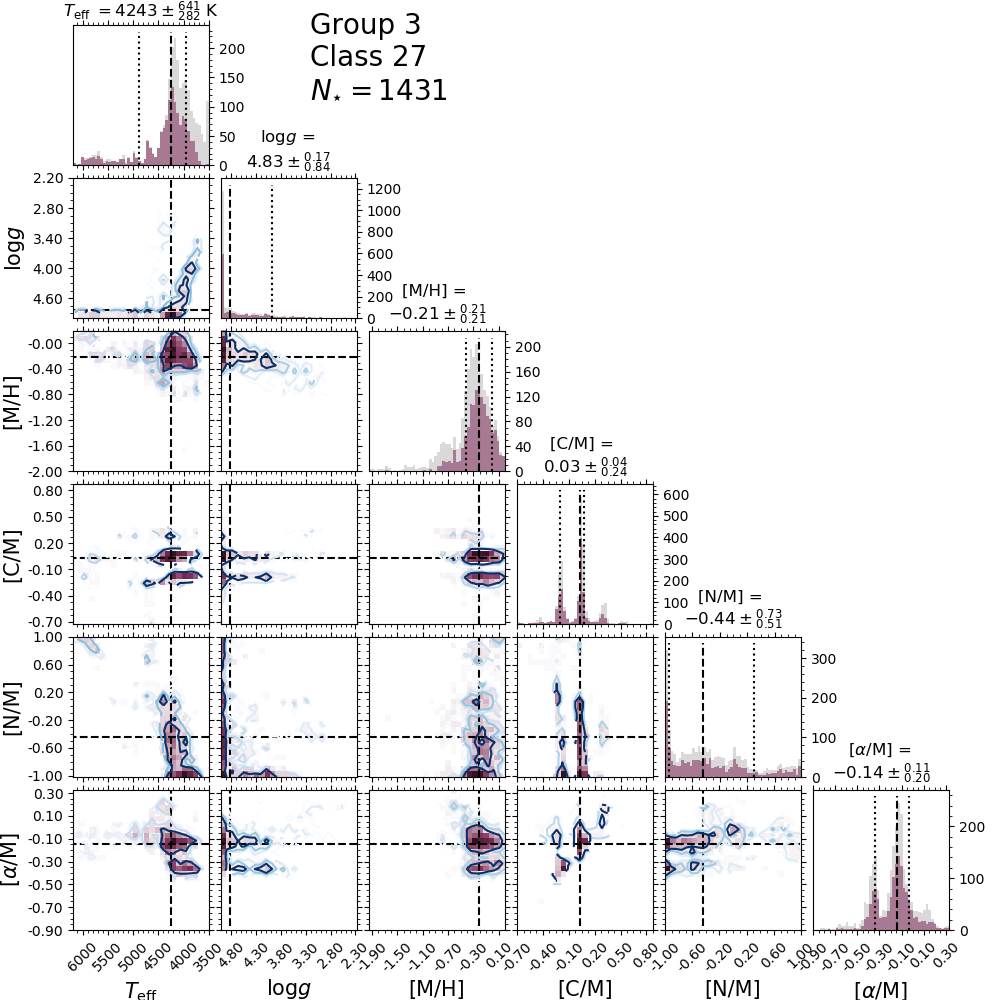}
    \caption{\label{class27} This figure folows the same pattern  from
    figure \ref{class00}. All the classes are described at Table  \ref{tab:desc}.}
    \end{figure*}

\begin{figure*}
    \centering
    \includegraphics[width=\textwidth]
        {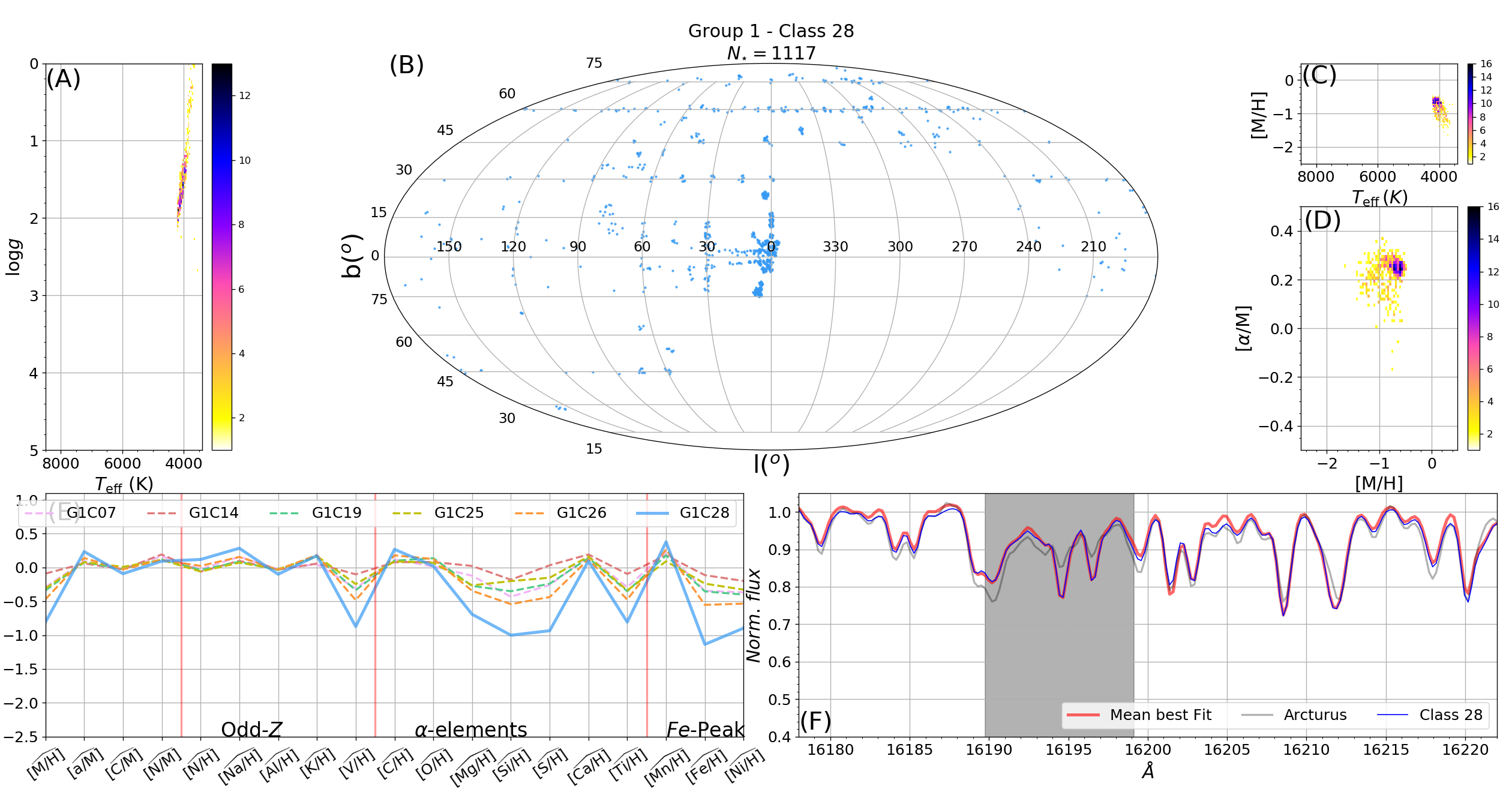}
    \includegraphics[width=0.7\textwidth]
        {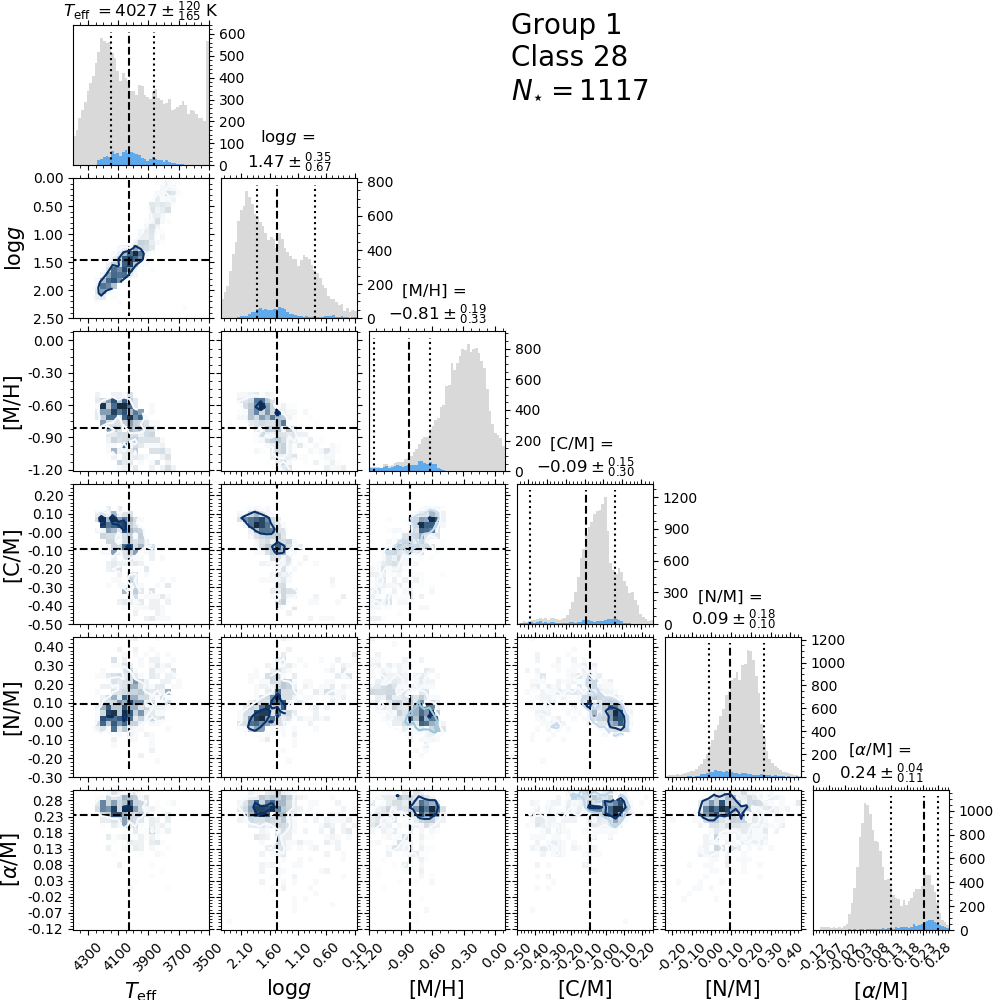}
    \caption{\label{class28} This figure folows the same pattern  from
    figure \ref{class00}. All the classes are described at Table  \ref{tab:desc}.}
    \end{figure*}

\begin{figure*}
    \centering
    \includegraphics[width=\textwidth]
        {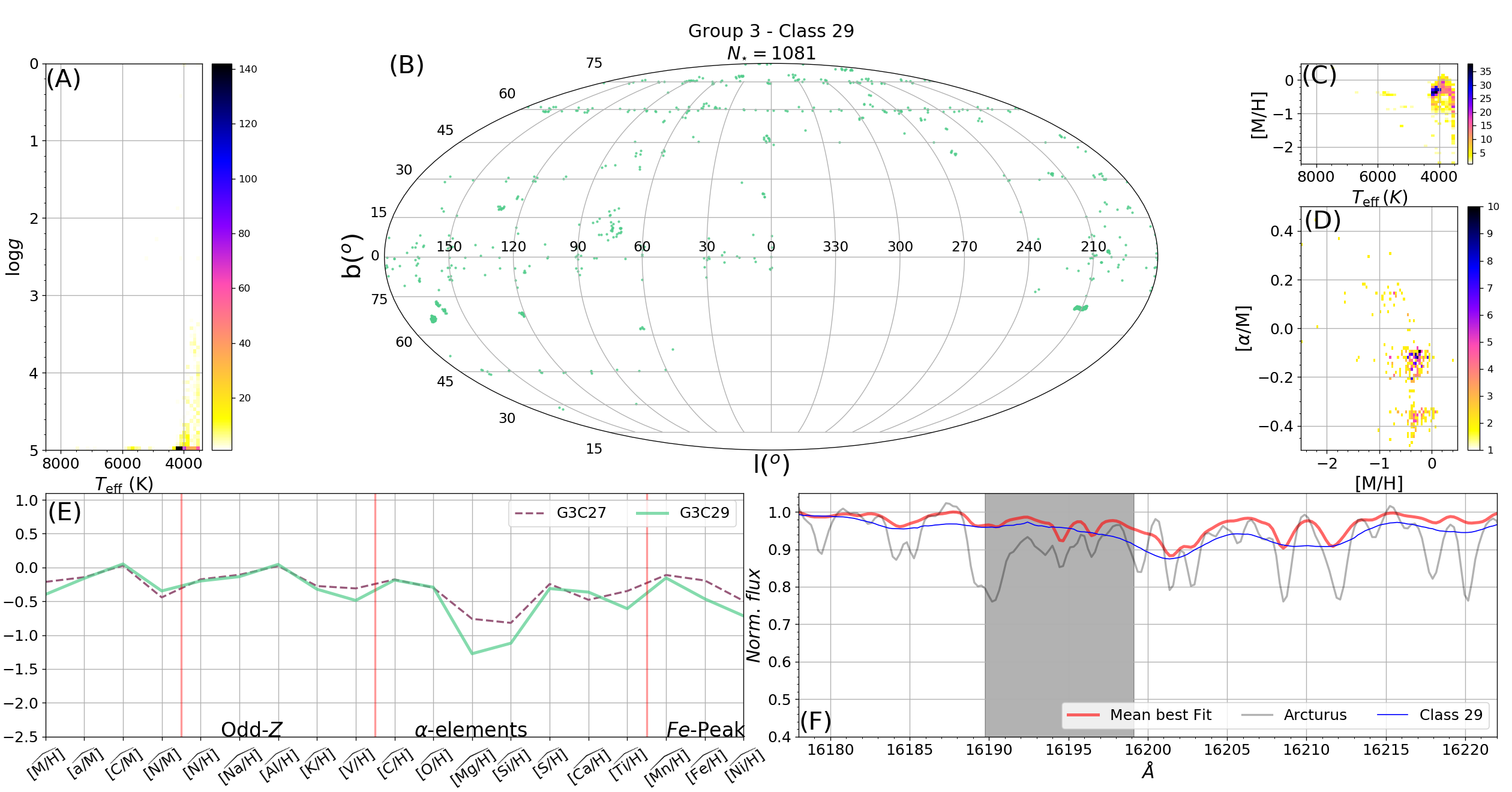}
    \includegraphics[width=0.7\textwidth]
        {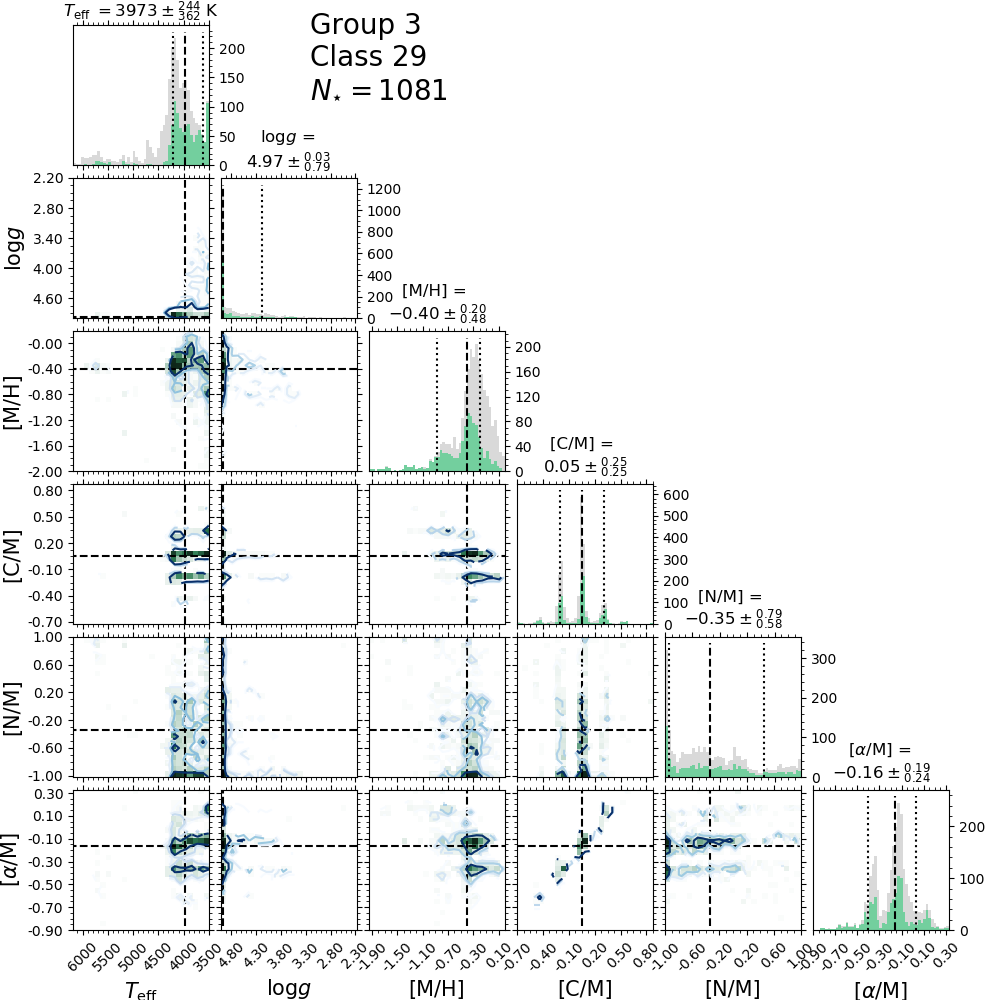}
    \caption{\label{class29} This figure folows the same pattern  from
    figure \ref{class00}. All the classes are described at Table  \ref{tab:desc}.}
    \end{figure*}

\begin{figure*}
    \centering
    \includegraphics[width=\textwidth]
        {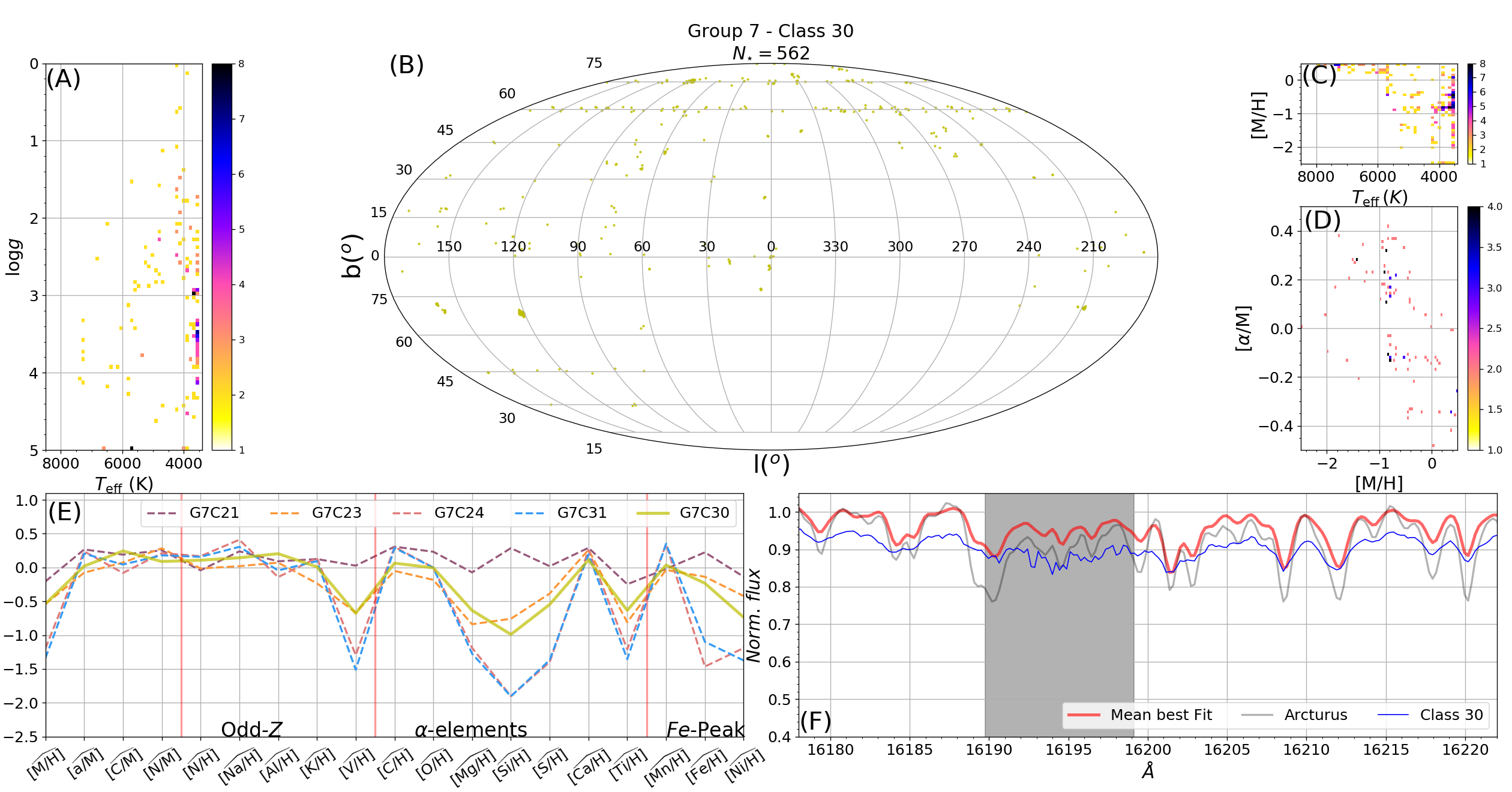}
    \includegraphics[width=0.7\textwidth]
        {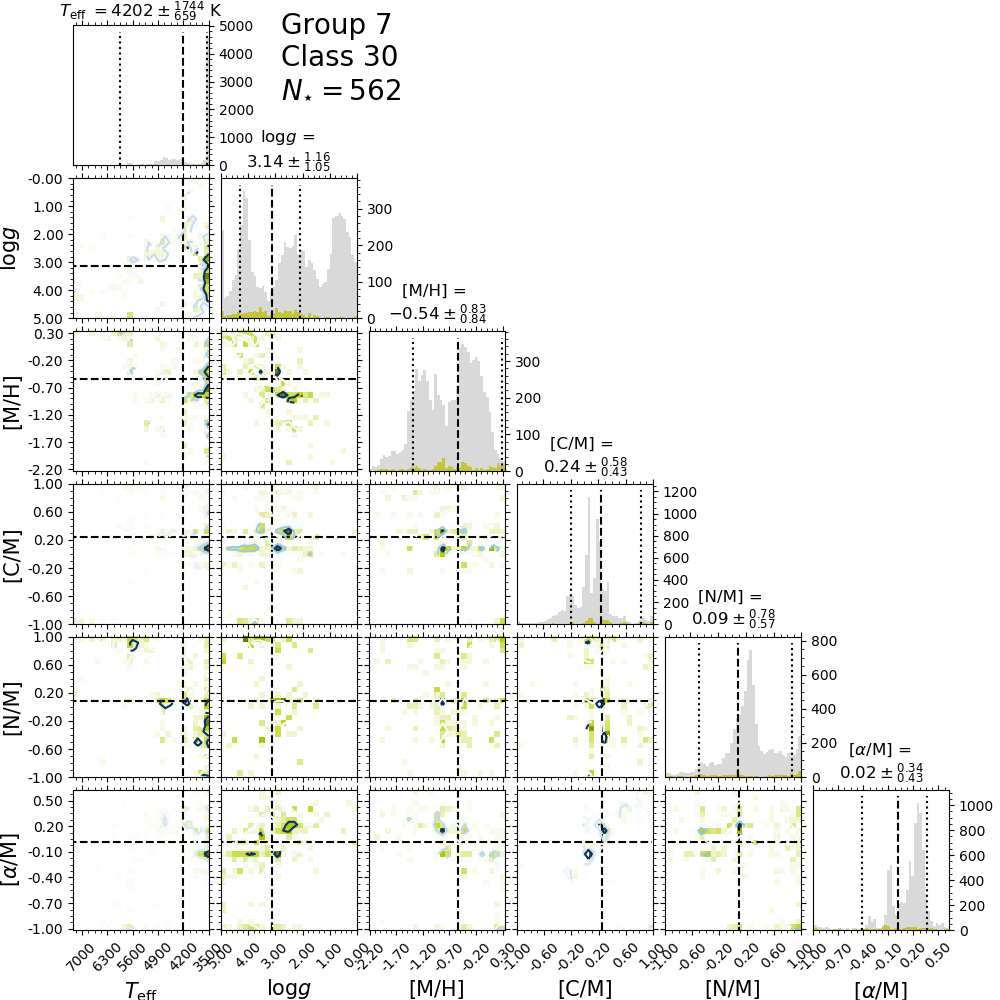}
    \caption{\label{class30} This figure folows the same pattern  from
    figure \ref{class00}. All the classes are described at Table  \ref{tab:desc}.}
    \end{figure*}

\begin{figure*}
    \centering
    \includegraphics[width=\textwidth]
        {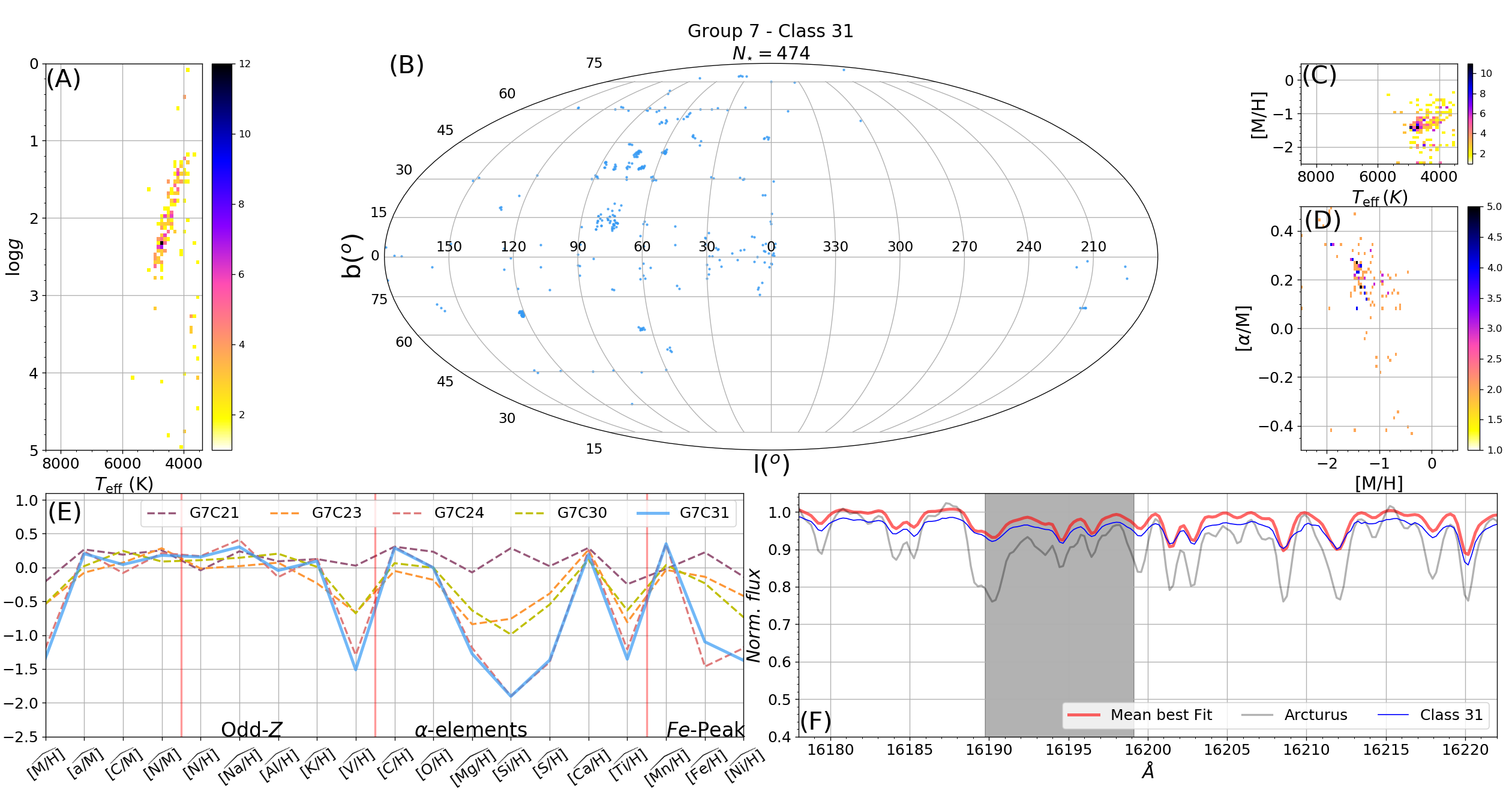}
    \includegraphics[width=0.7\textwidth]
        {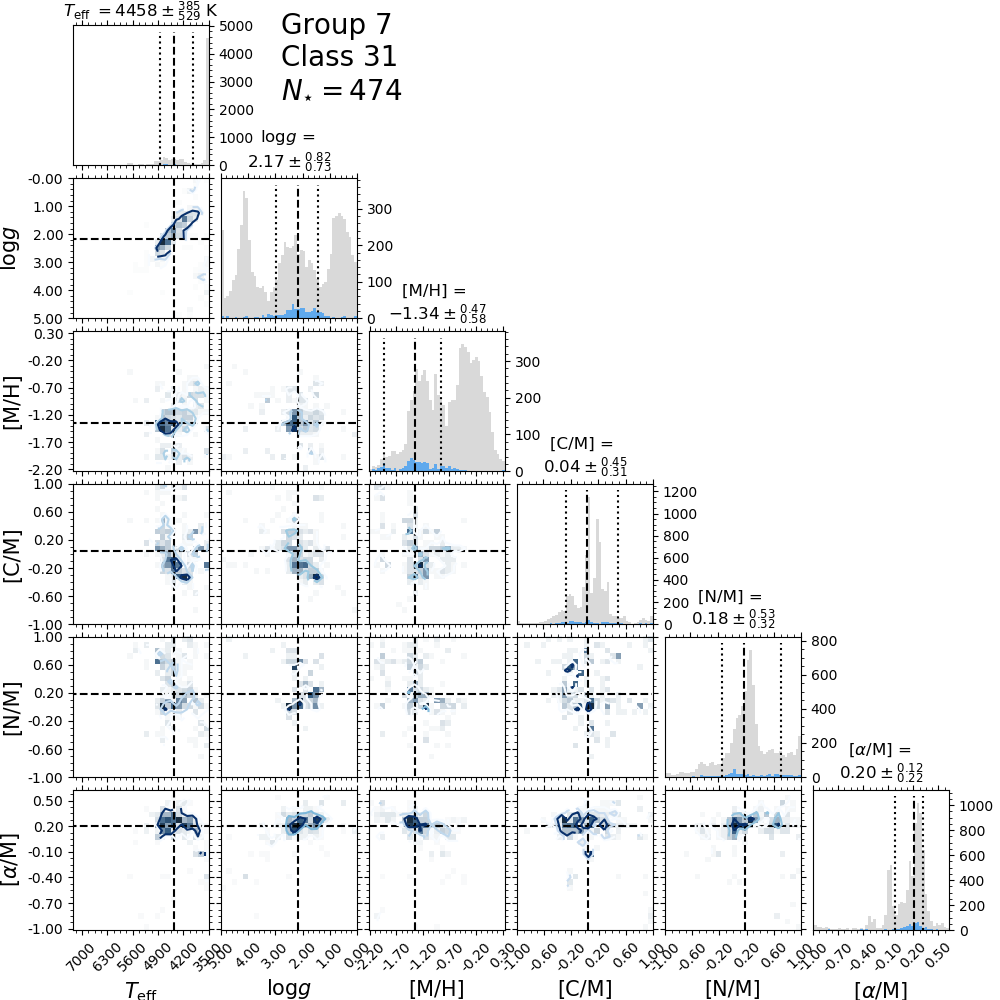}
    \caption{\label{class31} This figure folows the same pattern  from
    figure \ref{class00}. All the classes are described at Table  \ref{tab:desc}.}
    \end{figure*}

\section*{Acknowledgements}

\qquad We thank the help and comments by Diogo Souto and Dante Minniti.

We thank the anonymous referee for useful and precise comments that helped improve the readability of the paper.

We acknowledge financial support through grants, AYA2014-56359-P y AYA2017-86389-P and AYA2016-79724-C4-2-P (MINECO/FEDER). The research that leds to this article was partially funded by the Brazilian National Research Council (CNPq) through scholarship of the CSF program. CAP is thankful to the Spanish Government for funding for his research through grant AYA2014-56359-P. Funding for SDSS-III has been provided by the Alfred P. Sloan Foundation, the Participating Institutions, the National Science Foundation, and the U.S. Department of Energy Office of Science. The SDSS-III web site is http://www.sdss3.org/.

SDSS-III is managed by the Astrophysical Research Consortium for the Participating Institutions of the SDSS-III Collaboration including the University of Arizona, the Brazilian Participation Group, Brookhaven National Laboratory, Carnegie Mellon University, University of Florida, the French Participation Group, the German Participation Group, Harvard University, the Instituto de Astrofisica de Canarias, the Michigan State/Notre Dame/JINA Participation Group, Johns Hopkins University, Lawrence Berkeley National Laboratory, Max Planck Institute for Astrophysics, Max Planck Institute for Extraterrestrial Physics, New Mexico State University, New York University, Ohio State University, Pennsylvania State University, University of Portsmouth, Princeton University, the Spanish Participation Group, University of Tokyo, University of Utah, Vanderbilt University, University of Virginia, University of Washington, and Yale University.

\bibliographystyle{aa}
\bibliography{k-means}

\end{document}